\documentclass[11pt]{article}
\usepackage[colorlinks,
            linkcolor=red,
            anchorcolor=blue,
            citecolor=blue]{hyperref}
\usepackage{xcolor,amsmath,natbib,mathrsfs,geometry,float,graphicx,subfigure,booktabs,amssymb,multirow,indentfirst,tikz,bm}
\allowdisplaybreaks[4] 
\title{Likelihood Ratio test for Poisson graph}
\author{
  Shuyan Chen\textsuperscript{a}, 
  Xin Liu\textsuperscript{b}, 
    Shaoli Wang\textsuperscript{b} 
    \thanks{
    The corresponding authors.
    Email: 
    shufe.shaoli@gmail.com.\\
      All authors equally contribute to this paper and  are  ordered alphabetically.
    }
    \\
{\small$^{a}$School of Management,
University of Science and Technology of China
}
\\
{\small$^{b}$School of Statistics and Data Science,
Shanghai University of Finance and Economics
}
}
\newtheorem{thm}{\sc Theorem}[section]

\newtheorem{cor}{\sc Corollary}[section]
\newtheorem{lem}{\sc Lemma}
\newtheorem{rema}{\sc Remark}[section]
\newtheorem{assum}{\sc Assumption}[section]
\usetikzlibrary{positioning,shapes,arrows}
\newtheorem{prf}{Proof}
\begin{document}
\maketitle
\begin{abstract}
Directed acyclic graphs are widely used to describe the causal effects among random variables, and the inference of those causal effects has become an popular topic in statistics and machine learning, and has wide applications in neuroinformatics, bioinformatics and so on. However, most studies focus on the estimation or inference of the directional relations among continuous random variables, those among discrete random variables have not gained much attentions. In this article we focus on the inference of directed linkages and directed pathways in a Poisson directed graphical model. We employ likelihood ratio tests subject to non-convex acyclicity constraints, and derive the asymptotic distributions of the test statistic under the null hypothesis is true in high-dimensional situations. The power analysis and simulations suggest that the tests achieve the desired objectives of inference. An analysis of a basketball statistics dataset of NBA players during 2016-2017 season illustrates the utility of the proposed method to infer directed linkages and directed pathways in player's statistics network.

\textbf{Keywords:} Poisson directed acyclic graph, Constrained likelihood ratio test, Causal Inference, Non-convex optimization.
\end{abstract}

\section{Introduction}
The conditional independence relations or directional pairwise relations among a set of random variables can be displayed by the structure of a directed acyclic graph(DAG), as in Drug Response Networks of Tumor Cell Lines analysis \citep{Han_2016} and in T cell signaling network analysis \citep{Gu_2018}. The inference of directional connectivity has attract more and more attentions from many fields. In this paper, we focus on the  inference of directional pairwise relations for multivariate count data which are common in many fields such as social networks epidemics data, sports data and so on. 

The learning of DAG structures for both continuous and discrete has received more and more attentions over these years, Assuming a given natural ordering among the nodes, \cite{sho_2010} decomposed DAG estimation into a sequence of lasso regression problems. \cite{sch_2006} and \cite{sch_2007} employed $ \ell_1 $ regularization to learn the structure of DAGs. \cite{fu_2013} developed an $ \ell_1 $-penalized likelihood approach to learn the structure of sparse DAGs from Gaussian data without assuming a given ordering. This method has been further generalized to the use of concave penalties by \cite{ar_2015}. \cite{Han_2016} proposed Two-stage Adaptive Lasso method to Estimated Directed Acyclic Graphs where the first stage selects probabilistic neighborhood, and the second stage estimates the DAG within the identified probabilistic neighborhood. \cite{Gu_2018} developed a maximum penalized likelihood method to estimation the Directed Acyclic Graphs for Discrete Data. \cite{park_2019} developed MRS algorithm to learning High-Dimensional Poisson Structural Equation Model.
However, the study of significance testing of directed relations in DAG models remains largely unexplored, especially for discrete counting data.
 In this article, as \cite{chun_2020}, we propose a constrained likelihood ratio test subjecting to non-convex constraints to regulatory nuisance parameters that are not being tested.
 \cite{chun_2020} developed constrained likelihood ratio test for Gaussian random variables and formed linkage testing and directed pathways for directional pairwise relations.
 When it comes to the discrete data, there are more challenges emerge when deriving  the  sampling asymptotic distribution of likelihood ratio under high-dimensional setting. 

Another challenge facing in learning and making inference on DAGs are notoriously difficult problems due to the identifiability issue where the true graph may not be uniquely determined by the observed variables, since most Markov equivalence classes contain more
than one graph. Many works have been done to identify the identifiable of DAG
models by placing a different type of restrictions on the distributions. \cite{shimizu_2006} showed that linear non-Gaussian additive noise models can be identifiable. \cite{hoyer_2009} and \cite{peters_2011} proved the identifiability of nonlinear models.
\cite{park_2019} proved the identifiability of DAG
models where a higher order moment of the conditional distribution of each node given its parents is Poisson, and \cite{PB_2014} proved that (Gaussian) linear structural equation models (SEMs) with equal or known error variances are identifiable. More recently, \cite{Ghoshal_2018} and \cite{park_2020} showed
that Gaussian linear SEMs with heterogeneous error variances can be identifiable. 

In this work, we use constrained likelihood ratio test to make inference on the relations of interest through testing directed linkages and testing of directed pathways of Poisson networks under the acyclic constraints. We then give the identifiable of our proposed method under causal minimality condition. We employed an Quadratic approximation of our parameters to establish the asymptotic distribution of our likelihood ratio in high-dimension situations and an ADMM algorithm to compute the constrained likelihood ratio linkage and pathway tests.
 
 The rest of the article is organized as follows. Section 2 briefly decribes the Possion regression for DAG models and the acyclicity constraints related to DAGs. Section 3 develops two types of likelihood ratio test for graph linkages and directed pathways, and the assumptions needed to derive the sampling asymptotic distribution of test statistics. Section 4 develops in detail the Algorithm to our optimization problem under constraints. Section 5 reports numerical results and results on real networks. In the Appendix, we establish some asymptotic properties and power of the proposed likelihood ratio tests.
 
\section{Poisson Directed Acyclic Graph}
For a Graph $\mathcal{G}=(V,E)$ with a group of discrete random variables ${X}=\{X_1,...,X_p\}$. For variable $X_j,j=1\dots,p$, we use $V_j$ to represent the parents set of variable ${X_j}$, where $V_j=\{i: X_i\to X_j, i=1,\dots,p\}  $ then the conditional distribution of ${X_j}$ given $X_{V_{j}}$ is 
\begin{equation}\label{eq:generate}
X_j \mid \{X_i,i\in V_j\} \sim \text{ Poisson}(\lambda_j),
\end{equation}
where $\lambda_j$ depending on $X_{V_j}$ with the form
\[
\log \lambda_j=\log E\left( X_j\mid\{X_i=x_i,i\in V_j\}\right) =\beta_{0j}+\sum_{i \in V_j}x_i\beta_{ij}=\bm{x}\boldsymbol{\beta_{\cdot j}},
\]
where $ \bm{x}=(1,x_1,\dots,x_p) $ and $ \boldsymbol{\beta_{\cdot j}}=(\beta_{0j},\beta_{1j},\dots,\beta_{pj})^T $ is the coefficents for $ j^{th} $ variable with $ \beta_{kj}=0 $ for all $ k\notin V_j $.

Then the conditional distribution of ${X_j}$ given its parents set can be rewritten as :
\[
P(X_j=x_j \mid \{X_i,i\in V_j\})=\exp(-e^{\bm{x}\boldsymbol{\beta_{\cdot j}}})\frac{\exp\left[x_j(\bm{x}\boldsymbol{\beta_{\cdot j}})\right]}{x_j!},
\]
and the joint distribution can be represented by
\begin{equation}\label{joint}
   \mathcal{L}(X)=\prod_{j=1}^{p} P(X_j=x_j \mid \{X_i,i\in V_j\}).
\end{equation}
Given the sample $\{{X}_{hj}\}_{n\times (p+1)}$ with first column 1's and ${X}_{hj}$ representing the $h^{th}$ observation of variable $ X_j $, we have the likelihood function corresponding to graph $\mathcal{G}$:
\[
L(\boldsymbol{\beta})=\prod_{h=1}^{n}\prod_{j=1}^{p}P\left(X_{hj}=x_{hj} \mid \{X_{hi}=x_{hi} ,i\in V_j\}\right).
\]
Thus the log-likelihood fuction:
\begin{equation}\label{loglike}
l(\boldsymbol{\beta})=\sum_{h=1}^{n}\sum_{j=1}^{p}\left[-\exp(X_{h\cdot}\boldsymbol{\beta}_{\cdot j})+x_{hj} ({X_{h\cdot}}\boldsymbol{\beta}_{\cdot j})-\log(x_{hj}!)\right]. 
\end{equation}
where $ \boldsymbol{\beta} $ is estimated from \eqref{loglike} subject to the requirement that $ \boldsymbol{\beta} $ defines a DAG, or a directed graph without directed cycles. This enables us to identify all directional pairwise relations simultaneously by identifying nonzero off-diagonals of $ \boldsymbol{\beta} $.

According to \cite{yuan_2019}, $ \boldsymbol{\beta} $ renders a DAG if the following acyclicity constraints are satisfied:
\begin{equation}\label{acyclic}
\begin{array}{l}
\sum_{j_{1}=j_{L+1}:1 \leq i \leq L} \mathbb{I}\left(\boldsymbol{\beta}_{j_{i-1} j_{i}} \neq 0\right) \leq L-1\\
\text{for any }  \left(j_1,\cdots,j_L \right), L=2,\cdots,p,
\end{array}
\end{equation}
which ensures the local Markov property defining directional pairwise relations \citep{ed_2012}. It follows from Theorem of \cite{yuan_2019} that after introducing nonnegative dual variables $ (\gamma_{ij})_{p\times p} \in \mathbb{R}^{p\times p}$, we obtain an equivalent form of 	\eqref{acyclic},
\begin{equation}\label{equiv}
\gamma_{ki}+I(i\neq j)-\gamma_{ij}\geq I(\boldsymbol{\beta}_{jk}\neq 0) \quad (i,j,k=1,\dots,p,k\neq j). 
\end{equation}

To treat nonconvex constraints in \eqref{equiv}, we replace the corresponding indicator functions in \eqref{equiv} by its computational surrogate $ J_{\tau}(z)=\min(,\frac{|z|}{\tau},1)  $ in \cite{shen_2012}  where $ \tau $ is a small tuning parameter controlling the degree of approximation
in that $ J_{\tau}(z) $ approximates the indicator function as $ \tau\to 0^{+} $.
This yields that
\begin{equation}\label{cstr}
\gamma_{ki}+I(i\neq j)-\gamma_{ij}\geq J_{\tau}(\boldsymbol{\beta}_{jk}),
\end{equation}
and $ J_{\tau} $ can be divided into a difference of two convex functions
\[
J_{\tau}(z)=\frac{|z|}{\tau}-\max(\frac{|z|}{\tau}-1,0)\equiv f_1(z)-f_2(z).
\]
Then under $ H_0 $ and $ H_{\alpha} $ we can estimate $ \boldsymbol{\beta} $ by
\[
\begin{array}{l}
\hat{\boldsymbol{\beta}}^{H_0}=\arg\max l(\boldsymbol{\beta}), \text{ subject to }\eqref{cstr}, \boldsymbol{\beta}_{F}=0,\quad \sum_{(j,k)\in F^c} J_{\tau}(|\boldsymbol{\beta}_{jk}|)\leq \kappa;\\
\hat{\boldsymbol{\beta}}^{H_{\alpha}}=\arg\max l(\boldsymbol{\beta}), \text{ subject to  }\eqref{cstr},\quad \sum_{(j,k)\in F^c} J_{\tau}(|\boldsymbol{\beta}_{jk}|)\leq \kappa,
\end{array}
\]
where $ \kappa\geq 0 $ is an integer-valued tuning parameter and $ (\lambda_{ij})_{p\times p} $ is defined after \eqref{cstr}.

\begin{thm}\label{thm:iden}
   Let $\ell(\boldsymbol{\beta})$ be generated from model \eqref{eq:generate} with directed acyclic graph $\mathcal{G}_0$, then $\mathcal{G}_0$  is identifiable from $\ell(\boldsymbol{\beta})$ and the coefficients $\beta_{jk}$ can be reconstructed for all $j$ and $k\in V_j$. 
\end{thm}
\begin{rema}
The idea of the proof is to assume that there are two structural equation models with distinct graphs $\mathcal{G}$ and $\mathcal{G}_0$ that lead to the same joint distribution as displayed in Figure \ref{fig:h} with reverse edges between variables $X_1$ and $X_3$. We can  show that $X_3$ has different
variances in both graphs. This leads to a contradiction.
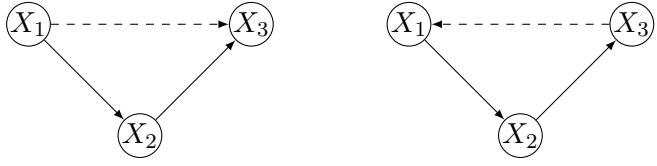
\begin{figure}
    \centering
    \begin{tikzpicture}[
          node distance=0.5cm,
          mynode/.style={circle, draw = black, inner sep = 0pt, minimum size = 0.5cm}
       ]
      \node[mynode] (C) {$X_2$};
      \node[mynode, above left= 1.5 cm
      of C] (A) {$X_1$};
      \node[mynode, above right= 1.5 cm of C] (D) {$X_3$};

      \node[mynode, right= 1.5 cm of D] (E) {$X_1$};

      \node[mynode, below right= 1.5 cm of E] (F) {$X_2$};

      \node[mynode, above right = 1.5 cm of F] (G) {$X_3$};
      \path [draw, -latex] (A) edge (C);
      \path [draw, -latex] (C) edge (D);
      \path [draw, -latex][dashed] (A) edge (D);

      \path [draw, -latex] (E) edge (F);
      \path [draw, -latex] (F) edge (G);
      \path [draw, -latex][dashed] (G) edge (E);

      \end{tikzpicture}
    \caption{\label{fig:h}Two Poisson graphs $\mathcal{G}$ and $\mathcal{G}^{\prime}$ with different edges.}
\end{figure}

\end{rema}

\section{Likelihood Ratio Tests for linkages and pathways}
Here we develop two types of tests concerning directional pairwise relations.
\subsection{Test of Graph Linkages}
Define an index set $ F \subseteq \{1,...,p\}^2 $, where an index $ (i,j)\in F $ represents a directed connection from variables $ X_i$ and $ X_j $, or node $ X_i$ is a parent of node $ X_j $ for the DAG $ \mathcal{G} $ we define in section 2. Given the index set F, the null hypothesis $ H_0 $ and the alternative $H_{\alpha}$ for testing linkages are
\[
H_0 : {\beta_{ij}}=0; \text{for all }(i,j)\in F~ ~\text{versus}~ ~ H_{\alpha} : \text{not}~ H_0. 
\]
The constrained likelihood ratio for $ H_0 $ versus $ H_{\alpha}$ is $lr=l(\hat{\boldsymbol{\beta}}^{H_{a}})-l(\hat{\boldsymbol{\beta}}^{H_{0}})$,
where 
\begin{equation}\label{linkh0}
\begin{aligned}
\hat{\boldsymbol{\beta}}^{H_{0}}=&\text{arg max}~ l ({\boldsymbol{\beta}})
\text{ subj to  }\eqref{cstr}, \boldsymbol{\beta}_{F}=0,\\ &\quad \sum_{(j,k)\in F^c} J_{\tau}(|\boldsymbol{\beta}_{jk}|)\leq \kappa;\\
\hat{\boldsymbol{\beta}}^{H_{\alpha}}=&\text{argmax } l ({\boldsymbol{\beta}}), \text{ subj to }\eqref{cstr},\\ &\quad \sum_{(j,k)\in F^c} J_{\tau}(|\boldsymbol{\beta}_{jk}|)\leq \kappa.
\end{aligned}
\end{equation}
To derive the asymptotic distribution of $ lr $ under $ H_0 $, we assume $ \boldsymbol{\beta}^0 $ is the ture parameter matrix, and  $ E^0 $ is the set of links in true graph $ \mathcal{G}^{0} $. A link $(i,j)\in F$ called testable when $ \{(i,j)\}\bigcup E^0 $ doesn't contain a directed cycle. Define $ D^0 $ a index set that consists of all testable likes in F.
\begin{assum}[ Degree of separation]\label{asm1}
 For some positive constant $ c_1 $,
\[
\begin{array}{c}C_{\min } \equiv \inf _{\left\{ S_{1} \cup S_{2} \nsupseteq E^{0},\left|S_{1}\right| \leq\left|E^{0} \backslash F\right|, S_{2} \subseteq F, S_{1} \cup S_{2} \text { forms DAG }\right\}} h^{2}\left(\boldsymbol{\beta}_{S_{1} \cup S_{2}}, \boldsymbol{\beta}^{0}\right) \\
\geq  c_{1} n^{-1} \max \left(\left|E^{0}\right|+|F|, \log p\right),\end{array}
\]
where $ h^2 $ is the Hellinger distance between $ \boldsymbol{\beta}$ and $ \boldsymbol{\beta}^0 $.
\[
h^2\left(\boldsymbol{\beta},\boldsymbol{\beta}^0\right)  \stackrel{\vartriangle}{=}1-\prod_{j=1}^{p}\exp\left(-\frac{1}{2}\left(e^{\frac{1}{2}X\boldsymbol{\beta}_{\cdot j}}-e^{\frac{1}{2}X\boldsymbol{\beta}^{0}_{\cdot j}} \right)^2 \right).
\]
\end{assum}
We have that 
\[
h^{2}\left(\boldsymbol{\beta}^{\tau}, \boldsymbol{\beta}^{0}\right)\geq h^{2}\left(\boldsymbol{\beta}, \boldsymbol{\beta}^{0}\right)-h^{2}\left(\boldsymbol{\beta}, \boldsymbol{\beta}^{\tau}\right),
\]
where $ \boldsymbol{\beta}^{\tau} $ is defined by $ \left( {\boldsymbol{\beta}}^{\tau}\right)_{jk}= {\beta}_{jk}\mathbb{I}(|{\beta}_{jk}|>\tau) $ for $ (j,k)\in F^c $ and $ \left( \boldsymbol{\beta}^{\tau}\right)_{jk}={\beta}_{jk}  $ for $ (j,k)\in F $.
\begin{assum}[Restriction of the index set]\label{asm2}
Assume that under $ H_0 $
\[
\frac{|D^0|^{1/2}|E^0\cup D^0|\log p}{n}\to 0, \text{ as } n\to \infty,
\]
where $ |\cdot| $ denotes the size of the set. This assuption permits $ p$ exceeding $ n $ provided that the number of testable linkages in $ F $ and nonzero nuisance links under $ H_0 $ are not too large
\end{assum}
\begin{assum}[Boundedness of the parameter space]\label{asm3}
	 For some constant $ c_2>0 $ and $ c_3>0 $,
	\[
	c_2\leq \frac{1}{n}\lambda^TX^T X\lambda\leq c_3,
	\]
	where $ \lambda\in \mathbb{R}^p $ with $ ||\lambda||^2_2=1 $
\end{assum}
\begin{assum}[Approximation]\label{asm4}
	For some positive constants c4–c6,
	\[
	c_{6} p \tau^{c_{5}}\geq c_{4} h^{2}\left(\boldsymbol{\beta}, \boldsymbol{\beta}^{0}\right)
	-h^{2}\left(\boldsymbol{\beta}^{\tau}, \boldsymbol{\beta}^{0}\right).
	\]	
\end{assum} 
We have that 
\[
\left|\boldsymbol{\beta}-\boldsymbol{\beta}_{\tau} \right|_1=\sum_{(j,k)\in F^c,|\beta_{jk}\leq \tau|} |\beta_{jk}|\leq \kappa \tau,
\]
\[
\left|h^{2}\left(\boldsymbol{\beta}, \boldsymbol{\beta}^{0}\right)
-h^{2}\left(\boldsymbol{\beta}^{\tau}, \boldsymbol{\beta}^{0}\right)\right|\leq \left|\sum_{(j,k)\in F^c}\frac{\partial h^{2}\left(\boldsymbol{\beta}, \boldsymbol{\beta}^{0}\right)}{\partial \beta_{jk}}\Big|_{\boldsymbol{\beta}=\boldsymbol{\beta}^{*}}\right|_{\max}\left|\boldsymbol{\beta}-\boldsymbol{\beta}_{\tau} \right|_1\leq \kappa\tau, 
\]	
where
\begin{align*}
\left|\frac{\partial h^{2}\left(\boldsymbol{\beta}, \boldsymbol{\beta}^{0}\right)}{\partial \beta_{jk}}\right|=&\left|\exp\left( \sum_{j=1}^{p}-\frac{1}{2}\left(e^{\frac{1}{2}X\boldsymbol{\beta}_{\cdot j}}-e^{\frac{1}{2}X\boldsymbol{\beta}^{0}_{\cdot j}} \right)^2 \right) \frac{1}{2}X_{k}e^{\frac{1}{2}X\boldsymbol{\beta}_{\cdot j}}\left(e^{\frac{1}{2}X\boldsymbol{\beta}_{\cdot j}}-e^{\frac{1}{2}X\boldsymbol{\beta}^{0}_{\cdot j}} \right)\right|\\
\leq&\left|\frac{1}{2}X_{k}e^{\frac{1}{2}X\boldsymbol{\beta}_{\cdot j}}\left(e^{\frac{1}{2}X\boldsymbol{\beta}_{\cdot j} }-e^{\frac{1}{2}X\boldsymbol{\beta}^{0}_{\cdot j}} \right)\right|.
\end{align*}
\begin{thm}[Sampling distribution of lr for testing linkages under $ H_0 $]\label{thm1}
Assume that assumptions \ref{asm1}, \ref{asm2}, \ref{asm3}, \ref{asm4} are met. Then there exists tuning parameters $ \kappa=|E^0\setminus F| $ and $ \tau<C_{\min}c_2/4p $ such that under $ H_0 $, as $ n\to \infty $,
\[
\begin{array}{ll}
(i)P(lr=0)\to 1,& \text{ if } |D^0|=0,\\
(ii)2lr\xrightarrow{d}\chi^2_{|D^0|},& \text{ if }|D^0|>0  \text{ is fixed},\\
(iii)(2|D^0|)^{-1/2}(2lr-|D^0|)\xrightarrow{d}N(0,1),&\text{ if }|D^0|\to\infty.
\end{array}
\]
\end{thm}
The degrees of freedom$ |D^0| $, as specified in Assumption \ref{asm4}, is usually unknown, and hence that we propose an estimate $ \hat{D}^0 $ of $ D^0 $, defined as the largest subset of $ F $ such that $ F\cup \hat{E}^0 $ forms a DAG. Here $ \hat{D}^0 $ is estimated based on an estimate $ \hat{\boldsymbol{\beta}}^0_{H_0} $ under $ H_0 $ from \eqref{linkh0}. The next corollary says that $ |\hat{D}0| $ consistently estimates $ |D^0| $.

\begin{cor}[Substitution of $ D^0 $ by $ \hat{D}^0 $]\label{cor1}
	Under the assumptions of Theorem \ref{thm1}, $ P(|\hat{D}^0| = |D^0|) \to 1 $, as $ n\to\infty$. Then the result of Theorem \ref{thm2} continues to hold with $ D^0 $ replaced by $ \hat{D}^0 $.
\end{cor} 

\subsection{Test of a Direct Pathway}
A direct pathway is specified by an index set $ F $ in a consecutive manner, where a common segment is shared by two consecutive indices of $ F =\left\lbrace(i_1,i_2),(i_2,i_3),\cdots,(i_{|F|},i_{|F|+1}) \right\rbrace $, for instance, $ (i_1,i_2) $ and $ (i_2,i_3) $ are shared by $ i_2 $. Now consider
\[
H_0:\boldsymbol{\beta}_{ij}=0; \text{ for some } (i,j) \in F \text{ versus } H_{\alpha} ,
\]
with unspeicified nuisance parameters $ \boldsymbol{\beta}_{jk} ;(j,k)\in F^c$. Rejection of $ H_0 $ suggests the presence of a directed pathway specified by $ F $.

In this situation, the estimation of parameters are different form that in linkages test. Now the constrained likelihood ratio statistic for $ H_0 $ versus $ H_{\alpha} $ is modified to account for a directed pathway at some indices: $ lr=l(\hat{\boldsymbol{\beta}}^{H_{\alpha}})-\max_{k=1}^{|F|}l\left( \hat{\boldsymbol{\beta}}^{H_{0}}(k)\right)  $, where
\begin{equation}\label{pathh0}
 \begin{aligned}
 \hat{\boldsymbol{\beta}}^{H_{0}}(k)=&\text{arg max}~ l ({\boldsymbol{\beta}})
 \text{ subject to  }\eqref{cstr}, \boldsymbol{\beta}_{i_k,i_{k+1}}=0;(i_k,i_{k+1})\in F,\\
 & \sum_{(j,k)\in F^c}J_{\tau}(\boldsymbol{\beta}_{jk})\leq \kappa,\\
 \hat{\boldsymbol{\beta}}^{H_{\alpha}}=&\text{arg max}~ l ({\boldsymbol{\beta}})
 \text{ subject to }\eqref{cstr},  \sum_{(j,k)\in F^c}J_{\tau}(\boldsymbol{\beta}_{jk})\leq \kappa.
 \end{aligned}
\end{equation}
Next Assumption is a modified version of Assumption \ref{asm4}.
\begin{assum}[Restriction of the index set $ E^0 $]\label{asm5}
	\[\frac{|E^0|\log p}{n}\to 0,\quad \text{ as } n\to \infty\].
\end{assum} 
 \begin{thm}[Sampling distribution of $ lr $ for testing a directed pathway under $ H_0 $]\label{thm2}
 	Under Assumptions \ref{asm1}, \ref{asm2}, \ref{asm3}, and \ref{asm5}, there exist tuning parameters $ \kappa= |E^0 \setminus F| $ and $ \tau \leq C_{\min}c_1/4p $ such that under $ H_0 $, as $ n\to\infty $,
 	\[
 	\begin{array}{ll}
 	(i)P(lr=0)\to 1& \text{ if } E^0 \cup F  \text{ does not form a DAG;}\\
 	(ii)2lr\xrightarrow{d}\min\{X_1,\cdots,X_d\}& \text{ if } E^0 \cup F  \text{ form a DAG and d is fixed;}\\
 	(iii) 2d^2lr\to \Gamma & \text{ if } E^0 \cup F \text{ does not form a DAG but } d\to \infty.
 	\end{array}
 	\]
 	where $ d=|\{(i,j)\in F:\boldsymbol{\beta}_{ij}^0=0\}| $ is the number of breakpoints in the hypothesized pathway, $ X_1,\cdots,X_d $ are independently identically distributed $ \chi^2_1 $ variables, and $ \Gamma $ is the generalized Gamma distribution with density $ \sqrt{\frac{1}{2\pi x}}\exp(-\sqrt{2x/\pi}) $ for $ x>0 $. 
 \end{thm}
The degrees of freedom $ d $ is estimated by $ \hat{d}=\max\left( \left|\left\{(j,k)\in F:\hat{\boldsymbol{\beta}}_{jk}=0\right\}\right|,1\right)  $, where $ \hat{\boldsymbol{\beta}} $ is the constrained maximum likelihood estimate under $ H_{\alpha} $ but with $ F =\emptyset $. The next corollary says that $ \hat{d} $ consistently estimates $ d $.

\begin{cor}[Substitution of d by $ \hat{d} $]\label{cor2}
	Under the assumptions of Theorem \ref{thm2}, $ P(\hat{d}= d)\to 1 $ as $n \to \infty$. Then the result of Theorem \ref{thm2} continues to hold with $ d $ replaced by $ \hat{d} $.
\end{cor}
\subsection{Power analysis}
For testing linkages, we consider a local alternative $ H_{\alpha}: \boldsymbol{\beta}_{ij}=\boldsymbol{\beta}^0_{ij}+\boldsymbol{\delta}_{ij}^n  $ for $ (i,j)\in F $, where $ \boldsymbol{\delta}^n=(\delta_{ij}^n)_{p\times p} $ satisfies: $ \boldsymbol{\delta}_{F^c}^n=0 $, and $ ||\boldsymbol{\delta}^n||_F=n^{-1/2}h $ if $ |D^0| $ is fixed and $ ||\boldsymbol{\delta}^n||_F=|D^0|^{1/4}n^{1/2}h $ if $ |D^0|\to \infty $, where $ ||X||_F\equiv \sqrt{\sum_i\sum_j x_{ij}^2} $ is the Frobenius norm of a matrix. When $ |D^0|=0 $, then for any $ E\subset E^0 $, $ E\cup E^0 $ is cyclic and we set $ \boldsymbol{\beta}^n=\boldsymbol{\beta}^0 $, and when $ |D^0| \neq 0$ we set $ \boldsymbol{\beta}^n=\boldsymbol{\beta}^0+\boldsymbol{\delta}^n$. Define the power function as $ \pi_n(h) $, then
\[
\begin{array}{ll}
(i)\pi_n(h)=1&\text{ if}|D^0|=0 \text{ and } h=0,\\
(ii)\pi_n(h)=P_{\boldsymbol{\beta}^n}\left(2lr\geq \chi^2_{|D^0|,1-\alpha} \right)& \text{ if }|D^0|>0 \text{ is fixed, } \\
(iii)\pi_n(h)=P_{\boldsymbol{\beta}^n}\left(2|D^0|^{-1/2}\left( 2lr-|D^0|\right) \geq z_{1-\alpha} \right)& \text{ if }|D^0|\to\infty.
\end{array}
\]

\begin{thm}[ Local limiting power for graph linkages]\label{thm3}	
For any $ \boldsymbol{\beta} $ satisfying Assumptions \ref{asm1}, \ref{asm2}, \ref{asm3}, \ref{asm4} such that $ \boldsymbol{\beta}^n $ indices a DAG, we have that $ \lim_{h\to \infty}\lim_{n\to \infty} \pi_n(h)=1$.
\end{thm}
For testing a directed pathway, we consider a local alternative $ H_{\alpha}: \boldsymbol{\beta}_{ij}=\boldsymbol{\beta}_{ij}^0+\delta_{ij}^n $ for $ (i,j)\in A $, where $ |\delta_{ij}^n|=\frac{h}{\sqrt{n}} $ for $ (i,j)\in A $ and $ A=\{(i,j)\in F:\boldsymbol{\beta}^0_{ij}=0\} $. Then the power function $ \tilde{\pi}_n(h) $ is
\[
\begin{array}{ll}
(i)\tilde{\pi}_n(h)=1&\text{ if }E\cup F \text{ is cyclic } \text{ and } h=0,\\
(ii)\tilde{\pi}_n(h)=P_{\boldsymbol{\beta}^n}\left(2lr\geq \Gamma_{d,1-\alpha} \right)& \text{ if } E\cup F \text{ is acyclic and } d \text{ is fixed,} \\
(iii)\tilde{\pi}_n(h)=P_{\boldsymbol{\beta}^n}\left( 2d^2lr\geq \Gamma_{1-\alpha} \right)& \text{ if }E\cup F \text{ is acyclic and } d\to \infty.
\end{array}
\]

\begin{thm}[Local limiting power for a directed pathway]\label{thm4}
	 For any $ \boldsymbol{\beta} $ satisfying Assumption \ref{asm1}, \ref{asm2}, \ref{asm3}, \ref{asm5} such that $ \boldsymbol{\beta}^n $ indiuces a DAG, we have that $ \lim_{h\to \infty}\lim\inf_{n\to \infty}\tilde{\pi}_n(h)=1 $.
\end{thm}

\section{Algorithm and optimization}
To solve the nonconvex constrained linkage test and directed pathway test, we relax the nonconvex contraints through a sequence of approximations involving convex constraints, then we solve each convex subproblem using ADMM (alternating direction method of multipliers). For ralaxation of contraints \eqref{acyclic} and $ \sum_{(j,k)\in F^c}J_{\tau}(\boldsymbol{\beta}_{jk})\leq \kappa $, we decomposition $ J_{\tau} $ into a difference of two convex functions 
\[ 
J_{\tau}(z)=S_1(z)-S_2(z)=\frac{|z|}{\tau}-\max(\frac{|z|}{\tau}-1,0), 
\]
then at iteration $ m $, we replace $ S_2 $, with 
\[ 
S_2(z^{(m-1)})+\triangledown S_2(z^{(m-1)})(z-z^{(m-1)}), 
\]
where 
\[
\triangledown S_2(z^{(m-1)})=\frac{\text{Sign}(z^{(m-1)})}{\tau}\mathbb{I}(|z^{(m-1)}|>\tau), 
\]
is a subgradient of $ S_2 $ at $ z^{(m-1)} $, this leads to a convex subproblem at the $ m^{th} $ iteration,
\begin{equation}\label{mth}
	\begin{aligned}
	\max_{\boldsymbol{\beta}}\quad l(\boldsymbol{\beta})=&\sum_{j=1}^{p}\sum_{h=1}^{n}\left[-\exp(\bm{x_{h\cdot}}\boldsymbol{\beta}_{\cdot j})+x_{hj} (\bm{x_{h\cdot}}\boldsymbol{\beta}_{\cdot j})-\log(x_{hj}!)\right]\\
	\text {subj to } \boldsymbol{\beta}_{E_1}&=0,\\
	\sum_{(j,k)\in F^c}|\beta_{jk}|\mathbb{I}&\left(|\beta_{jk}^{(m-1)}|\leq \tau\right)\leq  \tau \left(\kappa-\sum_{(j,k)\in F^c}|\mathbb{I}\left(|\beta_{jk}^{(m-1)}|> \tau\right) \right) ,\\
	\gamma_{ki}+I(i\neq j)&-\gamma_{ij}\geq\frac{|\beta_{jk}|}{\tau}\mathbb{I}\left(|\beta_{jk}^{(m-1)}|\leq \tau\right)+\mathbb{I}\left(|\beta_{jk}^{(m-1)}|> \tau\right)\\
	i,j,k=&1,\dots,p,k\neq j, 
	\end{aligned}
\end{equation}
under that
\begin{align*}
\sum_{(j,k)\in F^c}J_{\tau}(\beta_{jk})=&\sum_{(j,k)\in F^c}\frac{|\beta_{jk}|}{\tau}-\max(\frac{|\beta_{jk}|}{\tau}-1,0)\\
=&\sum_{(j,k)\in F^c}\frac{|\beta_{jk}|}{\tau}-\left(\frac{|\beta_{jk}^{(m-1)}|}{\tau}-1 \right) \mathbb{I}\left(|\beta_{jk}^{(m-1)}|>\tau \right) \\
&-\frac{\text{Sign}(\beta_{jk}^{(m-1)})}{\tau}\mathbb{I}\left(|\beta_{jk}^{(m-1)}|>\tau\right) \left(\beta_{jk}-\beta_{jk}^{(m-1)} \right) \\
=&\sum_{(j,k)\in F^c}\frac{|\beta_{jk}|}{\tau}\mathbb{I}\left(|\beta_{jk}^{(m-1)}|\leq \tau\right)+\mathbb{I}\left(|\beta_{jk}^{(m-1)}|> \tau\right)\\
\leq& \kappa  ,
\end{align*}
and constraint \eqref{cstr},
where $ {E_1}=F $ for $ H_0 $, and $ E_1=\emptyset $ for $ H_{\alpha} $ when testing linkages, $ E_1=\{(i_k,i_{k+1})\} $ for $ H_0 $, and $ E_1=\emptyset $ for $ H_{\alpha} $ when testing a directed pathway. $ \boldsymbol{\beta}^{(m-1)} $ is the solution of (4.1) at $ m^{th} $ iteration.\\
To solve \eqref{mth} we use ADMM algorithm. Note that \eqref{mth} can be transformed into minimization of 
\[
R(\boldsymbol{\beta})=\sum_{j=1}^{p}\sum_{h=1}^{n}\left[\exp({X_{h\cdot}}\boldsymbol{\beta}_{\cdot j})-x_{hj} ({X_{h\cdot}}\boldsymbol{\beta}_{\cdot j})\right].
\]
Now consider an equivalent form of \eqref{mth} by introducing a nonnegative multiplier $ \mu $
\begin{equation}\label{plusu}
\begin{aligned}
& \min_{\boldsymbol{\beta}}\quad R(\boldsymbol{\beta})+\mu\sum_{(j,k)\in F^c}|\beta_{jk}|\mathbb{I}\left(|\beta_{jk}^{(m-1)}|\leq \tau\right)\\
\text {subj to }&
\beta_{jk}\mathbb{I}\left(|\beta_{jk}^{(m-1)}|\leq \tau\right)\leq  \tau\gamma_{ki}+\tau I(i\neq j)-\tau \gamma_{ij}-\tau\mathbb{I}\left(|\beta_{jk}^{(m-1)}|> \tau\right)\\
&\beta_{jk}\mathbb{I}\left(|\beta_{jk}^{(m-1)}|\leq \tau\right)\geq  -\tau\gamma_{ki}-\tau I(i\neq j)+\tau \gamma_{ij}+\tau\mathbb{I}\left(|\beta_{jk}^{(m-1)}|> \tau\right)\\
&\boldsymbol{\beta}_{E_1}=0,\quad  i,j,k=1,\dots,p,k\neq j .
\end{aligned}
\end{equation}
To solve \eqref{plusu}, we introduce a decoupling matrix $ A_{p\times p} $ to seperate its differentiable form non-differentiable parts. In addition, we introduce slack variables $ \xi=\{\xi_{kji}^1,\xi_{kji}^2\}_{p\times p\times p} $ to convert inequality to equality constraints, which is
\begin{align*}
& \min_{\boldsymbol{\beta}}\quad R(\boldsymbol{\beta})+\mu\sum_{(j,k)\in F^c}|A_{jk}|\mathbb{I}\left(|\beta_{jk}^{(m-1)}|\leq \tau\right)\\
\text {subj to }& \boldsymbol{\beta}_{E_1}=0,\boldsymbol{\beta}-A=\bm{0}\\
&A_{jk}\mathbb{I}\left(|\beta_{jk}^{(m-1)}|\leq \tau\right)+\xi_{ijk}^1-  \tau\gamma_{ki}-\tau I(i\neq j)+\tau \gamma_{ij}+\tau\mathbb{I}\left(|\beta_{jk}^{(m-1)}|> \tau\right)=0\\
&A_{jk}\mathbb{I}\left(|\beta_{jk}^{(m-1)}|\leq \tau\right)-\xi_{ijk}^2+\tau\gamma_{ki}+\tau I(i\neq j)-\tau \gamma_{ij}-\tau\mathbb{I}\left(|\beta_{jk}^{(m-1)}|> \tau\right)=0 \\
& \xi_{ijk}^1,\xi_{ijk}^2>0, i,j,k=1,\dots,p,k\neq j .
\end{align*}
Following \cite{boyd_2011}, we introduce scale dual variables $ \boldsymbol{\alpha}=\{\alpha_{ijk}^1,\alpha_{ijk}^2\}_{p\times p\times p} $, and $ W=\{w_{jk}\}_{p\times p} $, which leads to an augmented Lagrangian,
\begin{equation}\label{admm}
	\begin{aligned}
	L_{\rho}&(\boldsymbol{\beta},A,\Lambda,\xi,\boldsymbol{\alpha},W)\\
	=&R(\boldsymbol{\beta})+\mu\sum_{(j,k)\in F^c}|A_{jk}|\mathbb{I}\left(|\beta_{jk}^{(m-1)}|\leq \tau\right)+\frac{\rho}{2}\left\| \boldsymbol{\beta}-A+W \right\|_F^2\\
	&+\frac{\rho}{2}\sum_{i}\sum_{k\neq j}\left(A_{jk}\mathbb{I}\left(|\beta_{jk}^{(m-1)}|\leq \tau\right)+\xi_{ijk}^1-  \tau\gamma_{ki}-\tau I(i\neq j)+\tau \gamma_{ij}+\tau\mathbb{I}\left(|\beta_{jk}^{(m-1)}|> \tau\right)+\alpha_{ijk}^1 \right)^2 \\
	&+\frac{\rho}{2}\sum_{i}\sum_{k\neq j}\left( A_{jk}\mathbb{I}\left(|\beta_{jk}^{(m-1)}|\leq \tau\right)-\xi_{ijk}^2+\tau\gamma_{ki}+\tau I(i\neq j)-\tau \gamma_{ij}-\tau\mathbb{I}\left(|\beta_{jk}^{(m-1)}|> \tau\right)+\alpha_{ijk}^2\right)^2.\\
	\end{aligned}
\end{equation}
We iterate over the six blocks until convergence. At iteration step $ s+1 $ of ADMM, we have
\begin{equation}\label{iter}
\begin{aligned}
& A^{(s+1)}=\arg\min_{A} L_{\rho}(A,\boldsymbol{\beta}^{(s)},\Lambda^{(s)},\xi^{(s)},\boldsymbol{\alpha}^{(s)},W^{(s)}),\\
& \boldsymbol{\beta}^{(s+1)}=\arg\min_{\boldsymbol{\beta}} L_{\rho}(A^{(s+1)},\boldsymbol{\beta},\Lambda^{(s)},\xi^{(s)},\boldsymbol{\alpha}^{(s)},W^{(s)}),\\
& (\Lambda^{(s+1)},\xi^{(s+1)})=\arg\min_{\Lambda,\xi} L_{\rho}(A^{(s+1)},\boldsymbol{\beta}^{(s+1)},\Lambda,\xi,\boldsymbol{\alpha}^{(s)},W^{(s)}),\\
& W^{(s+1)}=W^{(s)}+\boldsymbol{\beta}^{(s+1)}+A^{(s+1)},\\
& \alpha_{ijk}^{1(s+1)}=\alpha_{ijk}^{1(s)}+A_{jk}^{(s)}\mathbb{I}\left(|\beta_{jk}^{(s)}|\leq \tau\right)+\xi_{ijk}^{1(s)}-  \tau\gamma_{ki}^{(s)}-\tau I(i\neq j)+\tau \gamma_{ij}^{(s)}+\tau\mathbb{I}\left(|\beta_{jk}^{(s)}|> \tau\right),\\
& \alpha_{ijk}^{2(s+1)}=\alpha_{ijk}^{2(s)}+A_{jk}^{(s)}\mathbb{I}\left(|\beta_{jk}^{(s)}|\leq \tau\right)-\xi_{ijk}^{2(s)}+  \tau\gamma_{ki}^{(s)}+\tau I(i\neq j)-\tau \gamma_{ij}^{(s)}-\tau\mathbb{I}\left(|\beta_{jk}^{(s)}|> \tau\right).
\end{aligned} 
\end{equation}
The strategy for computing $ \boldsymbol{\beta}^{H_0} $ and $ \boldsymbol{\beta}^{H_{\alpha}} $ is summarized in Algorithm 1.\\
\rule[0pt]{16.6cm}{0.1em}\\
{\bf Algorithm1}:\\
\rule[5pt]{16.5cm}{0.05em}\\
{\bf Step 1: (Initialization)} Fix $ E_1 $ in \eqref{mth}, and initialized an estimate 
$ \boldsymbol{\beta}^{(0)},\Lambda^{(0)} $ satisfying \eqref{cstr}. Set $ E^{(0)} = \{(i, j): |\boldsymbol{\beta}_{ij}^{(0)}|>\tau\}$\\
{\bf Step 2: (ADMM)} At $ m^{th} $ iteration, compute $\boldsymbol{\beta}^{(m)},\Lambda^{(m)}   $by ADMM through \eqref{iter}.\\
{\bf Step 3: (Check for acyclicity)} Let $ E^{(m)} = \{(i, j): |\boldsymbol{\beta}_{ij}^{(m)}|>\tau\} $ If $ E^{(m)} $ constitutes a cycle in the graph,  then for $ (i,j)\in E^{(m)} $, sort $ |\boldsymbol{\beta}_{ij}^{(m)}| $ decreasingly; if $ E^{(m)}\setminus (i,j) $ induces a DAG, update $ E^{(m)} $ by $  E^{(m)}\setminus (i,j) $. Otherwise, keep $ E^{(m)} $ intact.\\
{\bf Step 4: (Termination)} Repeat Steps 2 and 3 until a termination criterion ismet, that is, $ R(\boldsymbol{\beta}^{(m)})- R(\boldsymbol{\beta}^{(m-1)})\leq \epsilon$. The final solution $ \boldsymbol{\beta}^{(m^{*})} $ is the corresponding estimate under $ H_0 $ or $ H_{\alpha} $, where $ m^{*} $ is the smallest index at termination.\\
\rule[8pt]{16.5cm}{0.1em}\\
Importantly, Step 3 in Algorithm 1 ensures that $ \boldsymbol{\beta}^{(m^{*})} $ satisfies the acyclicity condition by removing the weakest link in an existing cycle, hence that it yields a DAG.
Concerning Algorithm 1, we note that its complexity over six blocks in one iteration is roughly of order $ p^3+np^2 $. In terms of convergence speed, based on our limited numerical
experience, the ADMM component converges with a modest accuracy within a few thousand iterations, while the difference convex programming component usually terminates within ten
steps. \\
\begin{thm}[Convergence of Algorithm 1]\label{thm5}
For $ \rho>0 $ and
sufficiently small $ \tau>0 $, Algorithm 1 yields a local minimizer $ \hat{\boldsymbol{\beta}} $ which satisfies the optimality condition for some multipliers $ \mu>0 $ and $ \{v_{ijk}\}_{k\neq j} $,
\[
\partial_{ij}l(\boldsymbol{\beta})+\frac{\mu}{\tau}\partial_{ij} J_{\tau}(|\beta_{ij}|)+\frac{\sum_{1\leq i\leq p} v_{ijk}}{\tau}\partial_{ij} J_{\tau}(|\beta_{ij}|)=0;\quad
i,j=1,\dots,p,i\neq j,
\]
where $ \partial_{ji}  $ denotes the subgradient.
\end{thm} 
As indicated by Theorem \ref{thm5}, Algorithm 1 yields a local minimizer. However, as showed in Table 3, it can yield a global minimizer or a good localminimizer. In fact, the probability that
the solution of Algorithm 1 agrees with the oracle estimator is high, which is a global minimizer asymptotically, see Lemmas \ref{lm1} and \ref{lm2} in the Appendix. This aspect has been recognized \cite{Tao_2005} for a difference convex algorithm. Note, however, that a global minimizer can be attained if Breiman and Culter’s outer approximation method (a version of difference convex algorithm) can attain a global minimizer at the expense of slow convergence \cite{brei_1993}.

\section{Numerical Examples}
In this part we will check the property of the proposed tests and compares with oracle tests in size and power using simulated data and real data, where the oracle tests for linkages and directed pathway are operated under the assumption that the true graph structure would be known in advance. The corresponding likelihood ratios are
\[
\begin{array}{ll}
lr_{OR}=l(\hat{\boldsymbol{\beta}}_{E^0\cup F})-l(\hat{\boldsymbol{\beta}}_{E^0\setminus F})&\text{ for linkages test,}\\
lr_{OR}=l(\hat{\boldsymbol{\beta}}_{E^0\cup F})-\max_{k=1}^{|F|}l\left( \hat{\boldsymbol{\beta}}_{E^0\cup F\setminus (i_k,i_{k+1})}(k)\right) &\text{ for directed pathway test.}
\end{array}
\]
We consider random graphs and hub graphs for the linkages test and chain graphs for directed pathway test. Figure 1 displays the three types of DAGs. We show the size and the power of these two kinds of tests through 1000 simulations. For the size of a test, we calculate the average of times of rejecting $ H_0 $ under $ H_0 $ is true. For the power of a test, we choose three alternatives $ H_{\alpha} $ hypothesis and calculate the average of times of rejecting $ H_0 $ under $ H_{\alpha} $ is true. As for the tuning parameters $ (\mu,\tau) $, we use five-fold cross-validation to choose $ (\hat{\mu},\hat{\tau})  $ that maximizes log-likelihood $ \hat{l}(\mu,\tau) $, which is defined as 
\[
\sum_{l=1}^{5}\sum_{j=1}^{p}\left[-\exp(\beta_{j0}^{-l}(\mu,\tau)+\sum_{i\in V_j} \beta_{ij}^{-l}(\mu,\tau)x_{hi})+x_{hj} (\beta_{j0}^{-l}(\mu,\tau)+\sum_{i\in V_j} \beta_{ij}^{-l}(\mu,\tau)x_{hi})-\log(x_{hj}!),\right] 
\]
where $\beta_{ij}^{-l}(\mu,\tau)  $ are estimated under $ H_{\alpha} $ is true. The optimal tuning parameters $(\hat{\mu},\hat{\tau})=\arg\max  \hat{l}(\mu,\tau) $.
\begin{figure}[!ht]
	\centering
	\subfigure[random graph.]{
		\includegraphics[width=0.45\textwidth]{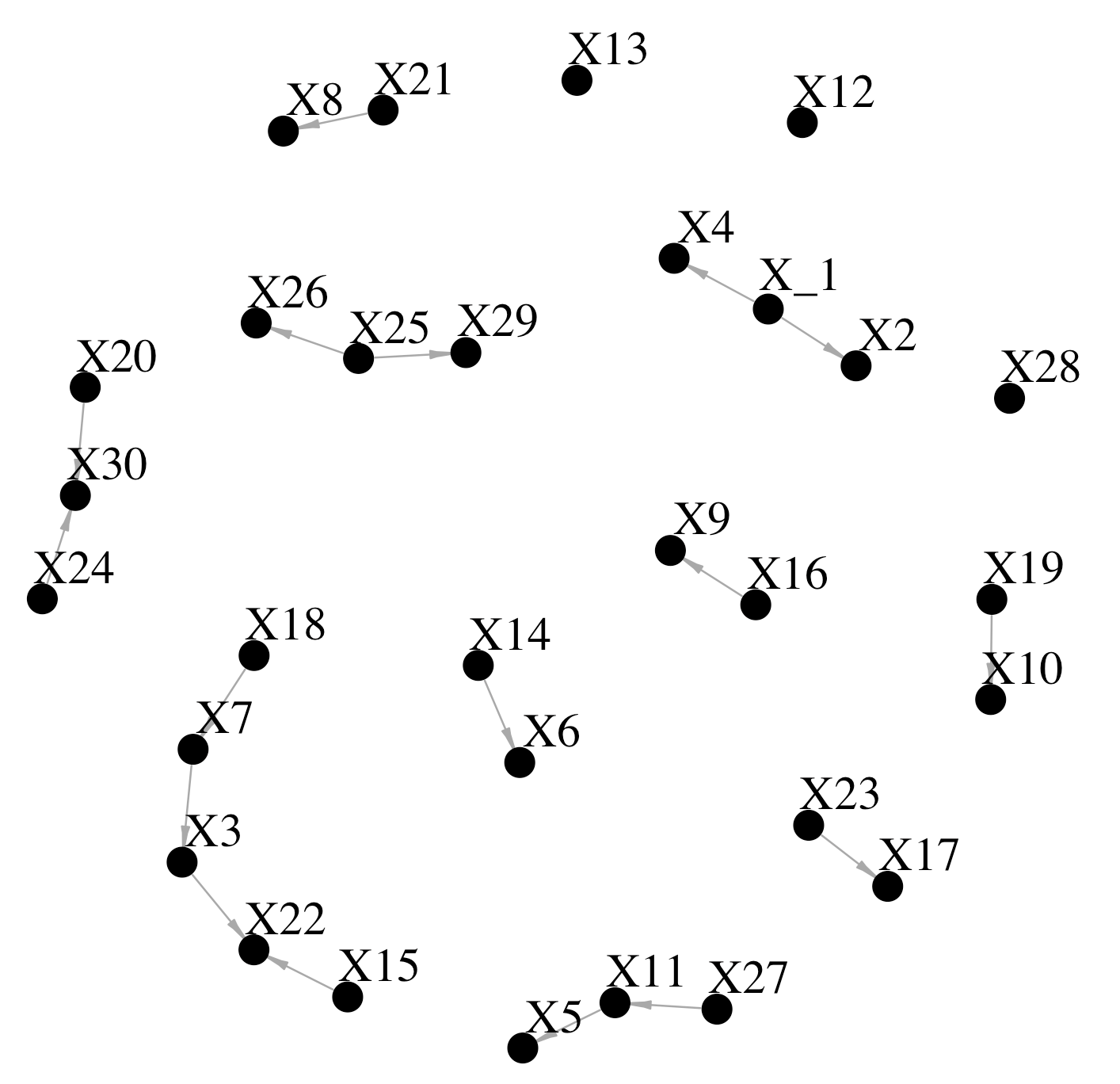}
	}
	\subfigure[hub graph.]{
		\includegraphics[width=0.45\textwidth]{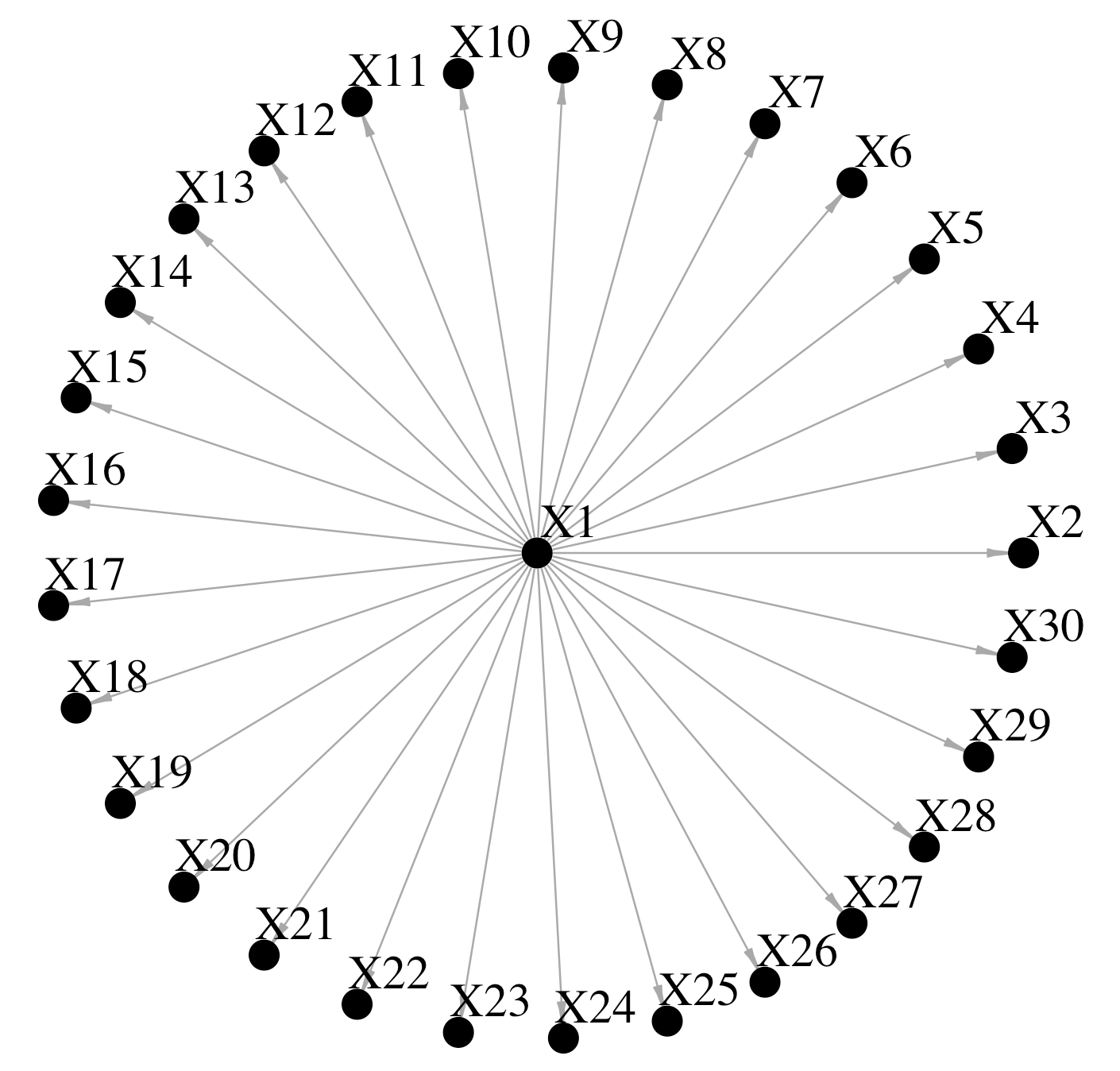}
	}
	\caption{\label{fg1}Two types of graphs}
\end{figure}
\subsection{Simulated Examples}
Example 1 (Test for linkage). Firstly, we consider a random DAG of $ p $ nodes with structure displayed in figure \ref{fg1}. For the adjacency matrix $ \boldsymbol{\beta} $, we generate the entries using the Bernoulli distribution with probability $ 1/p $ and set nonzero entries to be -0.5. For the likeages test, we use two different $ H_0 $ hypotheses: $(i) H_0: \beta_{ij}=0 ,\text{ for all }(i,j)\in F$ with $|F|=1$; $ (ii) H_0: \beta_{ij}=0,\text{ for all }(i,j)\in F$ with $|F|=30$. Here $ |D^0|=|F| $, where the elements of F are randomly chosen from the zero entries of true adjacency matrix without replacement. For the power analysis, we use three different alternatives $ H_{\alpha}: \beta_{ij}=-0.1l,l=1,2,3 $, with the elements of $ F $ randomly chosen from the nonzero entries of true adjacency matrix without replacement. \\
Example 2 (Test for linkage) We consider a hub graph of p nodes with structure displayed in figure \ref{fg1}. And the adjacency matrix $ \boldsymbol{\beta} $ is nonzero at $ \beta_{1i} $, with $ i=2,3,\dots,p $, and set the remaining entries to zero. Similarly to example 1, for the likeages test, we use two different $ H_0 $ hypotheses: $(i) H_0: \beta_{ij}=0, \text{ for all }(i,j)\in F $ with $|F|=1$; $ (ii) H_0: \beta_{ij}=0 \text{ for all }(i,j)\in F$ with $|F|=30$. Here $ |D^0|=|F| $, where the elements of F are randomly chosen from the zero entries of true adjacency matrix without replacement. For the power analysis, we use three different alternatives $ H_{\alpha}: \beta_{ij}=-0.1l,l=1,2,3 $, with the elements of $ F $ randomly chosen from the nonzero entries of true adjacency matrix without replacement.\\
Example 3 (Test for pathways) We consider a chain DAG of $ p $ nodes with structure displayed in figure \ref{fg1}. For the adjacency matrix $ \boldsymbol{\beta} $, we set $ \beta_{i(i+1)}=-0.5 ,i=1,2,\dots,p-1$  and $ \beta_{ij}=0 $ otherwise. For the pathway test, the elements of $ F $ are randomly chosen from the nonzero entries of true adjacency matrix without replacement with $ |D^0|=|F|=5 $. For the power analysis, we ramdomly sample $ d=5 $ edges from $ \{(i,i+1),i=1,2,\dots,p-1\} $ without replacement.we use three different alternatives $ H_{\alpha}: \beta_{i(i+1)}=-0.1l,(i,i+1)\in S;l=1,2,3 $.\\
Example 4 (Oracle rate of CLR) We compare the constrained likelihood ratio with the correspoding oracle constrained likelihood ratio $ lr_{OR} $ to see if our test can reconstruct the oracle test.
\subsection{Power and Size of proposed CLR Tests}
\begin{figure}[!ht] 
	\centering
	\subfigure[random graph.]{
		\includegraphics[width=0.3\textwidth]{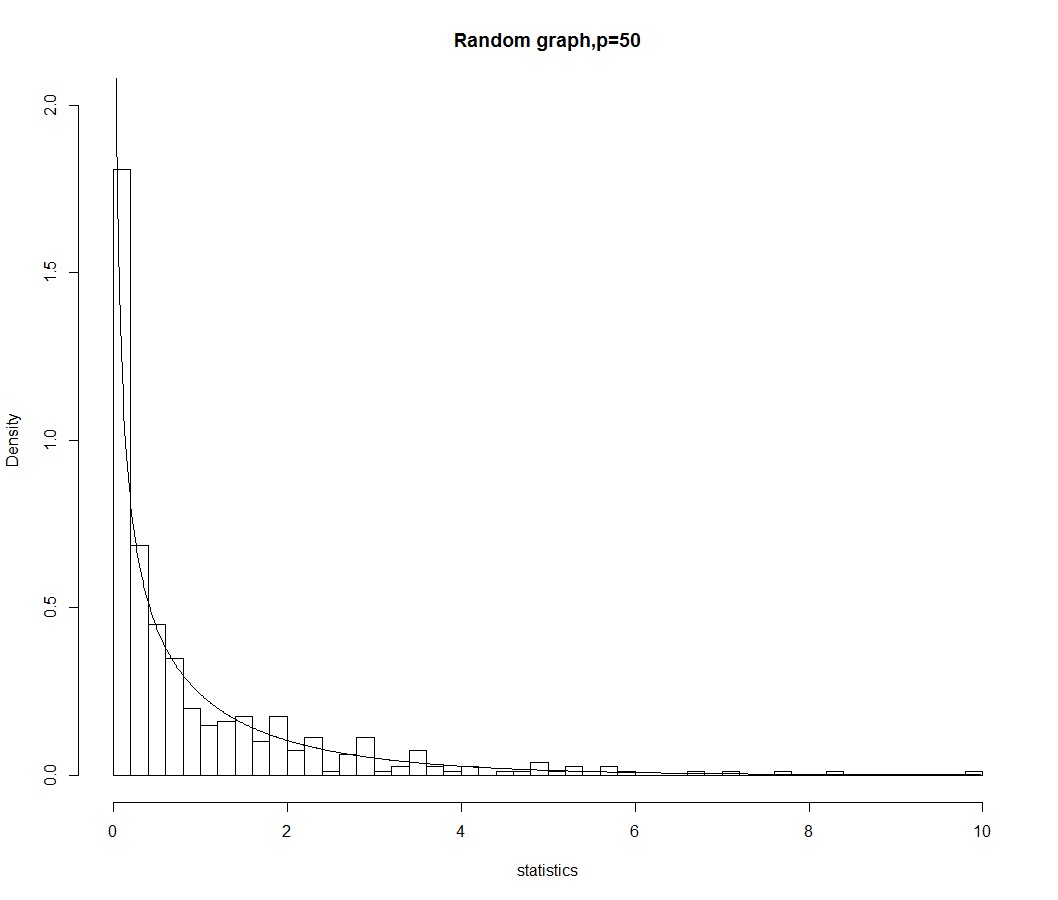}
	}
	\subfigure[random graph.]{
		\includegraphics[width=0.3\textwidth]{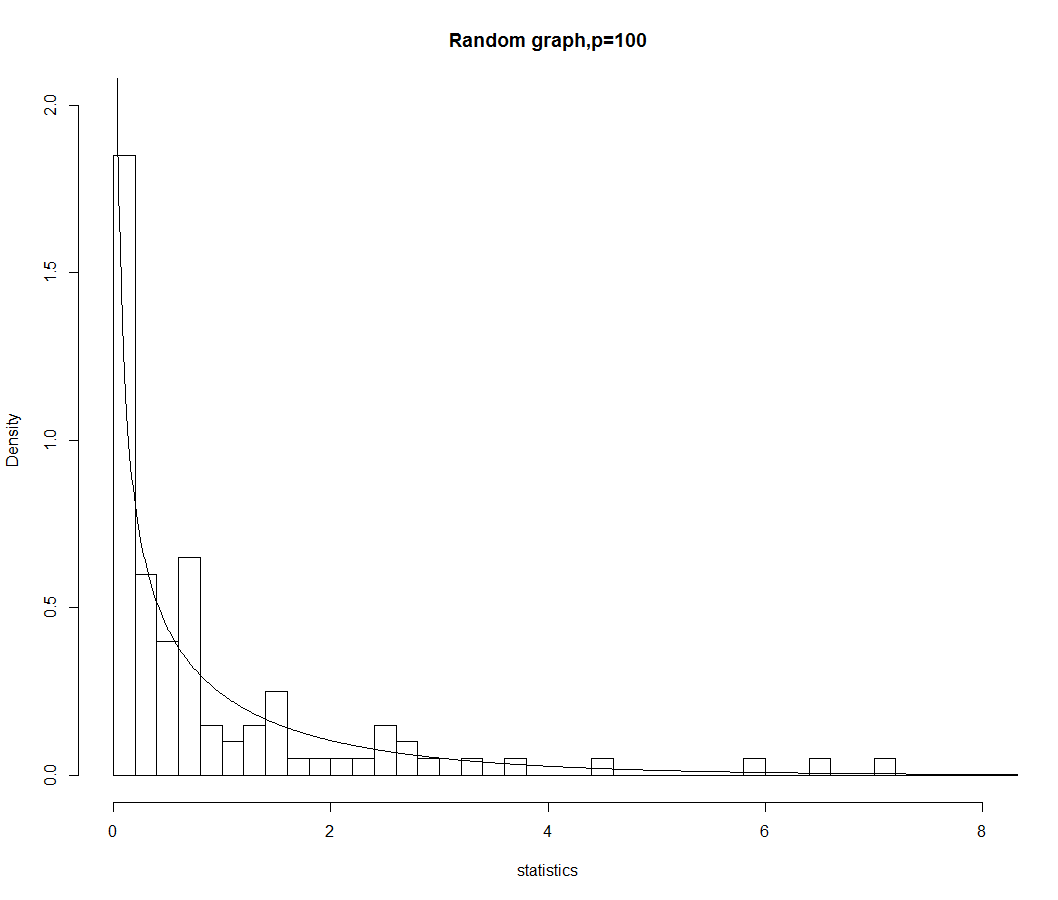}
	}
	\subfigure[random graph]{
		\includegraphics[width=0.3\textwidth]{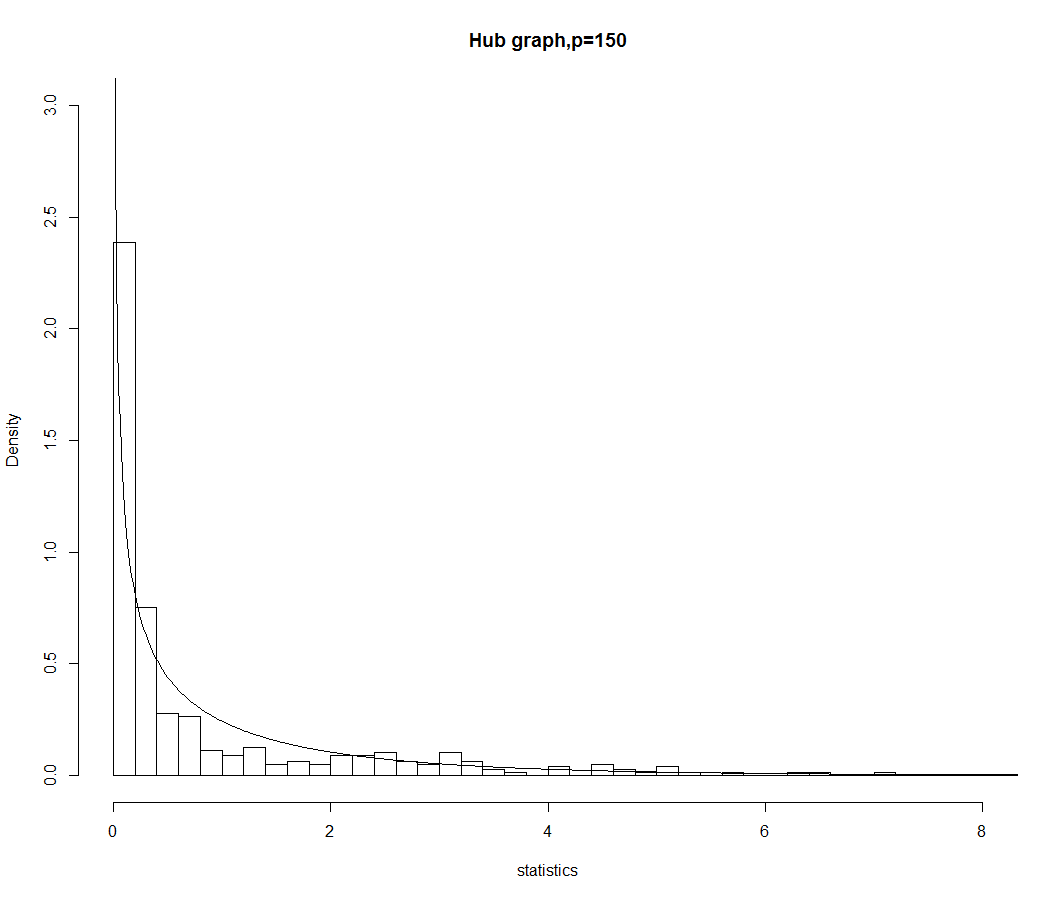}
	}
	\subfigure[hub graph.]{
		\includegraphics[width=0.3\textwidth]{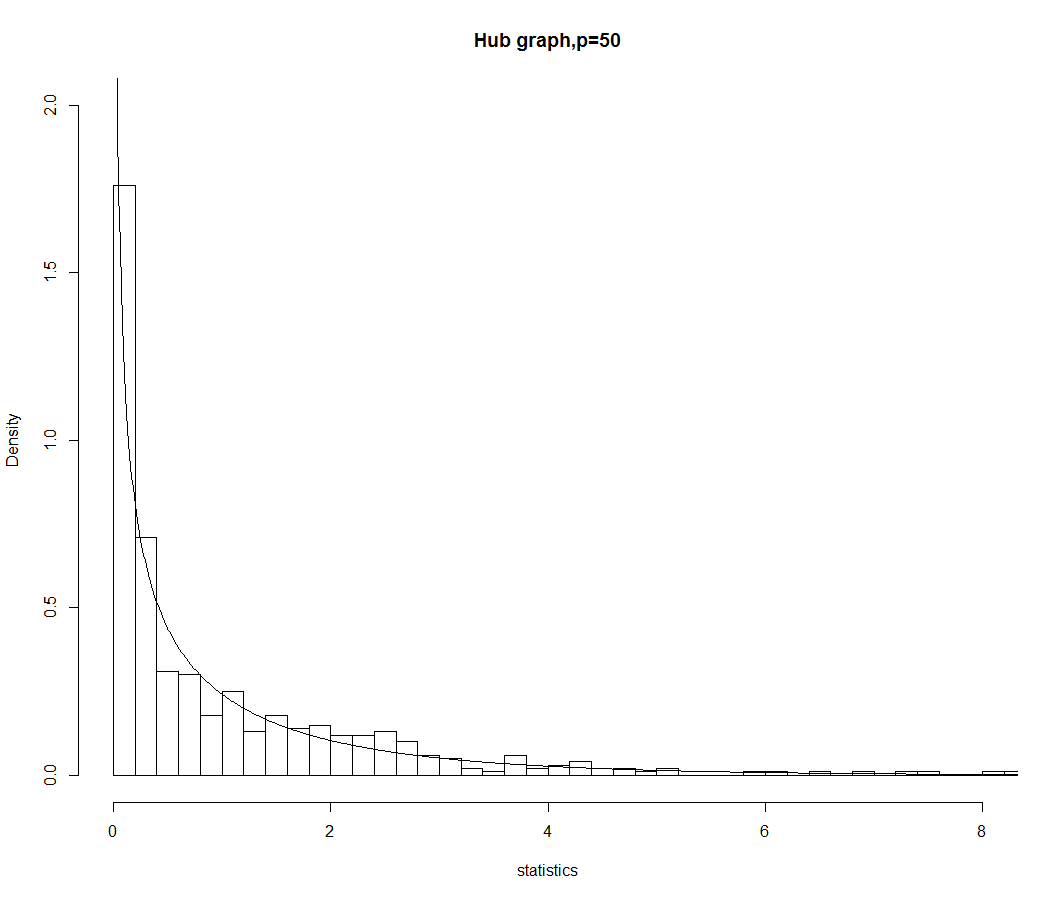}
	}
	\subfigure[hub graph.]{
		\includegraphics[width=0.3\textwidth]{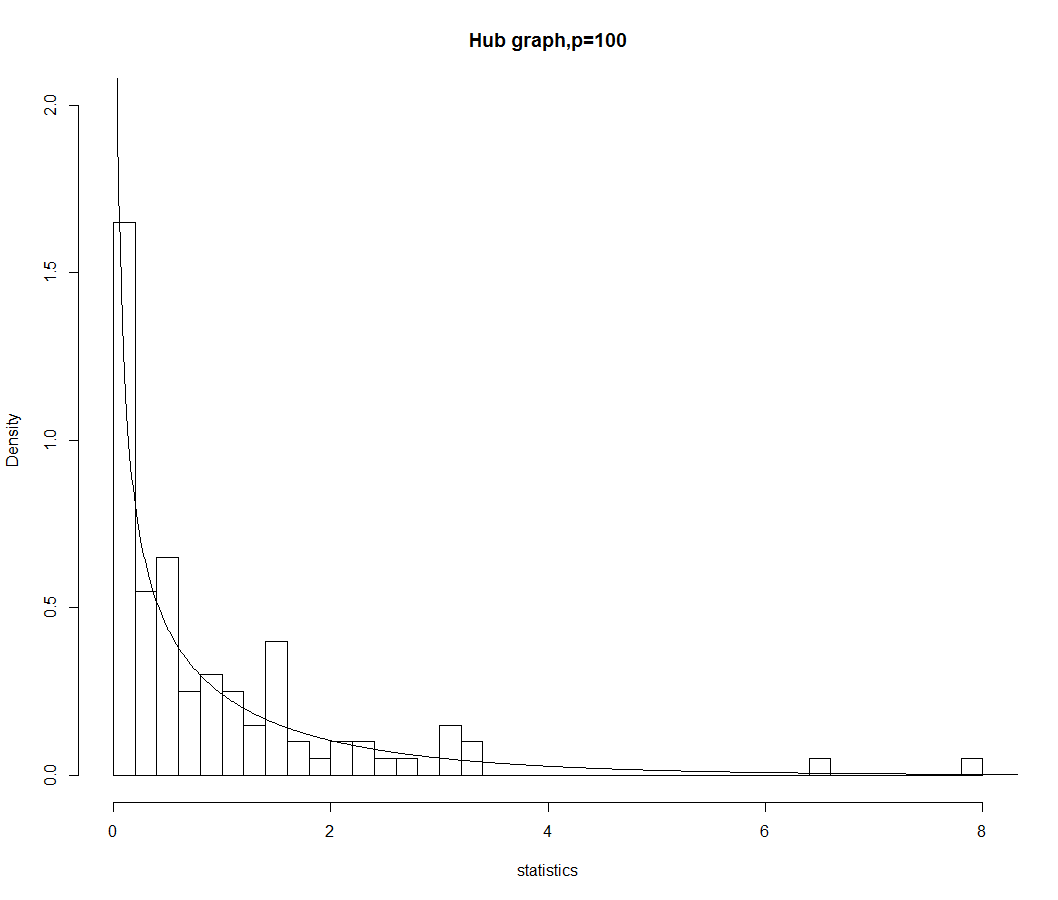}
	}
	\subfigure[hub graph]{
		\includegraphics[width=0.3\textwidth]{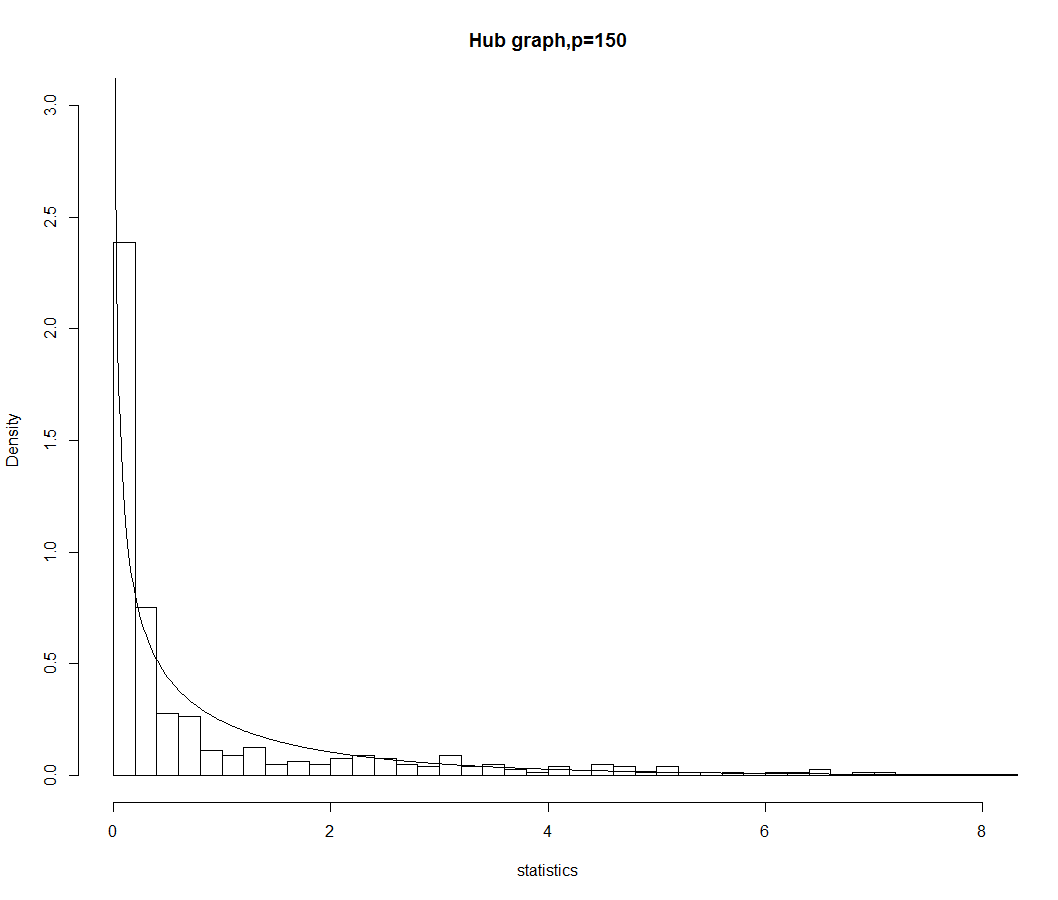}
	}
	\caption{\label{fg2} Null distribution of the proposed CLR linkages test with n=500 and $ |D^0|=1 $}
\end{figure}

\begin{figure}[!ht]
	\centering
	\subfigure[random graph.]{
		\includegraphics[width=0.3\textwidth]{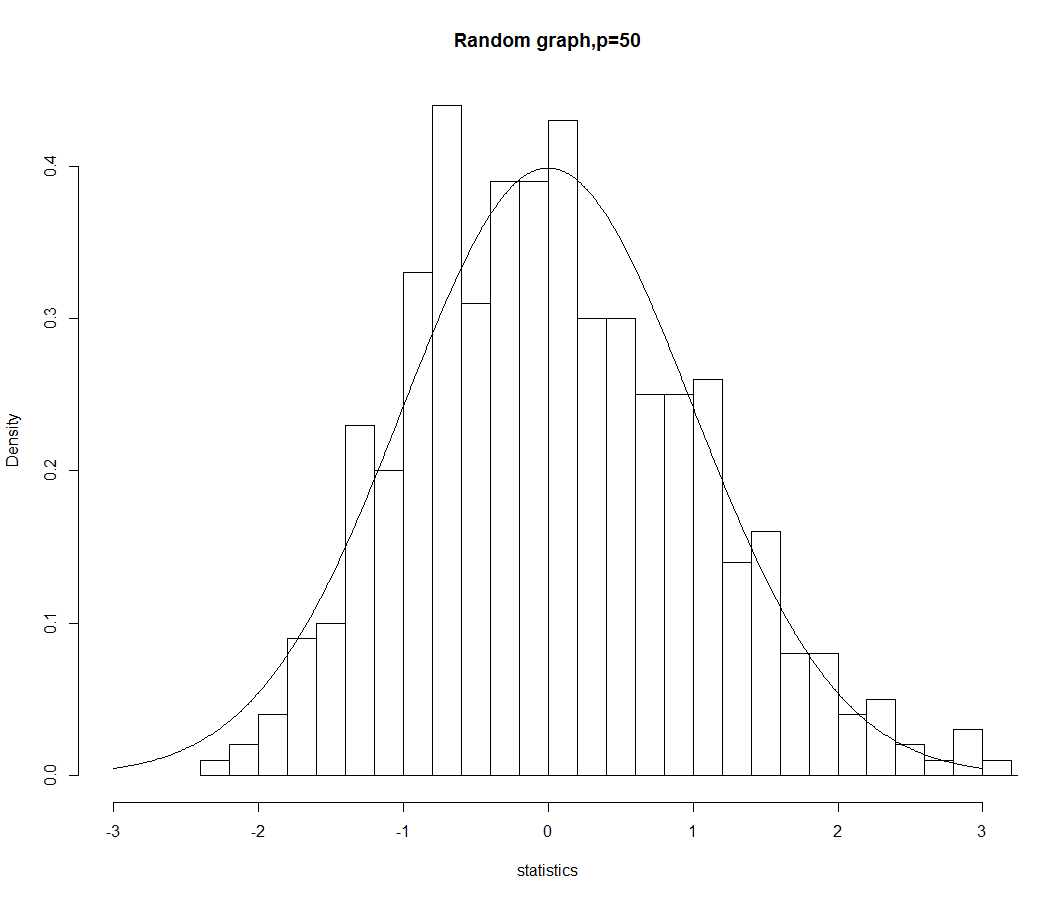}
	}
	\subfigure[random graph.]{
		\includegraphics[width=0.3\textwidth,height=4.2cm]{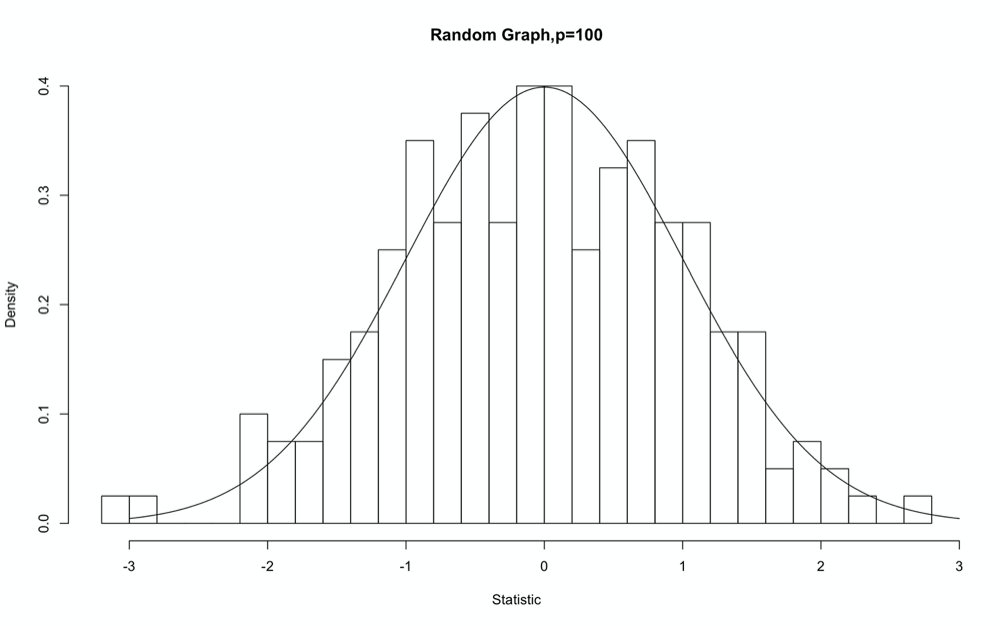}
	}
	\subfigure[random graph]{
		\includegraphics[width=0.3\textwidth]{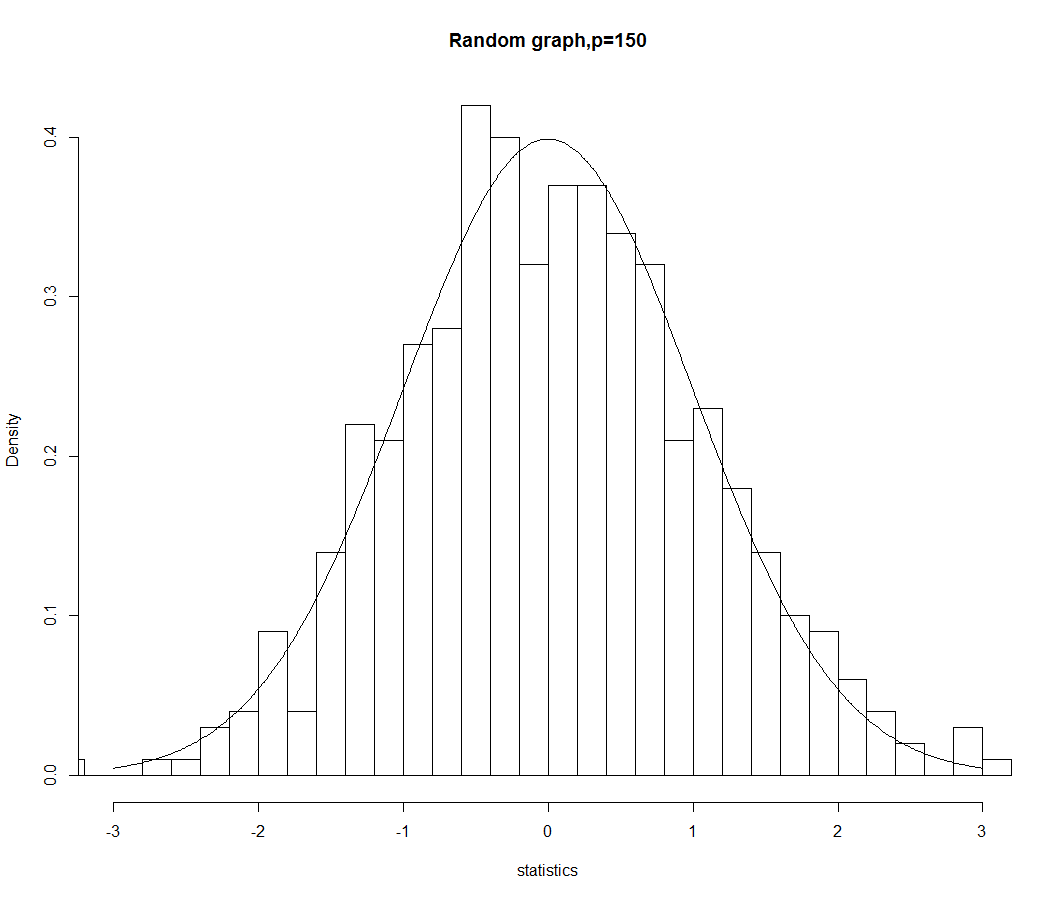}
	}
	\subfigure[hub graph.]{
		\includegraphics[width=0.3\textwidth]{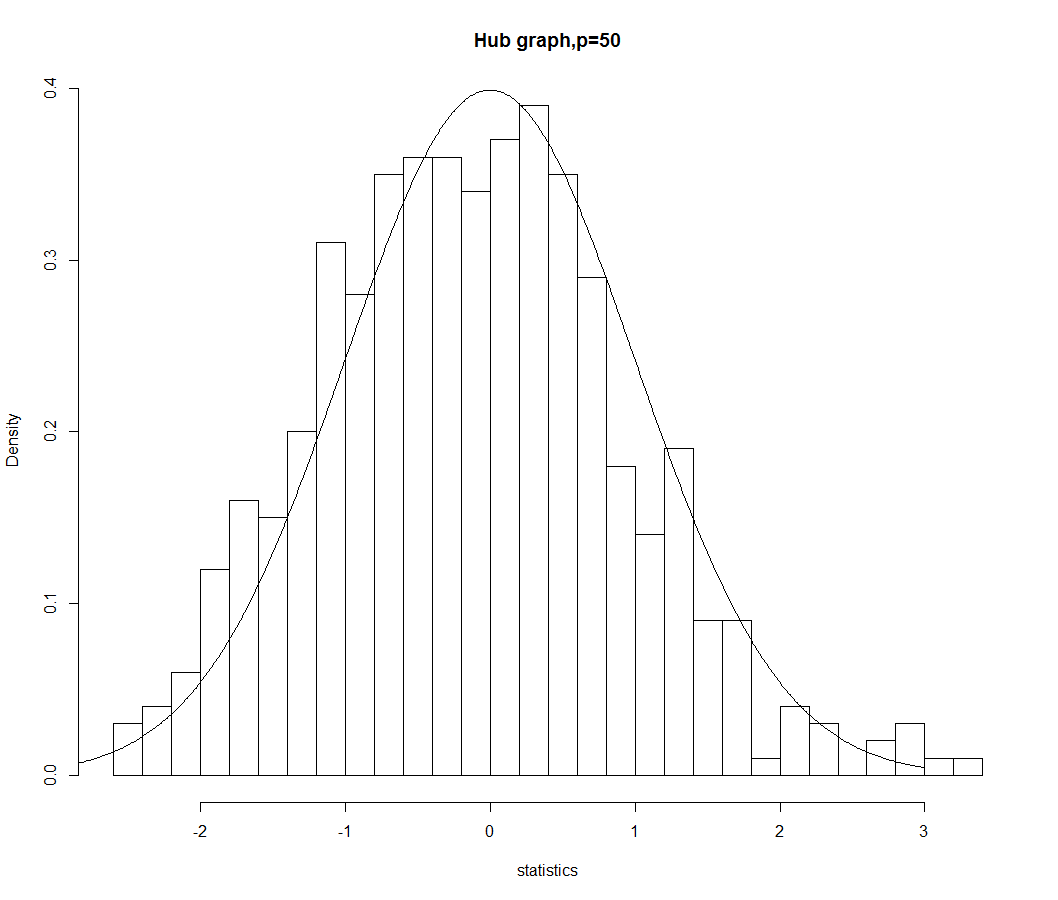}
	}
	\subfigure[hub graph.]{
		\includegraphics[width=0.3\textwidth]{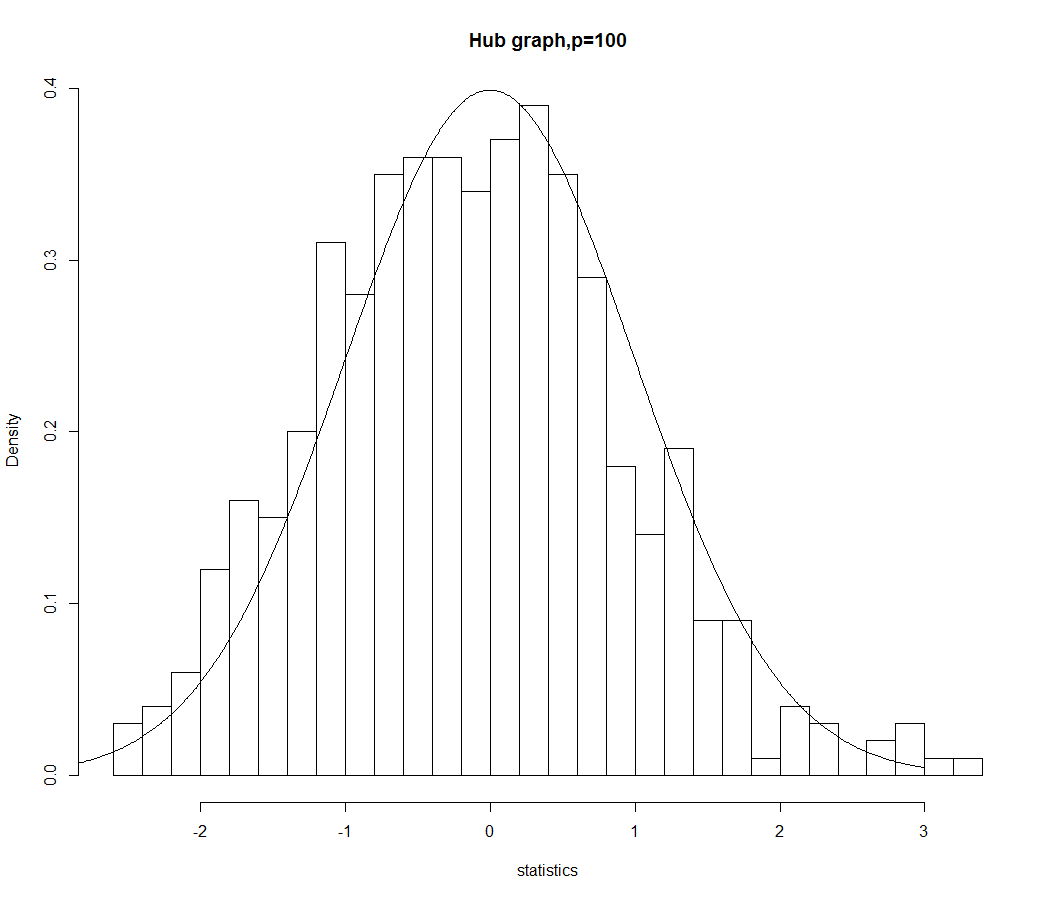}
	}
	\subfigure[hub graph]{
		\includegraphics[width=0.3\textwidth,height=4.3cm]{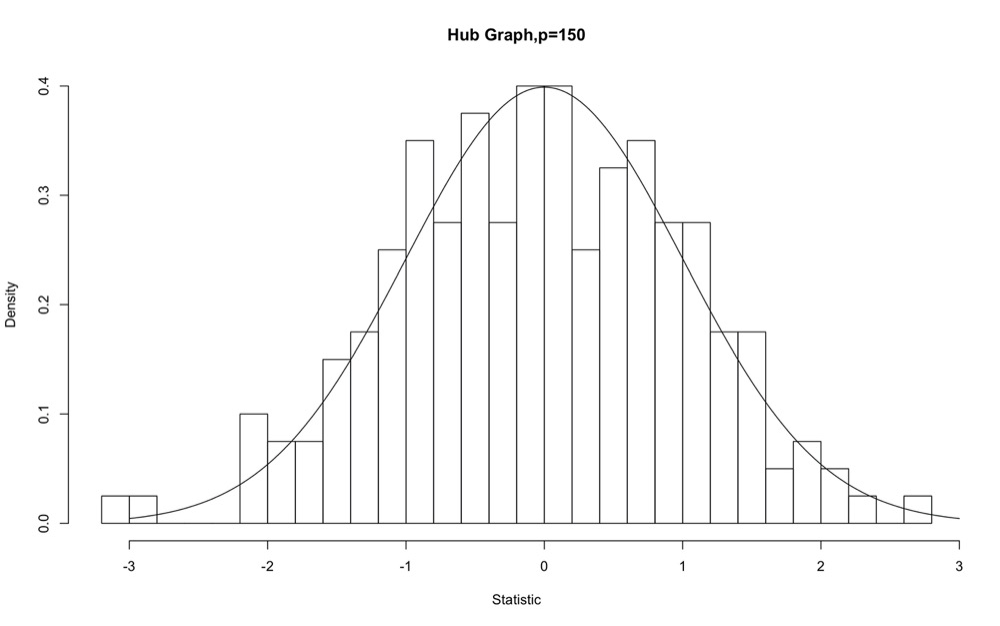}
	}
	\caption{\label{fg3} Null distribution of the proposed CLR linkages test with n=500 and $ |D^0|=30 $}
\end{figure}

\begin{figure}[!ht]
	\centering
	\subfigure[chain graph.]{
		\includegraphics[width=0.3\textwidth,height=4.3cm]{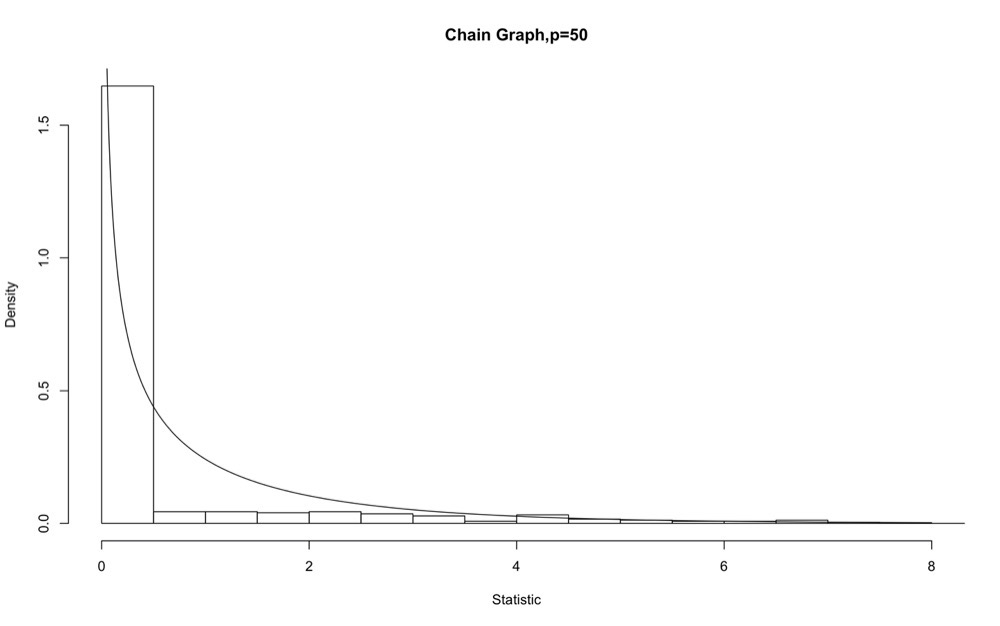}
	}
	\subfigure[chain graph.]{
		\includegraphics[width=0.3\textwidth,height=4.3cm]{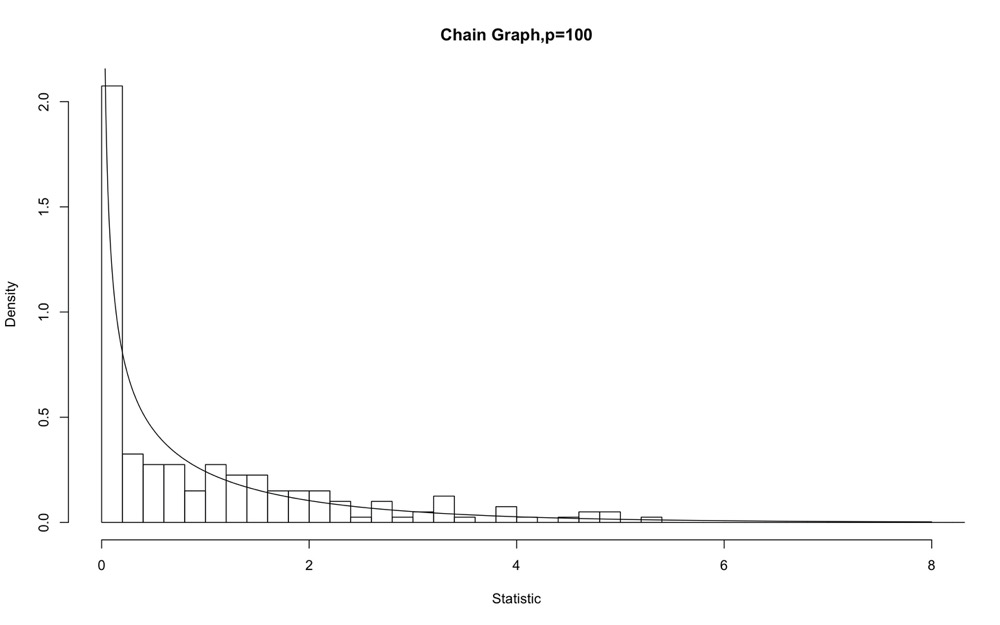}
	}
	\subfigure[chain graph]{
		\includegraphics[width=0.3\textwidth,height=4.3cm]{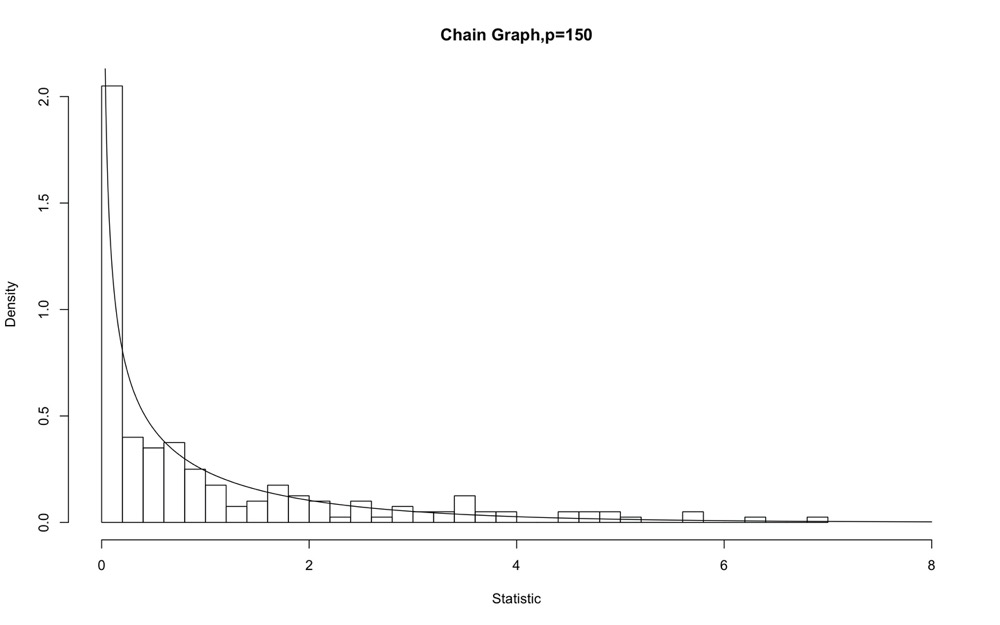}
	}
	\caption{\label{fg4} Null distribution of the proposed CLR directed pathway test with n=500 and $ d=5$}
\end{figure}

As suggested by table \ref{tb1} and table \ref{tb2} , the proposed CLR tests perform well both in testing linkages of a DAG in Example 1 and 2 and testing directed pathway of a DAG in Example 3. The sizes of the tests are colsed to the significance level 0.05 we preset, and the power is close to 1 when $ H_{\alpha} $ is true and the value of parameters are closed to the true values. As shown in figure \ref{fg2}, \ref{fg3}, \ref{fg4} for linkages tests, the empirical distribution of log-likelihood ratios of Example 1 and 2 are approximately chi-square distribution when $ |D^0|=1$ and normal distribution when $ |D^0|=30 $. For directed pathway tests, the empirical distribution of log-likelihood ratios of Example 3 are approximately the minimun of d chi-square variables when $ d=5$. This is consistent with the results of Theorem \eqref{thm1}, \eqref{thm2}, \eqref{thm3}, \eqref{thm4} which suggests the results of propsed CLR tests are as desired.

Also, as displayed in table \ref{tb3}, our constrained likelihood estimations of the structure of graphs we proposed based on Algorithm 1 has a good agreement rate with the true model, exceeding 90\% for the random graphs, and nearly 1 for the hub graphs.
\begin{table}[!ht]
	\centering
	\caption{\label{tb1} Size and power of the proposed CLR test for linkages in Example 1 and 2 with significance level 0.05.}
	\small
	\setlength{\tabcolsep}{4mm}{
		\begin{tabular}{ cccccccc}
			\toprule
			\multirow{4}{*}{} &	\multirow{4}{*}{}&	\multirow{4}{*}{}& \multicolumn{4}{c}{CLR} & \multirow{4}{*}{}    \\
			\cmidrule(r){4-8} 
			Graph&$ |D^0| $&(p,n)&Size&\multicolumn{3}{c}{Power}\\
			\cmidrule(r){5-8} 
			&&&$ \beta_{ij}=0$&$ \beta_{ij}=-0.1$&$ \beta_{ij}=-0.3$&$ \beta_{ij}=-0.5$&\\
			&&&$ (i,j)\in F $&$ (i,j)\in F $&$ (i,j)\in F $&$ (i,j)\in F $&\\
			\midrule
			Random & 1&(50,500)&0.045 & 0.16  &0.96&0.97\\
			& &(100,500)&0.05 &0.12  &0.95 &1\\
			& &(150,100)& 0.05&0.07  & 0.78&0.96\\
			\cmidrule(r){2-8} 
			& 30&(50,500)&0.01 &0.05& 1 &1\\
			& &(100,500)&0.04  &0.03 &1 &1\\
			& &(150,100)&0.044 &0.00 &0.57  &0.94\\
			\midrule
			Hub & 1&(50,500)&0.044 & 0.18&0.97   &1\\
			& &(100,500)&0.02  &0.15& 1 &1\\
			& &(150,100)& 0.048&0.06 &0.70  &0.95\\
			\cmidrule(r){2-8} 
			& 30&(50,500)&0.043 &1 &1 &1\\
			& &(100,500)&0.042 &0.99&1 &1\\
			& &(150,100)&0.05 &0.37 &1 &1\\
			\bottomrule
	\end{tabular}}
\end{table}

\begin{table}[!ht]
	\centering
	\caption{\label{tb2} Size and power of the proposed CLR test for directed pathway in Example 3 with significance level 0.05.}
	\small
	\setlength{\tabcolsep}{7mm}{
		\begin{tabular}{ cccccc}
			\toprule
			\multirow{4}{*}{}& \multicolumn{4}{c}{CLR} & \multirow{4}{*}{}    \\
			\cmidrule(r){2-6} 
			(p,n)&Size&\multicolumn{3}{c}{Power}\\
			\cmidrule(r){3-6} 
			&$ \beta_{ij}=0$&$ \beta_{ij}=-0.1$&$ \beta_{ij}=-0.3$&$ \beta_{ij}=-0.5$&\\
			&$ (i,j)\in F $&$ (i,j)\in F $&$ (i,j)\in F $&$ (i,j)\in F $&\\
			\midrule
			(50,500)&0.051 &00.16& 0.76 &1&\\
			(100,500)& 0.053&0 .15&0.68 &1&\\
			(150,100)& 0.065&0.12 &0.56  &1&\\
			\bottomrule
	\end{tabular}}
\end{table}


\begin{table}[!ht]
	\centering
	\caption{\label{tb3} Accuracy of estimation of the proposed constrained likelihood ratio method}
	\setlength{\tabcolsep}{18mm}{
		\begin{tabular}{ccc}
			\toprule
			(p,n) & Random & Hub\\
			\midrule
			(10,200) &0.9501 &0.997 \\
			(30,600)&0.962&1.000\\
			(50,1000) &0.987&1.000\\
			\bottomrule
	\end{tabular}}
\end{table}

\subsection{Real data}
We now apply our likelihood ratio test to data set that contained multivariate count data of basketball statistics for National Basketball Association (NBA) players during the 2016-17 regular season. The dataset consists of the statistics of 5 defensive players and 5 shooting guard. The data set contains 19 covariates: Home	(1 when playing home, 0 when playing away), Win(1 when winning the game, 0 otherwise), Minutes	(the playing time of the player), FG(Field goal), FGA(Field Goal Attempted), 3Pt(three point made), 3PA(three point Attempted), FT(Free Throw made), FTA(Free throw attempted), ORB(Offensive rebound), DRB(Defensive rebound), TRB(Total Rebound), AST(Assist made), STL(steal made), BLK(Block shot), TOV(Turn over), PF(Personal Foul), PTS(Total Points),GmSc(Game Score). 

We assumed each node’s conditional distribution given its parents is Poisson, this is reasonable because the NBA statistics, except for Minutes and GmSc, reflect the number of successes or attempts that were counted during the regular season. We use five-folds cross validation to choose the tuning parameters $ (\tau,\lambda) $, and we will give our eatimation of the network structure of Shooting Guards with 19 nodes and  392 observations and Defensive players with 19 nodes and 359 observations, respectively, to analysis the difference of the statistics network between the two groups.

\begin{figure}[htbp]
	\centering
	\subfigure[Defensive group.]{
		\includegraphics[width=0.45\textwidth]{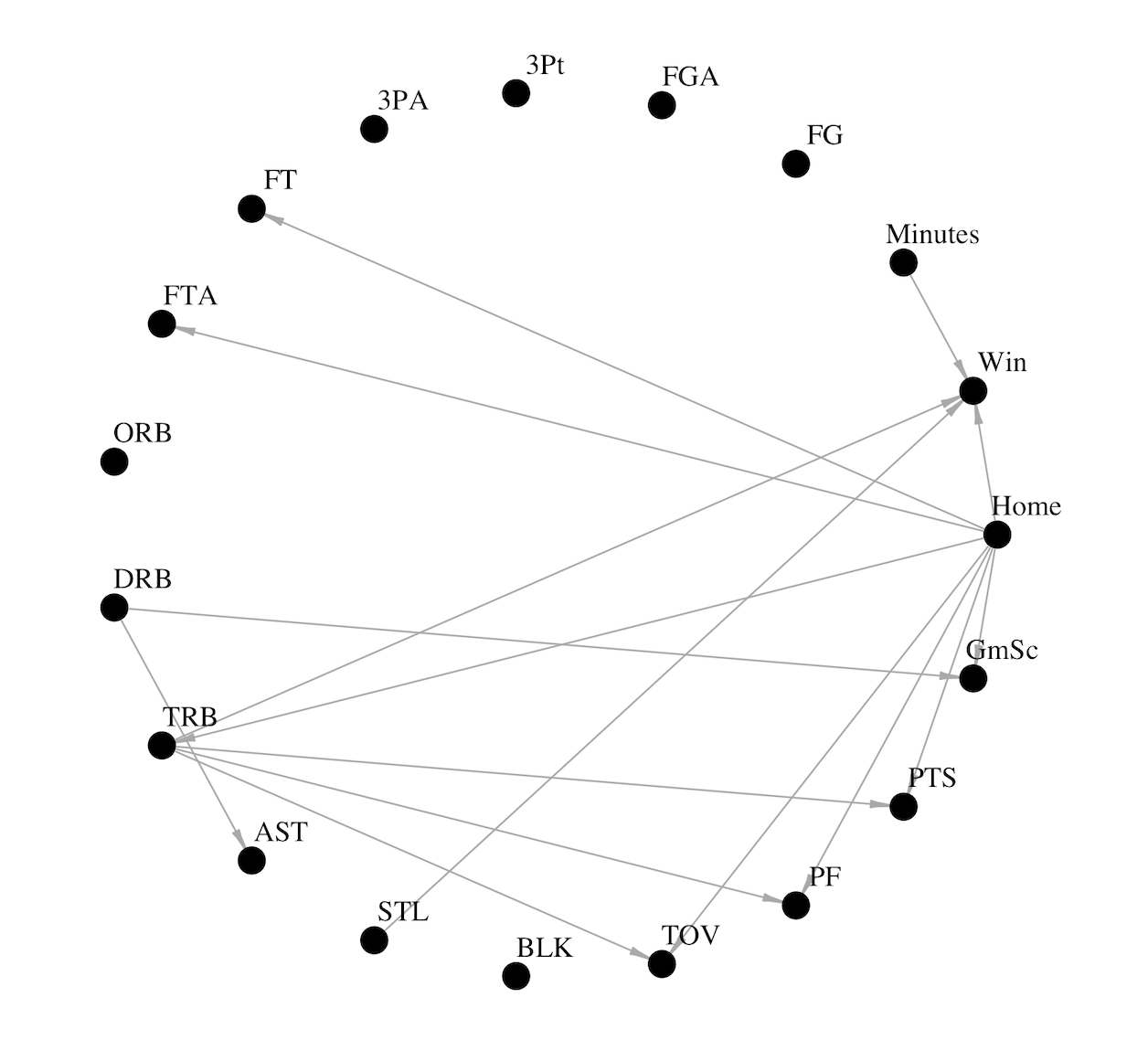}
	}
	\subfigure[Shooting group]{
		\includegraphics[width=0.45\textwidth]{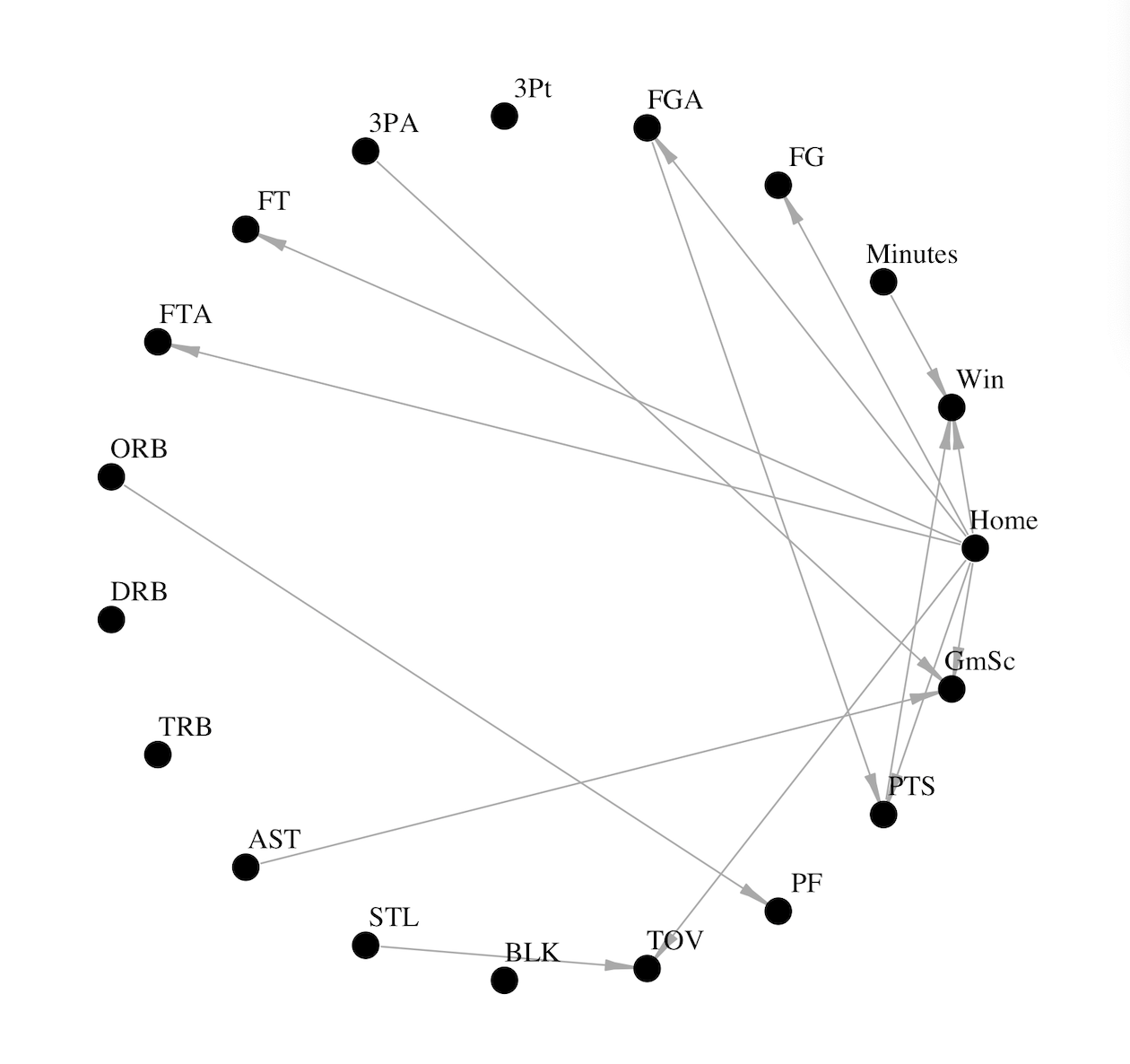}
	}
	
	\caption{\label{nba} Estimated networks for Defensive and Shooting Group.}
\end{figure}
From figure \ref{nba} we can see from the defensive and offensive statistics that playing home has impact on players performance for both of the groups. When playing at home, the players avoid the fatigue and have more energy during the game, the atmosphere at home is better, and with the support of the fans, the players will be more active, thus have better performance. Specifically, in defensive group there exists pathways Home$ \to $ FTA $ \to $ Win while in Shooting group there exists pathway Home$ \to $ PTS $ \to $ Win. Moreover, in defensive group, home has impact on points statistics including AST and PTS through defensive statistics including TRB, DRB , as well as on the turnover statistics when defensively is too strong. In shooting group, players tend to be more offensive at home and make more attempts, thus more points which contributes to winning of the game.

\section{Conclusion}
In this paper we propose a penalized maximum likelihood method based on Poisson regression for estimating and testing directed edges and pathways in discrete Bayesian networks. To ensure the acyclicity of the Bayesian network, truncated L1 penalties are applied to constrain cyclic pathways. We employ a likelihood ratio test with constraints for statistical inference of directed edges and pathways, and provides the asymptotic distributions of the test statistics under the null hypothesis and the alternative hypothesis in high-dimensional conditions. The conclusions of this paper are verified using simulated data and NBA player dataset.

\bibliographystyle{chicago}
\bibliography{ref} 

\section*{Appendix}
\subsection*{Analytical Updating Formulas for \eqref{iter}}
\renewcommand{\theequation}{A.\arabic{equation}}
\renewcommand{\thelem}{A.\arabic{lem}}
In this section we provide detailed analytic updating formulas for the alternating direction method of multipliers in \eqref{iter}. \\
{\bf A direction}: Elementwise minimization yields that
\[
A_{jk}^{(s+1)}=\left\{
\begin{array}{ll}
\text{Soft}_{\mu/\rho(2p+1)}(\theta_{jk}^{(s)}) & \text{ if }(j,k)\in F^c, |\beta_{jk}|\leq \tau,\\
\theta_{jk}^{(s)}  &  \text{ if }(j,k)\in F, |\beta_{jk}|\leq \tau,\\
\beta_{jk}^{(s)}+W_{jk}^{(s)}  &  \text{ if }|\beta_{jk}|> \tau,\\
\end{array}
\right.
\]
Where 
\[
\theta_{jk}^{(s)}=\frac{1}{2p+1}\left[ \beta_{jk}^{(s)}+W_{jk}^{(s)}-\sum_{i}\left(\xi_{ijk}^{1(s)}-\xi_{ijk}^{2(s)}+\alpha_{ijk}^{1(s)}-\alpha_{ijk}^{2(s)}\right),  \right] 
\]
and 
\[
\text{Soft}_{\mu/\rho(2p+1)}(\theta_{jk}^{(s)}) =\left(\theta_{jk}-\frac{\mu \text{ Sign}(\theta_{jk})}{\rho(2p+1)}\right)_{+}.
\]
{\bf $\boldsymbol{\beta}$ direction}: The minimizer satisfies $ \boldsymbol{\beta}_{\cdot j}^{(s+1)}=(\boldsymbol{\beta}_{V_j},\bm{0})^T ,j=1\dots,p$ with
\[
e^{(\boldsymbol{\beta}_{V_j}X_{V_j}^T)}X_{V_j}+\rho\boldsymbol{\beta}_{V_j}= X_{\cdot j}^TX_{V_j}+\rho\left(A^{(s+1)}_{V_j}-W^{(s)}_{V_j}\right),
\]
where $ V_j=\{k:k\neq j,(j,k)\notin E_1\} $.\\
$ (\Gamma,\xi)$ {\bf direction}: The updating formula of $ {\Gamma} $ is:
\[
\Gamma^{(s+1)}=M_{p\times p}L_{p\times p}^{(s+1)},
\]
where
\[
M_{p\times p}=\frac{1}{\tau}\left[ 
\begin{array}{ccccc}
1&0&0&\dots&0\\
1&\frac{2}{p}&\frac{1}{p}&\dots&\frac{1}{p}\\
1&\frac{1}{p}&\frac{2}{p}&\dots&\frac{1}{p}\\
\vdots&\vdots& &\ldots&\vdots\\
1&\frac{1}{p}&\frac{1}{p}&\dots&\frac{2}{p}\\
\end{array},
\right]
\]
and

\begin{align*}
L_{1j}^{(s+1)}=&1,\\
L_{ki}^{(s+1)}=& \frac{1}{2}\left[\tau+\frac{1}{2} \sum_{j}\left(2 \tau \mathbb{I}\left(\left|\hat{\boldsymbol{\beta}}_{jk}^{(s)}\right|>\tau\right)+\xi_{ijk}^{1(s)}+\xi_{ijk}^{2(s)}+\alpha_{ijk}^{1(s)}-\alpha_{ijk}^{2(s)}\right)\right. \\ 
&\left.-\frac{1}{2} \sum_{j}\left(2 \tau \mathbb{I}\left(\left|\hat{\boldsymbol{\beta}}_{kj}^{(s)}\right|>\tau\right)+\xi_{ikj}^{1(s)}+\xi_{ikj}^{2(s)}+\alpha_{ikj}^{1(s)}-\alpha_{ikj}^{2(s)}\right)\right] ;k\neq i,\\
L_{ii}^{(s+1)}=& \frac{1}{2}\left[-(p-1)\tau+\frac{1}{2} \sum_{j}\left(2 \tau \mathbb{I}\left(\left|\hat{\boldsymbol{\beta}}_{ji}^{(s)}\right|>\tau\right)+\xi_{i j i}^{1(s)}+\xi_{i j i}^{2(s)}+\alpha_{i j i}^{1(s)}-\alpha_{i j i}^{2(s)}\right)\right. \\ 
&\left.-\frac{1}{2} \sum_{j}\left(2 \tau \mathbb{I}\left(\left|\hat{\boldsymbol{\beta}}_{ij}^{(s)}\right|>\tau\right)+\xi_{  i i j}^{1(s)}+\xi_{ i i j }^{2(s)}+\alpha_{ i i j}^{1(s)}-\alpha_{i i j }^{2(s)}\right)\right].
\end{align*}
The updating formula of $ \xi $ is
\[
\xi_{ijk}^{1(s+1)}=\max\left( 0,-A^{(s+1)}_{jk}\mathbb{I}\left(|\beta_{jk}^{(s)}|\leq \tau\right)+ \tau\gamma_{ki}^{(s)}+\tau I(i\neq j)-\tau \gamma_{ij}^{(s)}-\tau\mathbb{I}\left(|\beta_{jk}^{(s)}|> \tau\right)-\alpha_{ijk}^{1(s)}\right), 
\]
and
\[
\xi_{ijk}^{2(s+1)}=\max\left( 0,A^{(s+1)}_{jk}\mathbb{I}\left(|\beta_{jk}^{(s)}|\leq \tau\right)+ \tau\gamma_{ki}^{(s)}+\tau I(i\neq j)-\tau \gamma_{ij}^{(s)}-\tau\mathbb{I}\left(|\beta_{jk}^{(s)}|> \tau\right)+\alpha_{ijk}^{2(s)}\right).
\]
\subsection*{ Analytical details}
A key quantity in the analysis is the Fisher information matrix, which is the Hessian of the node-conditional log-likelihood. In the following,  we show some conditions that gaurantee the limit distribution of likelihood ratio $ 2lr$. Firstly we calculate the Hessian matrix, the likelihood function can be rewritten as 
\begin{equation}
\begin{aligned}
l(\boldsymbol{\beta})=&\sum_{h=1}^{n}\sum_{j=1}^{p} \big[-\exp({X_{h\cdot}}\boldsymbol{\beta}_{\cdot j})+x_{hj}\cdot ({X_{h\cdot}}\boldsymbol{\beta}_{\cdot j})-\log(x_{hj}!)\big]\\
=&\sum_{j=1}^{p}l_j ,
\end{aligned}
\end{equation}
where $l_j=\sum_{h=1}^{n}\big[-\exp(X_{h\cdot}\boldsymbol{\beta}_{\cdot j})+x_{hj} {X_{h\cdot}}\boldsymbol{\beta}_{\cdot j}-\log(x_{hj}!)\big]$. Then for certain $ l_j $,  $ \frac{\partial l_j}{\partial \beta_{ik}}  =0$ if $ k\neq j $. When $ k=j $, we have
\begin{equation}\label{dir1}
 \frac{\partial l_j}{\partial \beta_{ij}} =
\sum_{h=1}^{n}x_{hi}\left[x_{hj}-\exp({X_{h\cdot}}\boldsymbol{\beta}_{\cdot j})\right],\quad   \text{if } i\neq j, 
\end{equation}
and $  \frac{\partial^2 l_j}{\partial \beta_{ik}\partial \beta_{ml}} =0 $ if $ k\neq j $ or $ l\neq j $, when $ k=l=j $, we have
\begin{equation}\label{dir2}
 \frac{\partial^2 l_j}{\partial \beta_{ij}\partial \beta_{mj}} = 
-\sum_{h=1}^{n}x_{hi}x_{hm}\exp({X_{h\cdot}}\boldsymbol{\beta}_{\cdot j}),\quad  \text{if }i,m\neq j, 
\end{equation}
Given a sample $ \{X_{hj}\}_{p\times p} $, We set
\[ X = \left[
\begin{array}{ccccc}
1&x_{11} & x_{12} & \ldots & x_{1p}\\
1&x_{21} & x_{22} & \ldots & x_{2p}\\
\vdots&\vdots & \vdots & \ddots & \vdots\\
1&x_{n1} & x_{n2} & \ldots & x_{np}\\
\end{array} \right] _{n\times (p+1)} \]
and a diagonal matrix $ \Gamma_j $ as 
\begin{equation}\label{gammaj}
 \Gamma_j = \text{diag}\left(
e^{({X_{1\cdot}}\boldsymbol{\beta}_{\cdot j})} , e^{({X_{2\cdot}}\boldsymbol{\beta}_{\cdot j})}, \dots ,e^{({X_{n\cdot}}\boldsymbol{\beta}_{\cdot j})} \right)_{n\times n} .
\end{equation}
Define $ \boldsymbol{\mu}_j =\left( e^{({X_{1\cdot}}\boldsymbol{\beta}_{\cdot j})},\dots,e^{({X_{n\cdot}}\boldsymbol{\beta}_{\cdot j})}\right)^T  $.
From \eqref{dir1} and \eqref{dir2} we have 
\[
\frac{\partial l_j}{\partial \boldsymbol{\beta}_{\cdot j}}=X^T\left(X_{\cdot j}-\boldsymbol{\mu}_j \right) ,\quad \frac{\partial^2 l_j}{\partial \boldsymbol{\beta}_{\cdot j}\partial\boldsymbol{\beta}_{\cdot j}}=-X^T\Gamma_jX,
\]
where  $ \Gamma_j $ are defined as \eqref{gammaj} .

Then we expand $ l_j $ at the ture value of the parameters $ \boldsymbol{\beta}^0_{\cdot j} $ as 
\begin{align*}
l_j=&\sum_{h=1}^{n}\left[-\exp(X_{h\cdot}\boldsymbol{\beta}_{\cdot j})+x_{hj} ({X_{h\cdot}}\boldsymbol{\beta}_{\cdot j})-\log(x_{hj}!)\right]\\
=&l_j^0+\left(X_{\cdot j}-\boldsymbol{\mu}_j^0 \right)^TX\left( \boldsymbol{\beta}_{\cdot j}-\boldsymbol{\beta}_{\cdot j}^0\right)-\frac{1}{2} \left( \boldsymbol{\beta}_{\cdot j}-\boldsymbol{\beta}_{\cdot j}^0\right)^TH_j\left( \boldsymbol{\beta}_{\cdot j}-\boldsymbol{\beta}_{\cdot j}^0\right)\\
&+\frac{1}{3!}G^{(3)}(\tilde{\boldsymbol{\beta}}_{\cdot j}^0)\times_1\left( \boldsymbol{\beta}_{\cdot j}-\boldsymbol{\beta}_{\cdot j}^0\right)\times_2\left( \boldsymbol{\beta}_{\cdot j}-\boldsymbol{\beta}_{\cdot j}^0\right)\times_3\left( \boldsymbol{\beta}_{\cdot j}-\boldsymbol{\beta}_{\cdot j}^0\right),
\end{align*}
where $  \boldsymbol{\mu}_j^0$ and $ H_j $ are the true value of $ \boldsymbol{\mu}_j $ and $ X^T\Gamma_jX $, $ G^{(3)}(\tilde{\boldsymbol{\beta}}_{\cdot j}^0) $ represents the third order derivative of $ l_j $ at $ \tilde{\boldsymbol{\beta}}_{\cdot j}^0 $ and $ \tilde{\boldsymbol{\beta}}_{\cdot j}^0 $ lies on the line joining $ \boldsymbol{\beta}_{\cdot j} $ and $\boldsymbol{\beta}_{\cdot j}^0  $.\\
Expanding $ \frac{\partial l_j}{\partial \boldsymbol{\beta}_{\cdot j}} $ at $ \boldsymbol{\beta}_{\cdot j}^{0} $ , we have 
\begin{equation}\label{dribj}
\frac{\partial l_j}{\partial \boldsymbol{\beta}_{\cdot j}}=X^T\left(X_{\cdot j}-\boldsymbol{\mu}_j^0 \right)-H_j\left( \boldsymbol{\beta}_{\cdot j}-\boldsymbol{\beta}_{\cdot j}^0\right)+\frac{1}{2}G^{(3)}(\tilde{\boldsymbol{\beta}}_{\cdot j}^0)\times_1\left( \boldsymbol{\beta}_{\cdot j}-\boldsymbol{\beta}_{\cdot j}^0\right)\times_2\left( \boldsymbol{\beta}_{\cdot j}-\boldsymbol{\beta}_{\cdot j}^0\right),
\end{equation}
where $\tilde{\boldsymbol{\beta}}_{\cdot j} $ is on the line segment joining $\boldsymbol{\beta}_{\cdot j}  $ and $ \boldsymbol{\beta}_{\cdot j}^0 $. \\
Set the right hand side $\frac{\partial l_j}{\partial \boldsymbol{\beta}_{\cdot j}}=0$, 
ignore the third order term, we set 
\begin{equation}\label{bols}
\hat{\boldsymbol{\beta}}_{\cdot j}^{ols} =\boldsymbol{\beta}_{\cdot j}^0+H_j^{-1}X^T\left(X_{\cdot j}-\bm{\mu}_j^0 \right),
\end{equation}
then we will see the difference between $ \hat{\boldsymbol{\beta}}_{\cdot j} \text{   and }\hat{\boldsymbol{\beta}}_{\cdot j}^{ols} $, From \eqref{dribj} we have 
\begin{align*}
\hat{\boldsymbol{\beta}}_{\cdot j}-\boldsymbol{\beta}_{\cdot j}^0&=H_j^{-1}X^T\left(X_{\cdot j}-\mu_j \right)+\frac{1}{2} H_j^{-1}G^{(3)}(\tilde{\boldsymbol{\beta}}_{\cdot j}^0)\times_1\left( \hat{\boldsymbol{\beta}}_{\cdot j}-\boldsymbol{\beta}_{\cdot j}^0\right)\times_2\left( \hat{\boldsymbol{\beta}}_{\cdot j}-\boldsymbol{\beta}_{\cdot j}^0\right) \\
&=\hat{\boldsymbol{\beta}}_{\cdot j}^{ols}-\boldsymbol{\beta}_{\cdot j}^0+\frac{1}{2} H_j^{-1}G^{(3)}(\tilde{\boldsymbol{\beta}}_{\cdot j}^0)\times_1\left( \hat{\boldsymbol{\beta}}-\boldsymbol{\beta}_{\cdot j}^0\right)\times_2\left( \hat{\boldsymbol{\beta}}-\boldsymbol{\beta}_{\cdot j}^0\right).
\end{align*}
Then we have
\[ 
\hat{\boldsymbol{\beta}}_{\cdot j}-\hat{\boldsymbol{\beta}}_{\cdot j}^{ols}=\frac{1}{2} H_j^{-1}G^{(3)}(\tilde{\boldsymbol{\beta}}_{\cdot j}^0)\times_1\left( \hat{\boldsymbol{\beta}}-\boldsymbol{\beta}_{\cdot j}^0\right)\times_2\left( \hat{\boldsymbol{\beta}}-\boldsymbol{\beta}_{\cdot j}^0\right). 
\]
According to the Hessian matrix of our model \eqref{gammaj}, we have that when $ k\neq j $ or $ l\neq j $, $ \frac{\partial^2 l_j}{\partial \beta_{ik}\partial \beta_{ml}} =0$. Thus, We can derive the asymptotic distribution of $ lr $ by dividing it into $p$ part,
\begin{align*}
lr=&\sum_{j=1}^{p}l\left( \hat{\boldsymbol{\beta}}_{\cdot j}^{H_{\alpha}}\right) -l\left( \hat{\boldsymbol{\beta}}_{\cdot j}^{H_{0}}\right)\\
=&\sum_{j=1}^{p}l\left( \hat{\boldsymbol{\beta}}_{\cdot j}^{H_{\alpha}}\right) -l\left( \boldsymbol{\beta}_{\cdot j}^{0}\right)-\left( l\left( \hat{\boldsymbol{\beta}}_{\cdot j}^{H_{0}}\right) -l\left( \boldsymbol{\beta}_{\cdot j}^{0}\right)\right) \\
=&\sum_{j=1}^{p}lr_j.
\end{align*}
By decomposing $\boldsymbol{\beta}_{\cdot j}$ into two parts as $\boldsymbol{\beta}_{\cdot j}=(\boldsymbol{\beta}_{j_1}^T,\boldsymbol{\beta}_{j_2}^T)^T,$ 
where $\boldsymbol{\beta}_{j_2}=\bf{0}$ under $ H_0 $, and the number of entries in $\boldsymbol{\beta}_{j_1}$ is $ p_1 $ and  the length of $\boldsymbol{\beta}_{j_2}$ is $ p_2$.
Thus the null and alternative hypotheses can be written as
\[
H_0: \boldsymbol{\beta}_{j_2}={\bf 0} \text{ v.s } H_{\alpha}:\boldsymbol{\beta}_{j_2}\neq \bf{0}.
\]
Since that our parameter matrix are sparse, we have that the number of nonzero entries in $ \boldsymbol{\beta}_{\cdot j}$, is smaller than $ p $ when $ p\rightarrow \infty $.\\
Then correspondingly we divide the sample into two parts, 
\[ 
X_{j_1}=(1,X_{\cdot 1},\dots,X_{\cdot p_1})_{n\times p_1} ,
\]
\[ 
X_{j_2}=(X_{\cdot p_1+1},\dots,X_{\cdot p})_{n\times p_2},
\]
and we rewrite $ H_j $ as
\[
H_j=\left[
\begin{array}{ll}
(H_j)_{11} & (H_j)_{12} \\
(H_j)_{21} & (H_j)_{22} \\
\end{array}
\right],
\]
where $ (H_j)_{11} $ is a $ p_1\times p_1 $ matrix. 
We rewrite \eqref{bols} as
\begin{equation}\label{hdecom}
\left[
\begin{array}{ll}
(H_j)_{11}&(H_j)_{12}\\
(H_j)_{12}^T&(H_j)_{22}
\end{array}
\right]
\left[
\begin{array}{c}
\hat{\boldsymbol{\beta}}_{j_1}^{ols}-\boldsymbol{\beta}_{j_1}^{0}\\
\hat{\boldsymbol{\beta}}_{j_2}^{ols}-\boldsymbol{\beta}_{j_2}^{0}
\end{array}
\right]
=\left[
\begin{array}{c}
X_{j_1}^T(X_{\cdot j}-\boldsymbol{\mu}_{ j})\\
X_{j_2}^T(X_{\cdot j}-\boldsymbol{\mu}_{ j})
\end{array}
\right].
\end{equation}
Under $ H_{\alpha} $, we expand $ l\left( \hat{\boldsymbol{\beta}}_{\cdot j}^{H_{\alpha}}\right)  $ at $ \boldsymbol{\beta}_{\cdot j}^0$,
\begin{equation}\label{hor}
\begin{aligned}
l\left( \hat{\boldsymbol{\beta}}_{\cdot j}^{H_{\alpha}}\right)=&l\left( \boldsymbol{\beta}_{\cdot j}^{0}\right)+\left(X_{\cdot j}-\boldsymbol{\mu}_j^0 \right)^TX\left( \hat{\boldsymbol{\beta}}_{\cdot j}^{H_{\alpha}}-\boldsymbol{\beta}_{\cdot j}^0\right)-\frac{1}{2} \left( \hat{\boldsymbol{\beta}}_{\cdot j}^{H_{\alpha}}-\boldsymbol{\beta}_{\cdot j}^0\right)^T H_{j} \left( \hat{\boldsymbol{\beta}}_{\cdot j}^{H_{\alpha}}-\boldsymbol{\beta}_{\cdot j}^0\right)\\
&+\frac{1}{3!}G^{(3)}(\bar{\boldsymbol{\beta}}_{\cdot j}^0)\times_1\left( \hat{\boldsymbol{\beta}}_{\cdot j}^{H_{\alpha}}-\boldsymbol{\beta}_{\cdot j}^0\right)\times_2\left( \hat{\boldsymbol{\beta}}_{\cdot j}^{H_{\alpha}}-\boldsymbol{\beta}_{\cdot j}^0\right)\times_3\left( \hat{\boldsymbol{\beta}}_{\cdot j}^{H_{\alpha}}-\boldsymbol{\beta}_{\cdot j}^0\right).
\end{aligned}
\end{equation}
Expanding $ \frac{\partial l_j}{\partial \hat{\boldsymbol{\beta}}_{\cdot j}} $ at $ \boldsymbol{\beta}_{\cdot j}^{0} $ , we have 
\[
\frac{\partial l\left( \hat{\boldsymbol{\beta}}_{\cdot j}^{H_{\alpha}}\right)}{\partial \boldsymbol{\beta}_{\cdot j}}=X^T\left(X_{\cdot j}-\boldsymbol{\mu}_j^0 \right)-H_j\left( \hat{\boldsymbol{\beta}}_{\cdot j}^{H_{\alpha}}-\boldsymbol{\beta}_{\cdot j}^0\right)+\frac{1}{2}G^{(3)}(\tilde{\boldsymbol{\beta}}_{\cdot j}^0)\times_1\left( \hat{\boldsymbol{\beta}}_{\cdot j}^{H_{\alpha}}-\boldsymbol{\beta}_{\cdot j}^0\right)\times_2\left( \hat{\boldsymbol{\beta}}_{\cdot j}^{H_{\alpha}}-\boldsymbol{\beta}_{\cdot j}^0\right),
\]
where $\bar{\boldsymbol{\beta}}_{\cdot j} $ is on the line segment joining $\boldsymbol{\beta}_{\cdot j}  $ and $ \boldsymbol{\beta}_{\cdot j}^0 $.
Then we have
\begin{align*}
\hat{\boldsymbol{\beta}}^{H_{\alpha}}_{\cdot j}-\boldsymbol{\beta}^{0}_{\cdot j}=&
\hat{\boldsymbol{\beta}}_{\cdot j}^{ols}- \boldsymbol{\beta}^{0}_{\cdot j}+\frac{1}{2} H_j^{-1}G^{(3)}(\bar{\boldsymbol{\beta}}_{\cdot j}^0)\times_1\left( \hat{\boldsymbol{\beta}}_{\cdot j}^{H_{\alpha}}-\boldsymbol{\beta}_{\cdot j}^0\right)\times_2\left( \hat{\boldsymbol{\beta}}_{\cdot j}^{H_{\alpha}}-\boldsymbol{\beta}_{\cdot j}^0\right),
\end{align*}
and 
\begin{equation}\label{rep}
X^T\left(X_{\cdot j}-\boldsymbol{\mu}_j^0 \right)=H_j\left( \hat{\boldsymbol{\beta}}_{\cdot j}^{H_{\alpha}}-\boldsymbol{\beta}_{\cdot j}^0\right)-\frac{1}{2}G^{(3)}(\tilde{\boldsymbol{\beta}}_{\cdot j}^0)\times_1\left( \hat{\boldsymbol{\beta}}_{\cdot j}^{H_{\alpha}}-\boldsymbol{\beta}_{\cdot j}^0\right)\times_2\left( \hat{\boldsymbol{\beta}}_{\cdot j}^{H_{\alpha}}-\boldsymbol{\beta}_{\cdot j}^0\right).
\end{equation}
Combine \eqref{hor} and \eqref{rep}, we have
\begin{align*}
l\left( \hat{\boldsymbol{\beta}}_{\cdot j}^{H_{\alpha}}\right)-l\left( \boldsymbol{\beta}_{\cdot j}^{0}\right)=&\left(X_{\cdot j}-\boldsymbol{\mu}_j^0 \right)^TX\left( \hat{\boldsymbol{\beta}}_{\cdot j}^{H_{\alpha}}-\boldsymbol{\beta}_{\cdot j}^0\right)-\frac{1}{2} \left( \hat{\boldsymbol{\beta}}_{\cdot j}^{H_{\alpha}}-\boldsymbol{\beta}_{\cdot j}^0\right)^TH_{j}^0 \left( \hat{\boldsymbol{\beta}}_{\cdot j}^{H_{\alpha}}-\boldsymbol{\beta}_{\cdot j}^0\right)\\
&+\frac{1}{3!}G^{(3)}(\bar{\boldsymbol{\beta}}_{\cdot j}^0)\times_1\left(\hat{\boldsymbol{\beta}}_{\cdot j}^{H_{\alpha}}-\boldsymbol{\beta}_{\cdot j}^0\right)\times_2\left( \hat{\boldsymbol{\beta}}_{\cdot j}^{H_{\alpha}}-\boldsymbol{\beta}_{\cdot j}^0\right)\times_3\left( \hat{\boldsymbol{\beta}}_{\cdot j}^{H_{\alpha}}-\boldsymbol{\beta}_{\cdot j}^0\right)\\
=& \left( \hat{\boldsymbol{\beta}}_{\cdot j}^{H_{\alpha}}-\boldsymbol{\beta}_{\cdot j}^0\right)^T H_{j}^0 \left( \hat{\boldsymbol{\beta}}_{\cdot j}^{H_{\alpha}}-\boldsymbol{\beta}_{\cdot j}^0\right)-\frac{1}{2} \left( \hat{\boldsymbol{\beta}}_{\cdot j}^{H_{\alpha}}-\boldsymbol{\beta}_{\cdot j}^0\right)^T H_{j}^0 \left( \hat{\boldsymbol{\beta}}_{\cdot j}^{H_{\alpha}}-\boldsymbol{\beta}_{\cdot j}^0\right)\\
&-\frac{1}{3}G^{(3)}(\bar{\boldsymbol{\beta}}_{\cdot j}^0)\times_1\left(\hat{\boldsymbol{\beta}}_{\cdot j}^{H_{\alpha}}-\boldsymbol{\beta}_{\cdot j}^0\right)\times_2\left( \hat{\boldsymbol{\beta}}_{\cdot j}^{H_{\alpha}}-\boldsymbol{\beta}_{\cdot j}^0\right)\times_3\left( \hat{\boldsymbol{\beta}}_{\cdot j}^{H_{\alpha}}-\boldsymbol{\beta}_{\cdot j}^0\right)\\
=&\frac{1}{2}\left( \hat{\boldsymbol{\beta}}_{{\cdot j}}^{ols}-\boldsymbol{\beta}_{{\cdot j}}^{0}\right)^T H_j \left( \hat{\boldsymbol{\beta}}_{{\cdot j}}^{ols}-\boldsymbol{\beta}_{{\cdot j}}^{0}\right)+o(\|\hat{\boldsymbol{\beta}}^{H_{\alpha}}_{\cdot j}-\boldsymbol{\beta}^0_{\cdot j}\|^2).
\end{align*}
Similarly, under $ H_0 $, we have $ \boldsymbol{\beta}_{\cdot j}^0=(\boldsymbol{\beta}_{j_1}^{0T},\bm{0}^T)^T $, we expand $ l\left(\hat{\boldsymbol{\beta}}_{\cdot j_1}^{H_{0}}\right)  $ at $ \boldsymbol{\beta}_{\cdot j_1}^0$,
\begin{align*}
l\left( \hat{\boldsymbol{\beta}}_{\cdot j_1}^{H_{0}}\right)=&l\left( \boldsymbol{\beta}_{\cdot j_1}^{0}\right)+\left(X_{\cdot j}-\boldsymbol{\mu}_j^0 \right)^TX_{j_1}\left( \hat{\boldsymbol{\beta}}_{\cdot j_1}^{H_{0}}-\boldsymbol{\beta}_{\cdot j_1}^0\right)-\frac{1}{2} \left( \hat{\boldsymbol{\beta}}_{\cdot j_1}^{H_{0}}-\boldsymbol{\beta}_{\cdot j_1}^0\right)^T\left( H_{j}\right)_{11} \left( \hat{\boldsymbol{\beta}}_{\cdot j_1}^{H_{0}}-\boldsymbol{\beta}_{\cdot j_1}^0\right)\\
&+\frac{1}{3!}G^{(3)}(\bar{\boldsymbol{\beta}}_{\cdot j}^0)\times_1\left( \hat{\boldsymbol{\beta}}_{\cdot j_1}^{H_{0}}-\boldsymbol{\beta}_{\cdot j_1}^0\right)\times_2\left( \hat{\boldsymbol{\beta}}_{\cdot j_1}^{H_{0}}-\boldsymbol{\beta}_{\cdot j_1}^0\right)\times_3\left( \hat{\boldsymbol{\beta}}_{\cdot j_1}^{H_{0}}-\boldsymbol{\beta}_{\cdot j_1}^0\right).
\end{align*}
Expanding  $\frac{\partial l\left( \hat{\boldsymbol{\beta}}_{\cdot j_1}^{H_{0}}\right) }{\partial \boldsymbol{\beta}_{\cdot j_1}}  $ at $ \boldsymbol{\beta}_{\cdot j_1}^0 $,
\begin{align*}
\frac{\partial l\left( \hat{\boldsymbol{\beta}}_{\cdot j_1}^{H_{0}}\right) }{\partial \boldsymbol{\beta_{\cdot j_1}}}=&X_{j_1}^T\left(X_{\cdot j}-\boldsymbol{\mu}_j^0 \right)-\left( H_{j}\right)_{11} \left( \hat{\boldsymbol{\beta}}_{\cdot j_1}^{H_{0}}-\boldsymbol{\beta}_{\cdot j_1}^0\right)+\frac{1}{2}G^{(3)}(\bar{\boldsymbol{\beta}}_{\cdot j}^0)\times_1\left( \hat{\boldsymbol{\beta}}_{\cdot j_1}^{H_{0}}-\boldsymbol{\beta}_{\cdot j_1}^0\right)\times_2\left( \hat{\boldsymbol{\beta}}_{\cdot j_1}^{H_{0}}-\boldsymbol{\beta}_{\cdot j_1}^0\right).
\end{align*}
Let $\frac{\partial l\left( \hat{\boldsymbol{\beta}}_{\cdot j_1}^{H_{0}}\right) }{\partial \boldsymbol{\beta_{\cdot j_1}}}=0  $, under \eqref{hdecom}, we have
\[
\begin{aligned}
\hat{\boldsymbol{\beta}}_{\cdot j_1}^{H_0}-\boldsymbol{\beta}_{\cdot j_1}^0&=(H_j)_{11}^{-1}X_{j_1}^T\left(X_{\cdot j}-\boldsymbol{\mu}_j^0 \right)+\frac{1}{2} (H_j)_{11}^{-1}G^{(3)}(\bar{\boldsymbol{\beta}}_{\cdot j}^0)\times_1\left( \hat{\boldsymbol{\beta}}_{\cdot j_1}^{H_{0}}-\boldsymbol{\beta}_{\cdot j_1}^0\right)\times_2\left( \hat{\boldsymbol{\beta}}_{\cdot j_1}^{H_{0}}-\boldsymbol{\beta}_{\cdot j_1}^0\right) \\
&=(H_j)_{11}^{-1}\left((H_j)_{11} \left( \hat{\boldsymbol{\beta}}_{j_1}^{ols}-\boldsymbol{\beta}_{j_1}^{0}\right)+(H_j)_{12}\hat{\boldsymbol{\beta}}_{j_2}^{ols}  \right) +\frac{1}{2} (H_j)_{11}^{-1}G^{(3)}(\bar{\boldsymbol{\beta}}_{\cdot j}^0)\times_1\left( \hat{\boldsymbol{\beta}}_{\cdot j_1}^{H_{0}}-\boldsymbol{\beta}_{\cdot j_1}^0\right)\times_2\left( \hat{\boldsymbol{\beta}}_{\cdot j_1}^{H_{0}}-\boldsymbol{\beta}_{\cdot j_1}^0\right) \\
&=\hat{\boldsymbol{\beta}}_{\cdot j_1}^{ols}-\boldsymbol{\beta}_{\cdot j_1}^0+ (H_j)_{11}^{-1}(H_j)_{12}\hat{\boldsymbol{\beta}}_{j_2}^{ols} +\frac{1}{2} H_j^{-1}G^{(3)}(\bar{\boldsymbol{\beta}}_{\cdot j}^0)\times_1\left( \hat{\boldsymbol{\beta}}_{\cdot j}^{H_{0}}-\boldsymbol{\beta}_{\cdot j}^0\right)\times_2\left( \hat{\boldsymbol{\beta}}_{\cdot j}^{H_{0}}-\boldsymbol{\beta}_{\cdot j}^0\right),
\end{aligned}
\]
and 
\begin{align*}
l\left( \hat{\boldsymbol{\beta}}_{\cdot j_1}^{H_{0}}\right)-l\left( {\boldsymbol{\beta}}_{\cdot j_1}^{0}\right)=&\left(X_{\cdot j}-\boldsymbol{\mu}_j^0 \right)^TX_{j_1}\left( \hat{\boldsymbol{\beta}}_{\cdot j_1}^{H_{0}}-\boldsymbol{\beta}_{\cdot j_1}^0\right)-\frac{1}{2} \left( \hat{\boldsymbol{\beta}}_{\cdot j_1}^{H_{0}}-\boldsymbol{\beta}_{\cdot j_1}^0\right)^T\left( H_{j}\right)_{11} \left( \hat{\boldsymbol{\beta}}_{\cdot j_1}^{H_{0}}-\boldsymbol{\beta}_{\cdot j_1}^0\right)\\
&+\frac{1}{3!}G^{(3)}(\bar{\boldsymbol{\beta}}_{\cdot j}^0)\times_1\left( \hat{\boldsymbol{\beta}}_{\cdot j_1}^{H_{0}}-\boldsymbol{\beta}_{\cdot j_1}^0\right)^T\times_2\left( \hat{\boldsymbol{\beta}}_{\cdot j_1}^{H_{0}}-\boldsymbol{\beta}_{\cdot j_1}^0\right)^T\times_3\left( \hat{\boldsymbol{\beta}}_{\cdot j_1}^{H_{0}}-\boldsymbol{\beta}_{\cdot j_1}^0\right)^T\\
=&\left( \hat{\boldsymbol{\beta}}_{\cdot j_1}^{H_{0}}-\boldsymbol{\beta}_{\cdot j_1}^0\right)^T\left( H_{j}\right)_{11} \left( \hat{\boldsymbol{\beta}}_{\cdot j_1}^{H_{0}}-\boldsymbol{\beta}_{\cdot j_1}^0\right)
 -\frac{1}{2} \left( \hat{\boldsymbol{\beta}}_{\cdot j_1}^{H_{0}}-\boldsymbol{\beta}_{\cdot j_1}^0\right)^T\left( H_{j}\right)_{11} \left( \hat{\boldsymbol{\beta}}_{\cdot j_1}^{H_{0}}-\boldsymbol{\beta}_{\cdot j_1}^0\right)\\
 &-\frac{1}{2}G^{(3)}(\bar{\boldsymbol{\beta}}_{\cdot j}^0)\times_1\left( \hat{\boldsymbol{\beta}}_{\cdot j_1}^{H_{0}}-\boldsymbol{\beta}_{\cdot j_1}^0\right)\times_2\left( \hat{\boldsymbol{\beta}}_{\cdot j_1}^{H_{0}}-\boldsymbol{\beta}_{\cdot j_1}^0\right)\times_3\left( \hat{\boldsymbol{\beta}}_{\cdot j_1}^{H_{0}}-\boldsymbol{\beta}_{\cdot j_1}^0\right)\\
&+\frac{1}{3!}G^{(3)}(\bar{\boldsymbol{\beta}}_{\cdot j}^0)\times_1\left( \hat{\boldsymbol{\beta}}_{\cdot j_1}^{H_{0}}-\boldsymbol{\beta}_{\cdot j_1}^0\right)\times_2\left( \hat{\boldsymbol{\beta}}_{\cdot j_1}^{H_{0}}-\boldsymbol{\beta}_{\cdot j_1}^0\right)\times_3\left( \hat{\boldsymbol{\beta}}_{\cdot j_1}^{H_{0}}-\boldsymbol{\beta}_{\cdot j_1}^0\right)\\
=&\frac{1}{2} \left( \hat{\boldsymbol{\beta}}_{\cdot j_1}^{H_{0}}-\boldsymbol{\beta}_{\cdot j_1}^0\right)^T\left( H_{j}\right)_{11} \left( \hat{\boldsymbol{\beta}}_{\cdot j_1}^{H_{0}}-\boldsymbol{\beta}_{\cdot j_1}^0\right)\\
&-\frac{1}{3}G^{(3)}(\bar{\boldsymbol{\beta}}_{\cdot j}^0)\times_1\left( \hat{\boldsymbol{\beta}}_{\cdot j_1}^{H_{0}}-\boldsymbol{\beta}_{\cdot j_1}^0\right)\times_2\left( \hat{\boldsymbol{\beta}}_{\cdot j_1}^{H_{0}}-\boldsymbol{\beta}_{\cdot j_1}^0\right)\times_3\left( \hat{\boldsymbol{\beta}}_{\cdot j_1}^{H_{0}}-\boldsymbol{\beta}_{\cdot j_1}^0\right)\\
=&\frac{1}{2}(\hat{\boldsymbol{\beta}}_{j_1}^{ols}-\boldsymbol{\beta}_{j_1}^{0})^T(H_j)_{11}(\hat{\boldsymbol{\beta}}_{{j_1}}^{ols}-\boldsymbol{\beta}_{{j_1}}^{0})+\frac{1}{2}(\hat{\boldsymbol{\beta}}_{{j_2}}^{ols})^T(H_j)^T_{12}(H_j)_{11}^{-1}(H_j)_{12}\hat{\boldsymbol{\beta}}_{{j_2}}^{ols}\\
&\quad+(\hat{\boldsymbol{\beta}}_{{j_1}}^{ols}-\boldsymbol{\beta}_{{j_1}}^{0})^T(H_j)_{12}\hat{\boldsymbol{\beta}}_{{j_2}}^{ols}+o(\|\hat{\boldsymbol{\beta}}^{H_0}_{\cdot j}-\boldsymbol{\beta}^0_{\cdot j}\|^2).
\end{align*}
Then under $ H_0 $, the $ j $-th part of log-likelihood ratio $ lr_j $ can be rewrite as
\begin{align*}
lr_j =&\frac{1}{2}\left( \hat{\boldsymbol{\beta}}_{{\cdot j}}^{ols}-\boldsymbol{\beta}_{{\cdot j}}^{0}\right)^T H_j \left( \hat{\boldsymbol{\beta}}_{{\cdot j}}^{ols}-\boldsymbol{\beta}_{{\cdot j}}^{0}\right)+o(\|\hat{\boldsymbol{\beta}}^{H_{\alpha}}_{\cdot j}-\boldsymbol{\beta}^0_{\cdot j}\|^2)\\
&-\frac{1}{2}(\hat{\boldsymbol{\beta}}_{j_1}^{ols}-\boldsymbol{\beta}_{j_1}^{0})^T(H_j)_{11}(\hat{\boldsymbol{\beta}}_{{j_1}}^{ols}-\boldsymbol{\beta}_{{j_1}}^{0})-\frac{1}{2}(\hat{\boldsymbol{\beta}}_{{j_2}}^{ols})^T(H_j)^T_{12}(H_j)_{11}^{-1}(H_j)_{12}\hat{\boldsymbol{\beta}}_{{j_2}}^{ols}\\
&\quad-(\hat{\boldsymbol{\beta}}_{{j_1}}^{ols}-\boldsymbol{\beta}_{{j_1}}^{0})^T(H_j)_{12}\hat{\boldsymbol{\beta}}_{{j_2}}^{ols}-o(\|\hat{\boldsymbol{\beta}}^{H_0}_{\cdot j}-\boldsymbol{\beta}^0_{\cdot j}\|^2)\\
=&\frac{1}{2} \left( \hat{\boldsymbol{\beta}}_{\cdot j_1}^{ols}-\boldsymbol{\beta}_{\cdot j_1}^0\right)^T\left( H_{j}\right)_{11} \left( \hat{\boldsymbol{\beta}}_{\cdot j_1}^{ols}-\boldsymbol{\beta}_{\cdot j_1}^0\right)+\left( \hat{\boldsymbol{\beta}}_{\cdot j_1}^{ols}-\boldsymbol{\beta}_{\cdot j_1}^0\right)^T(H_j)_{12}\hat{\boldsymbol{\beta}}_{\cdot j_2}^{ols}+\frac{1}{2} \left( \hat{\boldsymbol{\beta}}_{\cdot j_2}^{ols}\right)^T\left( H_{j}\right)_{22}  \hat{\boldsymbol{\beta}}_{\cdot j_2}^{ols}\\
&-\frac{1}{2}(\hat{\boldsymbol{\beta}}_{j_1}^{ols}-\boldsymbol{\beta}_{j_1}^{0})^T(H_j)_{11}(\hat{\boldsymbol{\beta}}_{{j_1}}^{ols}-\boldsymbol{\beta}_{{j_1}}^{0})-\frac{1}{2}(\hat{\boldsymbol{\beta}}_{{j_2}}^{ols})^T(H_j)^T_{12}(H_j)_{11}^{-1}(H_j)_{12}\hat{\boldsymbol{\beta}}_{{j_2}}^{ols}\\
&\quad-(\hat{\boldsymbol{\beta}}_{{j_1}}^{ols}-\boldsymbol{\beta}_{{j_1}}^{0})^T(H_j)_{12}\hat{\boldsymbol{\beta}}_{{j_2}}^{ols}-o(\|\hat{\boldsymbol{\beta}}^{H_0}_{\cdot j}-\boldsymbol{\beta}^0_{\cdot j}\|^2)\\
=&\frac{1}{2}(\hat{\boldsymbol{\beta}}_{{j_2}}^{ols})^T\left[ (H_j)_{22}-(H_j)^T_{12}(H_j)_{11}^{-1}(H_j)_{12}\right] \hat{\boldsymbol{\beta}}_{{j_2}}^{ols}+o(\|\hat{\boldsymbol{\beta}}^{H_{\alpha}}_{\cdot j}-\boldsymbol{\beta}^{H_0}_{\cdot j}\|^2) .
\end{align*}

\subsection*{Technical proofs}
\begin{prf}[Proof of Theorem \ref{thm:iden}]
In this part of
the proof, we don't assume faithfulness. Instead, we assume that the coefficients $\beta_{kj}$ does not vanish for any $k\in V_j$, i.e., if node $k$ is a parent of node $j$, $\beta_{kj}\neq 0$. Condition
and Proposition 2 in \cite{peters_2011} show that this condition implies causal minimality, which is a natural condition and in accordance with the intuitive understanding of a causal influence between variables.

The idea of the proof is to assume that there are two structural equation models as in equation \eqref{eq:generate} that both induce $\mathcal{L}(\boldsymbol{\beta})$, one with graph $\mathcal{G}$, the other with graph $\mathcal{G}^{\prime}$. We will show that $\mathcal{G}=\mathcal{G}^{\prime}$.

We start from  a simple case. Assume that there are two structural equation models with distinct graphs $\mathcal{G}$ and $\mathcal{G}_0$ that lead to the same joint distribution as displayed in Figure \ref{fig:h1} with reverse edges between variables $X_1$ and $X_3$. We can  show that $X_3$ has different
variances in both graphs. This leads to a contradiction.
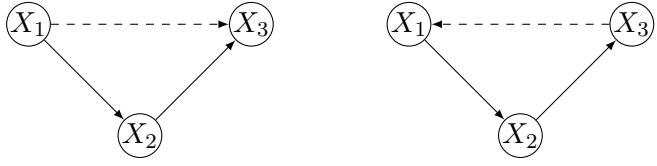
\begin{figure}
    \centering
    \begin{tikzpicture}[
          node distance=0.5cm,
          mynode/.style={circle, draw = black, inner sep = 0pt, minimum size = 0.5cm}
       ]
      \node[mynode] (C) {$X_2$};
      \node[mynode, above left= 1.5 cm
      of C] (A) {$X_1$};
      \node[mynode, above right= 1.5 cm of C] (D) {$X_3$};

      \node[mynode, right= 1.5 cm of D] (E) {$X_1$};

      \node[mynode, below right= 1.5 cm of E] (F) {$X_2$};

      \node[mynode, above right = 1.5 cm of F] (G) {$X_3$};
      \path [draw, -latex] (A) edge (C);
      \path [draw, -latex] (C) edge (D);
      \path [draw, -latex][dashed] (A) edge (D);

      \path [draw, -latex] (E) edge (F);
      \path [draw, -latex] (F) edge (G);
      \path [draw, -latex][dashed] (G) edge (E);

      \end{tikzpicture}
    \caption{\label{fig:h1}Two Poisson graphs $\mathcal{G}$ and $\mathcal{G}^{\prime}$ with different edges.}
\end{figure}
Assume that random variables $X_1,X_2,X_3$ are generated from model \eqref{eq:generate}. For variable $X_3$ in graph $\mathcal{G}$ we have
\begin{align*}
Var(X_3\mid X_2)&=E(Var(X_3 \mid  X_2)\mid X_1)+Var(E(X_3 \mid X_2)\mid X_1)\\
&\ge E(X_3\mid X_2),
\end{align*}
since that $E(Var(X_3 \mid  X_2)\mid X_1)= E(X_3\mid X_2)$ and $Var(E(X_3 \mid X_2)\mid X_1)>0$.
Similarly for variable $X_3$ in graph $\mathcal{G}^{\prime}$ we have that
\[
E(X_3 | X_2)=Var(X_3 | X_2),
\]
since $\{X_2\}$ is parent set of $X_3$. Node $X_3$  has different
variances in both graphs, the identifiability is satisfied.

Now we turn to the more general cases. For $(X_1,\dots,X_p)$ generated from model \eqref{eq:generate}, we have that for a variable $X_j, \; j=1,\dots,p$
\begin{itemize}
    \item[i:] $\operatorname{Var}(X_j\mid X_{-j})=\operatorname{Var}(X_j\mid V_j)\leq \operatorname{Var}(X_j)$.
    \item[ii:] For any variable $X_l,l\neq j$, if $X_l$ is not a descendent of $X_j$, i.e. $\beta_{jl}\neq 0$, we have that $X_l \perp \!\!\! \perp X_j\mid V_j$.
    \item[iii:] If variable $X_k$ is a parent of $X_j$, for all sets $S$ with
    $V_j\setminus \{X_k\} \subseteq S \subseteq ND_j\setminus \{X_k\}$, $X_j\not\perp \!\!\! \perp X_k\mid S$, where $ND_j=\{l:\beta_{jl}=0\}$ denote the non-descendent set of $X_j$.
\end{itemize}
Then follow the proof of Theorem 1 in \cite{PB_2014} with (i)-(iii) above, we have $\mathcal{G}=\mathcal{G}^{\prime}$ with same joint distribution $\mathcal{L}(\boldsymbol{\beta})$.
\end{prf}

\begin{lem}[Test for linkages]\label{lm1}
	Assume Assumptions are met. If $ \kappa=|E^{0}\setminus D| $ and $ \tau\leq (c_4C_{\min}/2pc_6)^{(1/c_5)}  $, then under $ H_0 $, as $ n\to \infty $,
\begin{align*}
&\max\left( P\left(\hat{ \boldsymbol{\beta}}_{H_0}\neq \hat{ \boldsymbol{\beta}}_{E^0} \right),P\left(\hat{ \boldsymbol{\beta}}_{H_{\alpha}} \neq \hat{ \boldsymbol{\beta}}_{E^0\cup D^0}\right) \right) \\
\leq & \exp \left(-d_{3} c_{4} n C_{\min } / 2+2 \log \left(\left(p^{2}-p+1\right)+|F| \log 2+1\right)\right) \to 0.
\end{align*}
\end{lem}
\begin{prf}[Proof of lemma \ref{lm1}]. 
Without loss of generality, we only prove the bound for $P\left( \boldsymbol{\beta}_{H_{\alpha}} \neq \boldsymbol{\beta}_{E^0\cup D^0} \right) $ as the proof for other case is similar.

Define a measure for the size of a space $ \mathcal{F} $. The bracketing Hellinger metric entrop of $ \mathcal{F} $, denoted by the function $ H(\cdot,\mathcal{F}) $, is defined by logarithm of the cardinality of the $ u- $bracketing of $ \mathcal{F} $ of smallest size. That is, for a bracketing covering $ S(\epsilon,m) =\{(f_1^l,f_1^u),\dots,(f_m^l,f_m^u)\} \subset \mathcal{L}_2$ satisfying $ \max_{1\leq j\leq m} \|f_j^u-f_j^l\|_2 \leq \epsilon$ for any $ f\in \mathcal{F} $, there exists a $ j $ such that $ f_j^l\leq f\leq f_j^u ,a.e. P$, then $ H(u,\mathcal{F}) $ is $ \log \left(\min \{m:S(u,m)\}\right) $, where $ \|f\|_2=\int f^2(z)du. $

Let $ S_1^{\tau}=\{(j,k)\in F^c: |\boldsymbol{\beta}_{jk}|\geq \tau\} $. When $ \kappa =|E^0\setminus F|=|E^0| $, $\sum_{(j,k)\in F^c} J_{\tau}(\boldsymbol{\beta}_{jk})\leq |E^0|$, so $ |S_1^{\tau}|\leq  |E^0| $. If $\hat{E}_{H_{\alpha}^{\tau}}\setminus F=E^0$, then $ \sum_{(j,k)\in F^c}\left| \hat{\boldsymbol{\beta}}_{jk}^{H_{\alpha}}\right|\mathbb{I}\left(\left|\hat{\boldsymbol{\beta}}_{jk}^{H_{\alpha}} \right|<\tau \right)=0$, then we obtain that $ \hat{\boldsymbol{\beta}}^{H_{\alpha}}=\hat{\boldsymbol{\beta}}_{E^0\cup D^0} $. Therefore we only need to show the case when $\hat{E}_{H_{\alpha}^{\tau}}\setminus F\neq E^0$.

For any $ S=S_1\cup S_2 $, where $ S_1 \subset F^c $ and $ S_2\subset F $.  we partition $ S_1 $ into two parts, where $ S_1=\left( S_1\cap E^0\right)\cup \left( S_1\setminus E^0\right)   $. Let $ {\bf k}=(k_1,k_2,k_3)  $, and 
\[ 
B_{\bf k} =\left\{ \boldsymbol{\beta}_{\tau} :S_1\neq E^0,\left|S_1 \cap E^0\right|=k_1,\left|S_1 \setminus E^0\right|=k_2,S_2\subset D,|S_2|=k_3, c_6 p\tau^{c_5} \geq c_4(|E^0|-k_1) C_{\min}-h^2(\boldsymbol{\beta}_{\tau},\boldsymbol{\beta}^0)  \right\};
\] 
we have that $ k_1=0,\cdots,|E^0|-1 , k_2=1,\cdots,|E^0|-k_1, k_3=0,\cdots,|F|$. Thus, $ B_{\bf k} $ consists of elements with $ \binom{|E^0|}{k_1}\binom{p(p-1)|E^0|}{k_2} \binom{|F|}{k_3} $ different supports. Then
\begin{align*}
P(\hat{\boldsymbol{\beta}}_{H_{\alpha}}\neq \hat{\boldsymbol{\beta}}_{E^0\cup D^0})&\leq P^{*}\left(\sup_{\boldsymbol{\beta}_{S_1\cup S_2}:S_1\neq E^0,|S_1|\leq |E^0|}l\left(\hat{\boldsymbol{\beta}}_{S_1\cup S_2}\right)-l\left(\hat{\boldsymbol{\beta}}_{E^0\cup D^0}\right) \geq 0\right)\\ 
&\leq P^{*}\left(\sup_{\boldsymbol{\beta}_{S_1\cup S_2}:S_1\neq E^0,|S_1|\leq |E^0|} l\left({\boldsymbol{\beta}}_{S_1\cup S_2}\right)-l\left({\boldsymbol{\beta}}^0\right) \geq 0 \right) \\
&\leq \sum_{k_1=0}^{|E^0|-1}\sum_{k_2=1}^{|E|^0-k_1}\sum_{k_3=0}^{|F|} P^{*}\left( \sup_{\boldsymbol{\beta}_{S_1\cup S_2}\in B_{\bf k}} l(\boldsymbol{\beta}_{S_1\cup S_2})-l(\boldsymbol{\beta}^0)\geq 0 \right) \equiv I,
\end{align*}
where $ P^{*}(\cdot) $ is the outer probability.
Let $ \mathcal{F}_k=\{f^{1/2}(\boldsymbol{\beta},\cdot): \boldsymbol{\beta}\in B_{\bf k}\} $, where $ f(\boldsymbol{\beta}) $ is the pdf of our model. Note that 
\[
f(\boldsymbol{\beta})=\prod_{j=1}^{p}P(X_j=x_j |\{X_i,i\in V_j\})=\prod_{j=1}^{p}e^{\bm{x}\boldsymbol{\beta_{\cdot j}}}\frac{\exp\left[x_j(\bm{x}\boldsymbol{\beta_{\cdot j}})\right]}{x_j!},
\]
Then 
\begin{align*}
&\frac{1}{2}\int \sup_{\bar{\boldsymbol{\beta}}:\|\bar{\boldsymbol{\beta}}-\boldsymbol{\beta}\|\leq \delta} \left|f^{1/2}(\bar{\boldsymbol{\beta}})-f^{1/2}({\boldsymbol{\beta}})   \right|^2dx\\
&=\frac{1}{2}\int \sup_{\bar{\boldsymbol{\beta}}:\|\bar{\boldsymbol{\beta}}-\boldsymbol{\beta}\|\leq \delta}
f(\bar{\boldsymbol{\beta}})+f({\boldsymbol{\beta}})-2\sqrt{f(\bar{\boldsymbol{\beta}})f({\boldsymbol{\beta}})} dx\\
&=1-\int \sup_{\bar{\boldsymbol{\beta}}:\|\bar{\boldsymbol{\beta}}-\boldsymbol{\beta}\|\leq \delta}
\sqrt{f(\bar{\boldsymbol{\beta}})f({\boldsymbol{\beta}})} dx\\
&=1-\int \sup_{\bar{\boldsymbol{\beta}}:\|\bar{\boldsymbol{\beta}}-\boldsymbol{\beta}\|\leq \delta}\sqrt{\prod_{j=1}^{p}e^{{\exp(\bm{x}\boldsymbol{\beta}_{\cdot j})}-\exp(\bm{x}\bar{\boldsymbol{\beta}}_{\cdot j})} e^{x_j(\bm{x}\bar{\boldsymbol{\beta}}_{\cdot j}-\bm{x}\boldsymbol{\beta}_{\cdot j})}} f(\boldsymbol{\beta})  dx\\
&=1-\sup_{\bar{\boldsymbol{\beta}}:\|\bar{\boldsymbol{\beta}}-\boldsymbol{\beta}\|\leq \delta}\prod_{j=1}^{p}\exp\left(-\frac{1}{2}\left(e^{\frac{1}{2}\bm{x}\boldsymbol{\beta}_{\cdot j}}-e^{\frac{1}{2}\bm{x}\bar{\boldsymbol{\beta}}_{\cdot j}} \right)^2 \right)\\
&\leq c p^2\delta^2 .
\end{align*}
According to lemma 2.1 of \cite{os_1987}. Let
$\{B_{\delta}(\boldsymbol{\beta}): \boldsymbol{\beta} \in B_k(\delta)\}$ be a collection of $ \delta $-neighborhoods of the members of $ B_k(\delta) $ which covers $ B_k$. Suppose that for each $ \boldsymbol{\beta} \in B_k(\delta) $
\[
E^{*}\left[\sup _{t \in B_{\delta}(s)}|f(s, V)-f(t, V)|^{2}\right]^{1 / 2} \leq g(\delta)
\]
for some strictly increasing continuous function $g : [0, \infty) \to [0,\infty)$. ($ E^{*} $ denotes
upper expectation), then for $ u >0$
\[
H^{B}(u,B_k, \rho) \leq H\left(g^{-1}(u / 2), B_k, \rho\right),
\]
where $ (B_k,\rho) $ is a metric space, $H(\cdot, B_k, \rho)$ metric entropy of $ B_k $ with respect to $ \rho $ and $H^B(\cdot, S, \rho)$ denotes the metric entropy with bracketing of $ f $ with respect to $ \rho $ in $ L_2 $. This implies $ H^B(u,\mathcal{F}_k)\leq H(u/pc^{'},B_{\delta}(\boldsymbol{\beta})) $ where $ \bar{\boldsymbol{\beta}}\in B_{\delta}(\boldsymbol{\beta}) $.\\
We have that the bracketing $ L_2 $ entropy is bounded by the $ L_2 $-metric entropy $ H(u,\mathcal{F}) $ of $ \mathcal{F}_k $. 
There are $ |S| $ nonzero entries of $ \boldsymbol{\beta} $ with $ \binom{p(p-1)}{|S|} $ possible locations. then for $ u>\epsilon^2 $, by \cite{ko_1959},
\begin{align*}
H(u,\mathcal{F}_k)&\leq c_0\left\{ \log\binom{p(p-1)}{|S|}+|S|\log \frac{c^{'}p^2\min(c_3^{1/2},1)}{u} \right\}\\
&\leq c_0\left\{ |S|\log \frac{ep(p-1)}{|S|}+|S|\log\frac{c^{'}p^2\min(c_3^{1/2},1)}{u} \right\}\\
&\leq c_0(|S|\log p\log(\frac{1}{u})),
\end{align*}
for that $ \binom{n}{m}\leq \left(\frac{en}{m} \right)^m  $.\\
According to Theorem 1 of \cite{wong_1995} There exists positive constants $ d_i ,i=1,2,3,4$, such that for any $ \epsilon>0 $ , if 
\[
\int_{\epsilon^2/2^8}^{\sqrt{2}\epsilon} H^{1/2}(u/d_1,\mathcal{F}_k)du\leq d_4n^{1/2}\epsilon^2,
\]
then
\[
P^{*}\left(\underset{\left\{\left\|p^{1 / 2}-p_{0}^{1 / 2}\right\|_{2} \geq \varepsilon, p \in \mathscr{F}_{n}\right\}}{\sup } \prod_{i=1}^{n} p\left(Y_{i}\right) / p_{0}\left(Y_{i}\right) \geq \exp \left(-d_{2} n \varepsilon^{2}\right)\right) \leq 4 \exp \left(-d_{3} n \varepsilon^{2}\right).
\]
Let $ \epsilon^{1+\delta}=\min (1,\sqrt{2c_0}d_4^{-1}\log(d_1/\sqrt{2})\sqrt{(|E^0|+|F|)/n}) $, for $ \delta>0 $, we have $ C_{\min}>\epsilon^2 $, provided that $ c_1^{-1/2}>(2c_0)^{1/2}d_4^{-1}\log(d_1/\sqrt{2}) $, then
\begin{align*}
\sup_{\bf k}\int_{\epsilon^2/2^8}^{\sqrt{2}\epsilon}& H^{1/2}(s/d_1,\mathcal{F}_k)ds\\
\leq& \sup_{\bf k}\int_{\epsilon^2/2^8}^{\sqrt{2}\epsilon}\sqrt{ c_0(|S|\log p\log(\frac{d_1}{s})) }ds\\
=&\sqrt{c_0 |S|\log p}\int_{\epsilon^2/2^8}^{\sqrt{2}\epsilon} \sqrt{\log d_1-\log s}ds\\
\leq& \sqrt{c_0 |S|\log p}\int_{\epsilon^2/2^8}^{\sqrt{2}\epsilon}\left(\log d_1-\log s \right)ds\\
=&\sqrt{c_0 |S|\log p}\left(s\log d_1+s-s\log s \right)\left.\right|_{\epsilon^2/2^8}^{\sqrt{2}\epsilon}\\
\leq &\sqrt{c_0 |S|\log p}\sqrt{2}\epsilon\log(c_1/\sqrt{2}) \\
\leq &d_4\sqrt{\log p}d_4^{-1}\sqrt{c_0(|E^0|+|F|)}\sqrt{2}\epsilon\log(c_1/\sqrt{2})\\
\leq& d_4n^{1/2}\epsilon^2.
\end{align*} 
By theorem 2.6 of \cite{Stanica_2001},
\[
\binom{b}{a}\leq \exp\{(a+1/2)\log(b/a)+a\}
\]
for any integers $ a<b $. Using that
\[
\sum_{j=0}^{|E^0|-k}\binom{p(p-1)-|E^0|}{j}\leq \{p(p-1)-|E^0|+1\}^{|E^0|-k},\quad \binom{|E^0|}{i}\leq |E^0|^{i}.
\]
We have that for some contant $ c_3>0 $,
\begin{align*} 
I \leq & 4 \sum_{k_{1}=0}^{\left|E^{0}\right|-1}\sum_{k_2=1}^{|E^0|-k_1} \sum_{k_{3}=0}^{|F|}\binom{\left|E^{0}\right|}{k_1} \binom{p(p-1)-|E^0|}{k_2}\binom{|F|}{k_3} \\
& \times \exp \left(-d_{3} n\left(c_{4}(|E^0|-k_1) C_{\min }-c_{6} p \tau^{c_{5}}\right)\right) \\ 
\leq & \sum_{k_1=1}^{\left|E^{0}\right|-1} 4 \exp \left(-d_{3} n\left( c_{4}(|E^0|-k_1) C_{\min }-c_{6} p \tau^{c_{5}}\right)+k_1\log \left|E^{0}\right|\right.\\
&\left.+(|E^0|-k_1)\left(\log \left(p(p-1)-\left|E^{0}\right|+1\right)\right)+|F|\log 2 \right) \\
\leq & \sum_{k=1}^{\left|E^{0}\right|} 4 \exp \left(-d_{3} n\left( c_{4} k C_{\min }-c_{6} p \tau^{c_{5}}\right)+k\log \left|E^{0}\right|+k\left(\log \left(p(p-1)-\left|E^{0}\right|+1\right)\right)+|F|\log 2 \right) \\
\leq & \sum_{k=1}^{\left|E^{0}\right|} 4 \exp \left(- d_{3} c_{4} n k C_{\min } / 2+k\left(\log \left|E^{0}\right|+\log \left(p(p-1)-\left|E^{0}\right|+1\right)\right)+|F| \log 2\right) \\
\leq &4R\left( - d_{3} c_{4} n  C_{\min } / 2+\left(\log \left|E^{0}\right|+\log \left(p(p-1)-\left|E^{0}\right|+1\right)\right)+|F| \log 2\right) ,
\end{align*}
where $ \log\{p(p-1)-|E^0|+1\}+\log|E^0|\leq 2\log[\{p(p-1)+1\}/2] $ and $ R(x)=\frac{e^x}{1-e^x} $. Then we have
\begin{align*}
 4R&(- d_{3} c_{4} n  C_{\min } / 2+\left(\log \left|E^{0}\right|+\log \left(p(p-1)-\left|E^{0}\right|+1\right)\right)+|F| \log 2)\\
&\leq 5\exp\left(- d_{3} c_{4} n  C_{\min } / 2+2\log[\{p(p-1)+1\}/2]+|F| \log 2\right)\\
&\leq \exp\left(- d_{3} c_{4} n  C_{\min } / 2+2\log\left(\{p(p-1)+1\}\right)+|F| \log 2+1\right).
\end{align*}
Hence 
\begin{equation}\label{concentration}
P(\hat{\boldsymbol{\beta}}_{H_{\alpha}}\neq \hat{\boldsymbol{\beta}}_{E^0\cup D^0}) \leq  \exp \left(-d_{3} c_{4} n C_{\min } / 2+2 \log \left(p^{2}-p+1\right)+|F| \log 2+1\right) .
\end{equation}
Similarly we bounded $ P(\hat{\boldsymbol{\beta}}_{H_{0}}\neq \hat{\boldsymbol{\beta}}_{E^0}). $ This completes the proof.
\end{prf}
\begin{prf}[Proof of theorem \ref{thm1} The asymptotic distribution of $ \boldsymbol{\beta}^{ols} $]
We use  Characteritic Function to derive the asymptotic distribution of $
\hat{\boldsymbol{\beta}}_{\cdot j}^{ols}-{\boldsymbol{\beta}}_{\cdot j}^{0} $. 
For any real $T=\left( t_1,\cdots,t_p\right) ^T$, the characteristic function $ \phi(T) $ of $ X^T(X_{\cdot j}-\mu_j)$ is
\begin{align*}
\phi_{X^T(X_{\cdot j}-\mu_j^0  )}(T)=&E\left[\exp\left(iT^T X^T\left( X_{\cdot j}-\boldsymbol{\mu}_j\right)  \right)  \right] \\
=&E\left[1+iT^T X^T\left( X_{\cdot j}-\boldsymbol{\mu}_j\right) +\frac{1}{2}\left(iT^T X^T\left( X_{\cdot j}-\boldsymbol{\mu}_j\right) \right)^2  \right.\\
&\left.\quad + \frac{1}{6} e^{\theta *}\left(iT^T X^T\left( X_{\cdot j}-\boldsymbol{\mu}_j\right) \right)^3
\right]\\
=&\exp\left(-\frac{1}{2}T^TH_jT+\frac{1}{6}e^{\theta *}\left(iT^T X^T\left( X_{\cdot j}-\boldsymbol{\mu}_j\right) \right)^3\right),
\end{align*}
Then we have 
\[
\phi_{H_j^{\frac{1}{2}}(\hat{\boldsymbol{\beta}}^{ols}_{\cdot j}-\boldsymbol{\beta}_{\cdot j}^0)}(T)=\exp\left(-\frac{1}{2}T^T T-\frac{i}{6}\frac{1}{n^{\frac{3}{2}}}\sum_{h=1}^{n}\lambda_{hj}\left(X_{h\cdot}(\frac{1}{n}H_j)^{-\frac{1}{2}}T\right)^3 e^{\theta ** }\right) 
\]
where $ \theta^{*} $ and $ \theta^{**} $ is on the line joining 0 and $ iT^T X^T\left( X_{\cdot j}-\boldsymbol{\mu}_j\right) $.\\
 The number of nonzero entries of $ \boldsymbol{\beta}_{\cdot j}^0 $ is $ |j_1| $, without loss of generality, we assume that the first $ |j_1| $ elements of $ \boldsymbol{\beta}_{\cdot j} $ are nonzeros and the rest are zeros. When $ p$ is fixed,
\[
(\hat{\boldsymbol{\beta}}^{ols}_{\cdot j}-{\boldsymbol{\beta}}_{\cdot j}^0)(\frac{1}{n}H_j)(\hat{\boldsymbol{\beta}}^{ols}_{\cdot j}-{\boldsymbol{\beta}}^0_{\cdot j})^T \ {\xrightarrow {\text{d}}}\ \chi^2_{|j_1|}.
\]
When $ p\to \infty $ as $ n\to \infty $, let $ (\hat{\theta}_1,\cdots ,\hat{\theta}_{p})^T=H_j^{\frac{1}{2}}(\hat{\boldsymbol{\beta}}^{ols}_{\cdot j}-{\boldsymbol{\beta}}^0_{\cdot j}) $. Our goal is to show that
\[
\hat{\Theta}_{p}=\frac{1}{\sqrt{2|j_1|}}(\hat{\theta}_1^2+\cdots +\hat{\theta}_{|j_1|}^2)-\sqrt{\frac{|j_1|}{2}}\ {\xrightarrow {\text{d}}}\ N(0,1).
\]
Define $ F_{(\hat{\theta}_1,\cdots,\hat{\theta}_{|j_1|})}\left( x_1,\cdots,x_{|j_1|}\right) \overset{\bigtriangleup}{=} F_n\left( x_1,\cdots,x_{|j_1|}\right)  $.
We need to show that for some $ \delta>0 $
\[
\left| F_n\left( x_1,\cdots,x_{|j_1|}\right) - F_{N({\bf 0},\mathcal{I}_{|j_1|})}\left( x_1,\cdots,x_{|j_1|}\right) \right| \leq O(n^{-\delta}) .
\]
Then we obtain that 
\begin{align*}
A&=P\left\lbrace \frac{1}{\sqrt{2|j_1|}}\left(\hat{\theta}_1^2+\cdots+\hat{\theta}_{|j_1|}^2 \right)-\sqrt{\frac{|j_1|}{2}}\leq y \right\rbrace \\
&= \underset{x_1^2+\cdots+x_{|j_1|}^2\leq\sqrt{2|j_1|}y+|j_1|}{\int\cdots\int} 1~ d~  F_n\left( x_1,\cdots,x_{|j_1|}\right),
\end{align*}
and
\[
\left|A-\underset{x_1^2+\cdots+x_{|j_1|}^2\leq\sqrt{2|j_1|}y+|j_1|}{\int\cdots\int} 1 ~d~  F_{N({\bf 0},\mathcal{I}_{|j_1|})}\left( x_1,\cdots,x_{|j_1|}\right) \right|  \leq O(n^{-\delta}) .
\]
Set $S_n=H_j^{\frac{1}{2}}\left(\hat{\boldsymbol{\beta}}^{ols}_{\cdot j}-{\boldsymbol{\beta}}_{\cdot j}^0 \right)= H_j^{-\frac{1}{2}} X^T\left(X_{\cdot j}-\boldsymbol{\mu}_j^0 \right)$ with $ \operatorname {E} (S_n)={\bf 0} $ and $ \Sigma= \mathcal{I}_{|j_1|}$
\[
\gamma=E {\left [\| H_j^{-\frac{1}{2}} X^T\left(X_{\cdot j}-\boldsymbol{\mu}_j^0 \right)\|_{2}^{3}\right ]}=\sum_{h=1}^{n}\lambda_{hj} ||H_j^{-1/2} X_{h\cdot}^T||_2^3.
\]
Then define $ F_{(n)} $ to be the CDF of $ H_j^{-\frac{1}{2}}X^T(X_{\cdot j}^T-\boldsymbol{\mu}_j)$, by \cite{go_1991} we have 
\[
\left| F_{(n)}\left( x_1,\cdots,x_p\right) - F_{N({\bf 0},\mathcal{I}_p)}\left( x_1,\cdots,x_p\right) \right| \leq C |j_1|^{1/4}\gamma=\frac{1}{n^{3/2}}C |\mathcal{A}_j|^{1/4}\sum_{h=1}^{n}\lambda_{hj}||(\frac{H_j}{n})^{-1/2} X_{h\cdot}^T||_2^3.
\]
Under the condition that the third order moment is bounded and $\frac{|E^0|}{n^{a}}\to 0$ with $ 0<a<1 $, then $ \beta_3 <\infty $  , we have 
\[
\left| F_{(n)}\left( x_1,\cdots,x_{|j_1|}\right) - F_{N({\bf 0},\mathcal{I}_{|j_1|})}\left( x_1,\cdots,x_{|j_1|}\right) \right| \leq O(n^{a/4-1/2}).
\]
Therefore,
\[
\left|A-\underset{x_1^2+\cdots+x_{|j_1|}^2\leq\sqrt{2|j_1|}y+|V_j|}{\int\cdots\int} 1 ~d~  F_{N({\bf 0},\mathcal{I}_{|j_1|})}\left( x_1,\cdots,x_{|j_1|}\right) \right|  \leq O(n^{a/4-1/2}), 
\]
which means 
\[ \sqrt{2|j_1|}\left(\hat{\boldsymbol{\beta}}^{ols}_{\cdot j}-{\boldsymbol{\beta}}_{\cdot j}^0 \right) H_j \left(\hat{\boldsymbol{\beta}}^{ols}_{\cdot j}-{\boldsymbol{\beta}}_{\cdot j}^0 \right)^T-\frac{|j_1|}{2} \xrightarrow{d} N(0,1), 
\]
when $ |j_1|\to \infty $ as $ n\to \infty $.
Next, we need to show that $ \hat{\boldsymbol{\beta}}^{ols}_{\cdot j} $ has same asymptotic as $ \hat{\boldsymbol{\beta}}^{tlp}_{\cdot j} $ from \eqref{linkh0}.\\
Firstly, we give some conditions to show that with probability tends to 1 that $ \hat{\boldsymbol{\beta}}^{ols}_{\cdot j} $ converges to true value $ {\boldsymbol{\beta}}^{0}_{\cdot j} $.\\
C1: There exists a constant $ c_q $, $ 0<c_q<1 $, such that $ |V_j|=O(n^{c_q}) $, where $ |V_j| $ is number of the nonzero parameters\\
C2: There exists a positive constant $ c_{\beta} $, $ c_q<c_{\beta}\leq 1 $, such that 
\[
n^{-\frac{1-c_{\beta}}{2}}=O(\min_{i\in V_j}|\beta_{ji}|).
\]
We first observe that the estimator $ \hat{\boldsymbol{\beta}}_{\cdot j}^{ols} $ has the form
\[
\hat{\boldsymbol{\beta}}_{\cdot j}^{ols} =\boldsymbol{\beta}_{\cdot j}^0+H_j^{-1}X\left(X_{\cdot j}-\mu_j^0 \right).
\]
Now we are to show that 
\[
n^{-\frac{1-c_{\beta}}{2}}=O_p(\min_{j\in\{1,\cdots,p\},i\in V_j}|\hat{{\beta}}_{ji}^{ols}|).
\]
Since $ |\hat{{\beta}}_{ij}^{ols}|\geq |{{\beta}}_{ij}^0|-|\hat{\boldsymbol{\beta}}_{ij}^{ols}-{{\beta}}_{ij}^0| $, by condition C2, we need to show that
\begin{equation}\label{ols}
\max_{j\in\{1,\cdots,p\},i\in V_j}|\hat{{\beta}}_{ij}^{ols}-{{\beta}}_{ij}^0| =o_p(n^{-\frac{1-c_{\beta}}{2}}).
\end{equation}
let $ z_{ij}=\sqrt{n}(\hat{{\beta}}_{ij}^{ols}-{{\beta}}_{ij}^0 )$.
Equation \eqref{ols} is equivalent to 
\[
\max_{j\in\{1,\cdots,p\},i\in V_j}|z_{ij}| =o_p(n^{-\frac{1-c_{\beta}}{2}}).
\]
Let $ \bm{z}_j=(z_{1j},\cdots,z_{pj})^T $ , we have $ \bm{z}_j=\frac{1}{\sqrt{n}}(H_j/n)^{-1}X^T\left(X_{\cdot j}-\mu_j^0 \right) $. Furthermore, assume our observation has finite 2r-th moment, which is $ E(X_j^{2r})<M_r ,j=1,\cdots,p$, then we have $ E(z_{ij}^{2r})<C_r $ for all $j=1,\cdots,p, i\in V_j $, then we have
\[
P(|z_{ij}|>t)=O(t^{-2r}).
\]
For any $ \alpha>0 $
\begin{align*}
&P(|z_{ij}|>\alpha n^{c_{\beta}/2} \text{ for some } i \in V_j)\\
\leq &\sum_{j=1}^{p} \sum_{i\in V_j} P(|z_i|>\alpha n^{c_{\beta}/2})\\
= &\sum_{j=1}^{p} \sum_{i\in V_j} O(\alpha^{-2r}n^{-c_{\beta}r})\\
=&  O(|E|\alpha^{-2r}n^{-c_{\beta}r})\\
\leq & C n^{-(c_{\beta}-c_q)r} \text{ for some }C>0,
\end{align*}
which leads to
\[
\delta_0=P(\max_{j\in\{1,\cdots,p\},i\in V_j}|\hat{{\beta}}_{ij}^{ols}-{{\beta}}_{ij}^0|>\alpha n^{-(1-c_{\beta})/2} )\leq C n^{-(c_{\beta}-c_q)r} .
\]
With condition 2 we have $n^{-(c_{\beta}-c_q)r}\to 0  $ as $ n\to \infty $, which leads to \eqref{ols}, thus implying that $ \hat{\boldsymbol{\beta}}^{ols} _{\cdot j}$ converges to $ \boldsymbol{\beta}^0_{\cdot j} $ with probability $ 1-\delta_0 $ tending to 1 as $ n\to \infty $.\\
Ignoring the penalty part, we have that 
\[
\frac{\partial l_j({\boldsymbol{\beta}},{\bf 0})}{\partial \boldsymbol{\beta}_{\cdot j}}=X^T(X_{\cdot  j}-\boldsymbol{\mu}_j),
\]
for all $ j\in\{1,\cdots p\} $, which means $ X^T\boldsymbol{\mu}_j(\hat{\boldsymbol{\beta}}^{oracle})= X^TX_{\cdot  j}$. Then similarly to \cite{fan_2014}, we
define a map $ T: D\to D$, satisfies that $ T(\Delta_j)=\left( T(\Delta_{j_1})^T ,\bm{0}\right) ^T $ with
$T(\Delta_{j_1})=\left( X_{j_1}^T\Gamma_j X_{j_1}\right)^{-1}X_{j_1}^T(X_{\cdot j}-\boldsymbol{\mu}_j({\boldsymbol{\beta}}_{\cdot j}^{0}+\Delta_j))+\Delta_{j_1}  $ and $ D=\left\lbrace \Delta_j \in \mathbb{R}^p:\| \Delta_{j_1}\|_{\max}\leq r,\Delta_{j_2}={\bf 0} \right\rbrace $ with $ r=2Q_1\cdot\left\| \frac{1}{n}X_{j_1}^T\left(X_{\cdot j}-\boldsymbol{\mu}_j({\boldsymbol{\beta}}_{\cdot j}^{0})\right)\right\|_{\max}$, where $ Q_1=\left\| \left(\frac{1}{n}H_j \right)^{-1} \right\|_{\ell_{\infty} } $.
We need to show that
\begin{equation}\label{TD}
T(D)\subset D,
\end{equation}
which leads to that there always exists a point $ \hat{\Delta_j}\in D $ such that $ T(\hat{\Delta_j})=\hat{\Delta_j} $. It follows that $ X^T_{j_1} \boldsymbol{\mu}_j({\boldsymbol{\beta}}_{\cdot j}^{0}+\hat{\Delta}_j)=X_{j_1}^T X$ and $ \hat{\Delta}_{j_2}=0 $, which leads to that ${\boldsymbol{\beta}}_{\cdot j}^{0}+\hat{\Delta_j}=\hat{\boldsymbol{\beta}}_{\cdot j}^{oracle}  $. \\
Thus we have
\[
||\hat{\boldsymbol{\beta}}_{\cdot j}^{oracle}-{\boldsymbol{\beta}}_{\cdot j}^{0}||_{\max}=||\hat{\Delta}_j||_{\max} \leq r.
\]
We now derive \eqref{TD}. Using Taylor expansion around $ \Delta=\bm{0} $, we have
\[
X_{j_1}^T\boldsymbol{\mu}_j\left({\boldsymbol{\beta}}_{\cdot j}^{0}+\Delta_j\right)=X_{j_1}^T\boldsymbol{\mu}_j\left({\boldsymbol{\beta}}_{\cdot j}^{0}\right)+X_{j_1}^T\Gamma_j\left({\boldsymbol{\beta}}_{\cdot j}^{0}\right)X_{j_1}\Delta_j+R_{j_1}(\tilde{\Delta_j})
\] 
Where
\[
||R_{j_1}(\tilde{\Delta}_j)||_{\max}\leq \max_k \Delta_{k_1}X_{k_1}^T \text{ diag}\{X_{\cdot k}\circ\Gamma_j (\bar{\boldsymbol{\beta}_{\cdot j}} )\} X_{j_1}\Delta_{j_1}
\]
for $ \bar{\boldsymbol{\beta}_{\cdot j}} $ being on the line segment joining $ \boldsymbol{\beta}_{\cdot j}^{0}  $ and $ \boldsymbol{\beta}_{\cdot j}^0+\tilde{\Delta_j} $.
Set $ Q_2=\max_j\lambda_{\max}\left( \frac{1}{n}X_{j_1}^T\text{ diag }\{|X_{(j)}|\}X_{j_1}\right) $, we have that
\[
R_{j_1}(\tilde{\Delta_j})\leq nQ_2||\Delta_{j_1}||^2\leq nQ_2|j_1|r^2.
\]
Note that 
\begin{align*}
T(\Delta_{j_1})&=\left( X_{j_1}^T\Gamma_j X_{j_1}\right)^{-1}X_{j_1}^T(X_{\cdot j}-\boldsymbol{\mu}_j({\boldsymbol{\beta}}_{\cdot j}^{0}+\Delta_j))+\Delta_{j_1}\\
&=\left( X_{j_1}^T\Gamma_j X_{j_1}\right)^{-1}X_{j_1}^T\left(X_{\cdot j}-\boldsymbol{\mu}_j({\boldsymbol{\beta}}_{\cdot j}^{0})-R_{j_1}(\tilde{\Delta_j}) \right).
\end{align*}
Therefore
\begin{align*}
||T(\Delta_{j_1})||_{\max}=&\left\|\left( X_{j_1}^T\Gamma_j X_{j_1}\right)^{-1}X_{j_1}^T\left(X_{\cdot j}-\boldsymbol{\mu}_j({\boldsymbol{\beta}}_{\cdot j}^{0})-R_{j_1}(\tilde{\Delta_j}) \right)\right\|_{\max}\\
\leq& Q_1 \left( \left\|\frac{1}{n}X_{j_1}^T\left(X_{\cdot j}-\boldsymbol{\mu}_j({\boldsymbol{\beta}}_{\cdot j}^{0}) \right)\right\|_{\max}+\frac{1}{n} \left\|R_{j_1}(\tilde{\Delta_j}) \right\|_{\max}  \right) \\
\leq & \frac{r}{2}+ Q_1Q_2|j_1|r^2.
\end{align*}
Then when 
\[
\left\|\frac{1}{n}X_{j_1}^T\left(X_{\cdot j}-\boldsymbol{\mu}_j({\boldsymbol{\beta}}_{\cdot j}^{0}) \right)\right\|_{\max}\leq \frac{1}{4Q_1^2Q_2|j_1|}.
\]
We have $ ||T(\Delta_{j_1})||_{\max}\leq r $ which leads to \eqref{TD}. Under the events $ r \leq c ||\boldsymbol{\beta}_{\cdot j}^0||_{\min}$ for $ c<1 $ we have $ ||\boldsymbol{\beta}_{\cdot j}^{oracle}||_{\min}\geq (1-c)||\boldsymbol{\beta}_{\cdot j}^0||_{\min}$ 
Then under the event that $ \left\|\frac{1}{n}X_{j_1}^T\left(X_{\cdot j}-\boldsymbol{\mu}_j({\boldsymbol{\beta}}_{\cdot j}^{0}) \right)\right\|_{\max}\leq \frac{1}{Q_1^2Q_2|j_1|} $ and $ r \leq c ||\boldsymbol{\beta}_{\cdot j}^0||_{\min}$, by hoeffding's bound, set $ \delta_1=P(||\hat{\boldsymbol{\beta}}_{\cdot j}^{oracle}-{\boldsymbol{\beta}}_{\cdot j}^{0}||_{\max}\geq c ||\boldsymbol{\beta}_{\cdot j}^0||_{\min}  ) $we have
\begin{align*}
\delta_1&=P\left(\left\|\frac{1}{n}X_{j_1}^T\left(X_{\cdot j}-\boldsymbol{\mu}_j({\boldsymbol{\beta}}_{\cdot j}^{0}) \right)\right\|_{\max}>\min\left\{ \frac{1}{4Q_1^2Q_2|j_1|},\frac{c||\boldsymbol{\beta}_{\cdot j}^0||_{\min}}{2Q_1}  \right\}  \right)\\
&=P\left(\left\|\frac{1}{n}(H_j)^{-1/2}X_{j_1}^T\left(X-\boldsymbol{\mu}_j({\boldsymbol{\beta}}_{\cdot j}^{0}) \right)\right\|_{\max}>\min\left\{ \frac{\|H_j^{-1/2}\|_{\max}}{4Q_1^2Q_2|j_1|},\frac{c\|H_j^{-1/2}\|_{\max}||\boldsymbol{\beta}_{\cdot j}^0||_{\min}}{2Q_1}  \right\}  \right)\\
&=P\left(\left\|(H_j)^{-1/2}X_{j_1}^T\left(X_{\cdot j}-\boldsymbol{\mu}_j({\boldsymbol{\beta}}_{\cdot j}^{0}) \right)\right\|_{\max}>\min\left\{ \frac{n^{1/2}\|{H_j/n}^{-1/2}\|_{\max}}{4Q_1^2Q_2|j_1|},\frac{cn^{1/2}\|(H_j/n)^{-1}\|_{\max}||\boldsymbol{\beta}_{\cdot j}^0||_{\min}}{2Q_1}  \right\}  \right)\\
&\leq  2|j_1|\exp\left( -n\min\left\{ \frac{1}{8Q_1^2Q_2^2|j_1|^2},\frac{c^2||\boldsymbol{\beta}_{\cdot j}^0||^2_{\min}}{2}  \right\}\right).
\end{align*}
Thus we have that $\hat{ \boldsymbol{\beta}}_{\cdot j}^{oracle} $ converges to $ \boldsymbol{\beta}_{\cdot j}^0 $  with probability $ 1-\delta_1 $ tending to 1 as $ n\to \infty $.
Combine with that $\hat{ \boldsymbol{\beta}}_{\cdot j}^{tlp} $ converges to $\hat{ \boldsymbol{\beta}}_{\cdot j}^{oracle} $ with probability $ 1-\delta_2 $ tending to 1 as $ n\to \infty $, where $ \delta_2 $ defined as \eqref{concentration},
now we have that $ \hat{\boldsymbol{\beta}}^{tlp} $ converges to $ \hat{\boldsymbol{\beta}}^{ols} $ with probability $ 1-\delta_0-\delta_1-\delta_2 $ tending to 1 as $ n\to \infty $.
In our likelihood-ratio test, $ E^{0} $ is the set of edges of ture graph $ \mathcal{G}^{0} $ and $ D^0 $ is the set of all testable edges in F. Set $ V_{0} =E^0$ and $ V_{\alpha}=E^0 \cup D^0 $, when $ 
\min\{n,p\}\to\infty $
\begin{align*}
lr_j \approx &\frac{1}{2}(\hat{\boldsymbol{\beta}}_{D_j^0}^{ols})^T\left[ (H_j)_{22}-(H_j)^T_{12}(H_j)_{11}^{-1}(H_j)_{12}\right] \hat{\boldsymbol{\beta}}_{D_j^0}^{ols}.
\end{align*}
For a matrix has the form that 
\[
\left[
\begin{array}{cc}
A_{m\times m}&B{m\times n}\\
C_{n\times m}&D_{n\times n}
\end{array}
\right],
\]
we have 
\[
\left[
\begin{array}{cc}
A_{m\times m}&B{m\times n}\\
C_{n\times m}&D_{n\times n}
\end{array}
\right]^{-1}
=
\left[
\begin{array}{cc}
A^{-1}+A^{-1}B(D-CA^{-1}B)^{-1}CA^{-1}&-A^{-1}B(D-CA^{-1}B)^{-1}\\
(D-CA^{-1}B)^{-1}CA^{-1}&(D-CA^{-1}B)^{-1}
\end{array}
\right].
\]
Then when $ |D^0|>0 $ is fixed, 
\[
2lr_j\to \chi^2_{|D^0_j|},
\]
and
\[
2lr\to \chi^2_{|D^0|}.
\]
When $ |D^0|\to \infty $,with probability tending to 1,
\[
\sqrt{2|D^0|}(2lr-|D^0|)\to N(0,1).
\]
We complete the proof.
\end{prf}
\begin{lem}[Test of pathway]\label{lm2}
 Assume Assumptions \ref{asm1}, \ref{asm2}, \ref{asm3} are met. Let $ A=\{(i_k,i_{k+1})\in F: \boldsymbol{\beta}_{i_{k+1}i_k}=0\} $ and $ F_{-k}= F\setminus\{(i_k,i_{k+1})\} $. If $ \kappa=|E^0\setminus F| $ and $ \tau=(c_4C_{\min}/2pc_6)^{(1/c_5)} $, then under $ H_0 $,
\begin{align*}
&P\left(\max_{k:(i_k,i_{k+1})\notin A}l(\hat{ \boldsymbol{\beta}}(k))-l(\boldsymbol{\beta}^0)\geq 0 \right) \to 0,\\
&\max\left(P\left(\exists k:\hat{E}_{H_0}(k)\neq E^0\cup F_{-k};(i_k,i_{k+1})\in A \right) ,P\left( \hat{E}_{H_{\alpha}}\neq E^0\cup F\right)  \right) \to 0.
\end{align*}
\end{lem}
\begin{prf}[Proof of lemma \ref{lm2}]
For any $ S $ that forms a DAG, we partition $ S $ into four parts, $ S=E_1\cup E_2\cup E_3\cup E_4 $, where $ E_1=S_1\cap E^0,E_2=S_1\setminus E^0, E_3=S_2\cap E^0,E_4=S_2\setminus E^0 $, where $ S_1 $ and $ S_2 $ are defined the same as in Assumption 1. Let $ \kappa_1=|E^0\setminus F| $ and $ \kappa_2=|E^0\cap F| $. Let $ {\bf k}=(k_1,k_2,k_3,k_4) $ and \[
B_{\bf k}=\{\boldsymbol{\beta}_{\tau}:S_1\cup S_2\nsupseteq E^0, |S_1|\leq |E^0\setminus F|, |E_i|=k_i;c_4C_{\min}-c_6p\tau^{c_5}\leq h^2(\boldsymbol{\beta}_{\tau},\boldsymbol{\beta}^0)\}  ,
\]
where $ 0\leq k_1\leq \kappa_1, 0\leq k_2\leq \kappa_1-k_1,0\leq k_3\leq \kappa_2,0\leq k_4\leq |F|-\kappa_2 $ and $ k_1+k_3\leq \kappa_1+\kappa_2=|E^0| $. Thus $ B_{\bf k} $ contains elements with $ \binom{\kappa_1}{k_1}\binom{p(p-1)-|F|-\kappa_1}{k_2}\binom{\kappa_2}{k_3}\binom{|F|-\kappa_2}{k_4} $ different supports. Then
\begin{align*}
&P\left(\max_{k:(i_k,i_{k+1})\notin A}l(\hat{ \boldsymbol{\beta}}(k))-l(\boldsymbol{\beta}^0)\geq 0 \right)\leq \sum_k P^{*}(\sup_{\boldsymbol{\beta}_{S_1\cup S_2}\in B_{\bf k}} l(\boldsymbol{\beta}_{S_1\cup S_2})-l(\boldsymbol{\beta}^0)\leq 0)\equiv I^{'},\\
&P(\hat{E}_{H_{\alpha}}\neq E^0\cup F)\leq I^{'},\\
&P\left(\exists k:\hat{E}_{H_0}(k)\neq E^0\cup F_{-k};(i_k,i_{k+1})\in A \right)\leq I^{'}.
\end{align*}
Then similar as Lemma \ref{lm1}, let $ \epsilon^{1+\delta}=\min(1,\sqrt{2c_0}c_4^{-1}\log(c_3/\sqrt{2})\sqrt{(|E^0|+|F|)/n}) $, we have 
\begin{align*}
I^{'}\leq& \sum_{k_1=0}^{\kappa_1}\sum_{k_2=0}^{\kappa_1-k_1}\sum_{k_3=0}^{\kappa_2}\sum_{k_4=0}^{|F|-\kappa_2}\binom{\kappa_1}{k_1}\binom{p(p-1)-|F|-\kappa_1}{k_2}\binom{\kappa_2}{k_3}\binom{|F|-\kappa_2}{k_4}\\
&\times 4\exp\left( -d_3n(c_4C_{\min}-c_6p\tau^{c_5})\right) \\
\leq & 4\sum_{k=1}^{|E^0|}\exp\left( -d_3c_4nC_{\min}/2 +k\left( \log\left(p(p-1)-|F| \right)+\log |E^0|\right) +|F|\log 2\right) \\
\leq &\sum_{k=1}^{|E^0|}2^{|F|+2}\exp\left( -d_3c_4nC_{\min}/2+k\left(\log |E^0|+2\log\left(p(p-1)-|F| \right)  \right) \right) \\
\leq & \exp \left(-d_3c_4nC_{\min}/2+\log |E^0|+2\log\left(\{p(p-1)+1\}\right)+|F|\log 2+1 \right) \to 0.
\end{align*}
We completes the proof.
\end{prf}
\begin{prf}[Proof of theorem \ref{thm2}]
	 It follows from lemma \ref{lm2} that $ P(\hat{E}_{H_{\alpha}}\setminus F=E^0\setminus F)\to 1 $ and $P( \hat{E}_{H_0}(k)\setminus F=E^0\setminus F)\to 1 $ for $ k $ such that $ \boldsymbol{\beta}^0_{i_{k+1}i_k}=0 $. Now consider that event $ \{\hat{E}_{H_{\alpha}}\setminus F=E^0\setminus F\} $.\\
	For (i), suppose $ E^0 \cup F$ dosen't form a DAG, then there exists $ 1\leq k\leq |F| $ such that $ (\hat{\boldsymbol{\beta}}_{H_{\alpha}})_{i_{k+1}i_k}=0 $, thus $ l(\hat{ \boldsymbol{\beta}}_{H_{\alpha}})=l(\hat{ \boldsymbol{\beta}}_{H_0}(k)) $, (i) is established.
	
	For (ii) and (iii), suppose $ E^0 \cup F$ forms a DAG, then for $ (i_k,i_{k+1}) \in E^0\cap F$ and $ (i_{k^{'}},i_{k^{,}+1}) \in F\setminus E^0$ we have $ P\left( l( \hat{ \boldsymbol{\beta}}_{H_0}(k)) <l( \hat{ \boldsymbol{\beta}}_{H_0}(k^{'}))\right)   \to 1$ . Consider the edge $ A=\{(i_k,i_{k+1})\in F:\boldsymbol{\beta}^0_{i_{K+1}i_k}=0\} $, similar as the proof of theorem 1, for any $ (i_k,i_{k+1})\in A $,we have
	\begin{align*}
	l(\hat{\boldsymbol{\beta}}_{H_{\alpha}})-l(\hat{\boldsymbol{\beta}}_{H_0}(k))=& \frac{1}{2}\beta_{i_{k+1}i_k}\left[(H)_{22}-(H)^T_{12}(H)^{-1}_{11}(H)_{12} \right]\beta_{i_{k+1}i_k},
	\end{align*}
	where $(H^0)_{11}=(H^0)_{i_{k+1}i_{k+1}} $. It follows that for each $ (i_k,i_{k+1})\in A $
	\[
	2\left( l(\hat{\boldsymbol{\beta}}_{H_{\alpha}})-l(\hat{\boldsymbol{\beta}}_{H_0}(k))\right)\xrightarrow{d}\chi^2_1 .
	\]
	Let $\{X_1,\dots,X_d\}$ be a series random variables following chi-square distribution with degree of freedom 1. 
	When $ d=|A| $ is fixed, $ 2lr\xrightarrow{d}\Gamma_d $ has same distribution as $ \min \{X_1,\cdots,X_d\}$, where $ X_i,i\in 1,\cdots,d $ are independent identically distributed $
	\chi^2_1$ variables.\\
	When $ d\to\infty $, let $ F_{\chi^2} $ be the distribution function of $ \chi^2_1 $ random variable, then for every $ x>0 $, 
	\[
 \lim_{d\to\infty} d\log(1-F_{\chi^2}(x/d^2))=-\lim_{d\to\infty}d\sum_{k=1}^{\infty}\frac{F^k_{\chi^2}(x/d^2)}{k}=-\sqrt{\frac{2x}{\pi }},
 \]
	which is 
	\[
	F_{\chi^2}(x/d^2)=1-\exp\left(- \sqrt{\frac{2x}{\pi d^2 }}\right). 
	\]
	Under that $ 2lr=\min\{X_1,X_2,\dots\} $, we set $ T=2d^2lr $ and $ F_{\chi^2}(t) $ be the CDF of $ \chi^2_1 $,
	\begin{align*}
	F_T(t)&=P(T\leq t)\\
	&=P(2d^2lr\leq t)\\
	&=P\left(\min\{d^2X_1,d^2X_2,\dots,d^2X_d\}\leq t\right)\\
	&=1-P\left(\min\{X_1,X_2,\dots,X_d\}\geq t/d^2 \right)\\
	&=1-\left( 1-F_{\chi^2}(t/d^2) \right)^d\\
	&=1-\exp\left(- \sqrt{\frac{2x}{\pi }}\right),
	\end{align*}
	hence we have 
	\[
	\lim_{d\to\infty} f_{2d^2lr}(x)=\sqrt{\frac{1}{2\pi x}}\exp(-\sqrt{2x/\pi})\mathbb{I}(x>0),
	\]
	which means
	\[
	2d^2lr\xrightarrow{d}\Gamma,
	\]
	where $ \Gamma $ is the generalized Gamma distribution, We complete the proof.
\end{prf}
\begin{prf}[Proof of theorem \ref{thm3}]
	Without loss of generality, assume $ X_1\Leftrightarrow\cdots\Leftrightarrow X_p $ is the partial order of DAG $ \mathcal{G}^0 $.  Now, we derive the likelihood ratio under $ H_{\alpha} $ where $ \boldsymbol{\beta}_{\cdot j}^0=(\boldsymbol{\beta}_{j_1}^{0T},\boldsymbol{\beta}_{j_2}^{0T})^T $. Expanding $	l\left( \hat{\boldsymbol{\beta}}_{\cdot j}^{H_{0}}\right)$ at $\boldsymbol{\beta}_{\cdot j}^{0}$,
	\begin{align*}
	l\left( \hat{\boldsymbol{\beta}}_{\cdot j}^{H_{0}}\right)-l\left( \boldsymbol{\beta}_{\cdot j}^{0}\right)=&\left(X_{\cdot j}-\boldsymbol{\mu}_j^0 \right)^TX\left( \hat{\boldsymbol{\beta}}_{\cdot j}^{H_{0}}-\boldsymbol{\beta}_{\cdot j}^0\right)-\frac{1}{2} \left( \hat{\boldsymbol{\beta}}_{\cdot j}^{H_{0}}-\boldsymbol{\beta}_{\cdot j}^0\right)^TH_{j}^0 \left( \hat{\boldsymbol{\beta}}_{\cdot j}^{H_{0}}-\boldsymbol{\beta}_{\cdot j}^0\right)\\
	&+\frac{1}{3!}G^{(3)}(\bar{\boldsymbol{\beta}}_{\cdot j}^0)\times_1\left(\hat{\boldsymbol{\beta}}_{\cdot j}^{H_{0}}-\boldsymbol{\beta}_{\cdot j}^0\right)\times_2\left( \hat{\boldsymbol{\beta}}_{\cdot j}^{H_{0}}-\boldsymbol{\beta}_{\cdot j}^0\right)\times_3\left( \hat{\boldsymbol{\beta}}_{\cdot j}^{H_{0}}-\boldsymbol{\beta}_{\cdot j}^0\right)\\
	=& \left(X_{\cdot j}-\boldsymbol{\mu}_j^0 \right)^TX_{j_1}\left( \hat{\boldsymbol{\beta}}_{\cdot j_1}^{H_{0}}-\boldsymbol{\beta}_{\cdot j_1}^0\right)-\left(X_{\cdot j}-\boldsymbol{\mu}_j^0 \right)^TX_{j_2}\boldsymbol{\beta}_{\cdot j_2}^0\\
	& -\frac{1}{2} \left( \hat{\boldsymbol{\beta}}_{\cdot j_1}^{H_{0}}-\boldsymbol{\beta}_{\cdot j_1}^0\right)^T(H_{j}^0)_{11} \left( \hat{\boldsymbol{\beta}}_{\cdot j_1}^{H_{0}}-\boldsymbol{\beta}_{\cdot j_1}^0\right)\\
 &-\frac{1}{2} \left( \boldsymbol{\beta}_{\cdot j_2}^0\right)^T(H_{j}^0)_{22} \boldsymbol{\beta}_{\cdot j_2}^0+ \left( \hat{\boldsymbol{\beta}}_{\cdot j_1}^{H_{0}}-\boldsymbol{\beta}_{\cdot j_1}^0\right)^T(H_{j}^0)_{12} \boldsymbol{\beta}_{\cdot j_2}^0\\
	&-\frac{1}{3!}G^{(3)}(\bar{\boldsymbol{\beta}}_{\cdot j}^0)\times_1\left(\hat{\boldsymbol{\beta}}_{\cdot j}^{H_{0}}-\boldsymbol{\beta}_{\cdot j}^0\right)\times_2\left( \hat{\boldsymbol{\beta}}_{\cdot j}^{H_{0}}-\boldsymbol{\beta}_{\cdot j}^0\right)\times_3\left( \hat{\boldsymbol{\beta}}_{\cdot j}^{H_{0}}-\boldsymbol{\beta}_{\cdot j}^0\right).
	\end{align*}
Expanding  $\frac{\partial l\left( \hat{\boldsymbol{\beta}}_{\cdot j_1}^{H_{0}},\bm{0}\right) }{\partial \boldsymbol{\beta}_{\cdot j_1}}  $ at $ \boldsymbol{\beta}_{\cdot j}^0 $,
\begin{align*}
\frac{\partial l\left( \hat{\boldsymbol{\beta}}_{\cdot j_1}^{H_{0}},\bm{0}\right) }{\partial \boldsymbol{\beta_{\cdot j_1}}}=&X_{j_1}^T\left(X_{\cdot j}-\boldsymbol{\mu}_j^0 \right)-\left( H_{j}\right)_{11} \left( \hat{\boldsymbol{\beta}}_{\cdot j_1}^{H_{0}}-\boldsymbol{\beta}_{\cdot j_1}^0\right)+(H_j)_{12}\boldsymbol{\beta}_{\cdot j_2}^0\\
&+\frac{1}{2}G^{(3)}(\bar{\bar{\boldsymbol{\beta}}}_{\cdot j}^0)\times_1\left( \hat{\boldsymbol{\beta}}_{\cdot j}^{H_{0}}-\boldsymbol{\beta}_{\cdot j}^0\right)\times_2\left( \hat{\boldsymbol{\beta}}_{\cdot j}^{H_{0}}-\boldsymbol{\beta}_{\cdot j}^0\right).
\end{align*}
Let $\frac{\partial l\left( \hat{\boldsymbol{\beta}}_{\cdot j_1}^{H_{0}},\bm{0}\right) }{\partial \boldsymbol{\beta_{\cdot j_1}}}=0  $, we have
\[
\begin{aligned}
\hat{\boldsymbol{\beta}}_{\cdot j_1}^{H_0}-\boldsymbol{\beta}_{\cdot j_1}^0&=(H_j)_{11}^{-1}X_{j_1}^T\left(X_{\cdot j}-\boldsymbol{\mu}_j^0 \right)+(H_j)_{11}^{-1}(H_j)_{12}\boldsymbol{\beta}_{\cdot j_2}^0\\
&+\frac{1}{2} (H_j)_{11}^{-1}G^{(3)}(\bar{\bar{\boldsymbol{\beta}}}_{\cdot j}^0)\times_1\left( \hat{\boldsymbol{\beta}}_{\cdot j}^{H_{0}}-\boldsymbol{\beta}_{\cdot j}^0\right)\times_2\left( \hat{\boldsymbol{\beta}}_{\cdot j}^{H_{0}}-\boldsymbol{\beta}_{\cdot j}^0\right) ,
\end{aligned}
\]
and under 
\[
\begin{array}{l} 
X_{j_1}^T(X_{\cdot j}-\boldsymbol{\mu}^0_{ j})=(H_j)_{11}\left(\hat{\boldsymbol{\beta}}_{{\cdot j_1}}^{ols}-\boldsymbol{\beta}_{{\cdot j_1}}^{0}\right)+(H_j)_{12}\left( \hat{\boldsymbol{\beta}}_{\cdot {j_2}}^{ols}-\boldsymbol{\beta}_{{\cdot j_2}}^{0}\right) \\ 
X_{j_2}^T(X_{\cdot j}-\boldsymbol{\mu}^0_{ j})=(H_j)_{21}(\hat{\boldsymbol{\beta}}_{{\cdot j_1}}^{ols}-\boldsymbol{\beta}_{{\cdot j_1}}^{0})+(H_j)_{22}\left( \hat{\boldsymbol{\beta}}_{{\cdot j_2}}^{ols}-\boldsymbol{\beta}_{{\cdot j_2}}^{0}\right),
\end{array}
\]
we have that
\[
\hat{\boldsymbol{\beta}}_{\cdot j_1}^{H_0}-\boldsymbol{\beta}_{\cdot j_1}^0=\left(\hat{\boldsymbol{\beta}}_{{\cdot j_1}}^{ols}-\boldsymbol{\beta}_{{\cdot j_1}}^{0}\right)+(H_j)^{-1}_{11}(H_j)_{12}\left( \hat{\boldsymbol{\beta}}_{{\cdot j_2}}^{ols}\right)+\frac{1}{2} (H_j)_{11}^{-1}G^{(3)}(\bar{\boldsymbol{\beta}}_{\cdot j}^0)\times_1\left( \hat{\boldsymbol{\beta}}_{\cdot j_1}^{H_{0}}-\boldsymbol{\beta}_{\cdot j_1}^0\right)\times_2\left( \hat{\boldsymbol{\beta}}_{\cdot j_1}^{H_{0}}-\boldsymbol{\beta}_{\cdot j_1}^0\right) \\
\]
Then 
\begin{align*}
&l\left( \hat{\boldsymbol{\beta}}_{\cdot j}^{H_{0}}\right)-l\left( \boldsymbol{\beta}_{\cdot j}^{0}\right)\\
=& \left(X_{\cdot j}-\boldsymbol{\mu}_j^0 \right)^TX_{j_1}\left( \hat{\boldsymbol{\beta}}_{\cdot j_1}^{H_{0}}-\boldsymbol{\beta}_{\cdot j_1}^0\right)-\left(X_{\cdot j}-\boldsymbol{\mu}_j^0 \right)^TX_{j_2}\boldsymbol{\beta}_{\cdot j_2}^0\\
& -\frac{1}{2} \left( \hat{\boldsymbol{\beta}}_{\cdot j_1}^{H_{0}}-\boldsymbol{\beta}_{\cdot j_1}^0\right)^T(H_{j}^0)_{11} \left( \hat{\boldsymbol{\beta}}_{\cdot j_1}^{H_{0}}-\boldsymbol{\beta}_{\cdot j_1}^0\right)-\frac{1}{2} \left( \boldsymbol{\beta}_{\cdot j_2}^0\right)^T(H_{j}^0)_{22} \boldsymbol{\beta}_{\cdot j_2}^0+ \left( \hat{\boldsymbol{\beta}}_{\cdot j_1}^{H_{0}}-\boldsymbol{\beta}_{\cdot j_1}^0\right)^T(H_{j}^0)_{12} \boldsymbol{\beta}_{\cdot j_2}^0\\
&-\frac{1}{3}G^{(3)}(\bar{\boldsymbol{\beta}}_{\cdot j}^0)\times_1\left(\hat{\boldsymbol{\beta}}_{\cdot j}^{H_{0}}-\boldsymbol{\beta}_{\cdot j}^0\right)\times_2\left( \hat{\boldsymbol{\beta}}_{\cdot j}^{H_{0}}-\boldsymbol{\beta}_{\cdot j}^0\right)\times_3\left( \hat{\boldsymbol{\beta}}_{\cdot j}^{H_{0}}-\boldsymbol{\beta}_{\cdot j}^0\right)\\
=&\left(\hat{\boldsymbol{\beta}}_{{j_1}}^{ols}-\boldsymbol{\beta}_{{j_1}}^{0}\right)^T(H_j)_{11}\left( \hat{\boldsymbol{\beta}}_{\cdot j_1}^{H_{0}}-\boldsymbol{\beta}_{\cdot j_1}^0\right)+\left( \hat{\boldsymbol{\beta}}_{{j_2}}^{ols}-\boldsymbol{\beta}_{{j_2}}^{0}\right)^T(H_j)_{21}\left( \hat{\boldsymbol{\beta}}_{\cdot j_1}^{H_{0}}-\boldsymbol{\beta}_{\cdot j_1}^0\right)\\
&-\left(\hat{\boldsymbol{\beta}}_{{j_1}}^{ols}-\boldsymbol{\beta}_{{j_1}}^{0}\right)^T(H_j)_{12}\boldsymbol{\beta}_{\cdot j_2}^0-\left( \hat{\boldsymbol{\beta}}_{{j_2}}^{ols}-\boldsymbol{\beta}_{{j_2}}^{0}\right)^T(H_j)_{22}\boldsymbol{\beta}_{\cdot j_2}^0\\
& -\frac{1}{2} \left( \hat{\boldsymbol{\beta}}_{\cdot j_1}^{H_{0}}-\boldsymbol{\beta}_{\cdot j_1}^0\right)^T(H_{j}^0)_{11} \left( \hat{\boldsymbol{\beta}}_{\cdot j_1}^{H_{0}}-\boldsymbol{\beta}_{\cdot j_1}^0\right)-\frac{1}{2} \left( \boldsymbol{\beta}_{\cdot j_2}^0\right)^T(H_{j}^0)_{22} \boldsymbol{\beta}_{\cdot j_2}^0+ \left( \hat{\boldsymbol{\beta}}_{\cdot j_1}^{H_{0}}-\boldsymbol{\beta}_{\cdot j_1}^0\right)^T(H_{j}^0)_{12} \boldsymbol{\beta}_{\cdot j_2}^0\\
&+o(\|\hat{\boldsymbol{\beta}}^{H_{0}}_{\cdot j}-\hat{\boldsymbol{\beta}}^{0}_{\cdot j}\|^2)\\
=&\left(\hat{\boldsymbol{\beta}}_{{j_1}}^{ols}-\boldsymbol{\beta}_{{j_1}}^{0}\right)^T(H_j)_{11}\left( \left(\hat{\boldsymbol{\beta}}_{{j_1}}^{ols}-\boldsymbol{\beta}_{{j_1}}^{0}\right)+(H_j)^{-1}_{11}(H_j)_{12}\left( \hat{\boldsymbol{\beta}}_{{j_2}}^{ols}\right)\right)\\
&+\left( \hat{\boldsymbol{\beta}}_{{j_2}}^{ols}-\boldsymbol{\beta}_{{j_2}}^{0}\right)^T(H_j)_{12}\left( \left(\hat{\boldsymbol{\beta}}_{{j_1}}^{ols}-\boldsymbol{\beta}_{{j_1}}^{0}\right)+(H_j)^{-1}_{11}(H_j)_{12}\left( \hat{\boldsymbol{\beta}}_{{j_2}}^{ols}\right)\right)\\
&-\left(\hat{\boldsymbol{\beta}}_{{j_1}}^{ols}-\boldsymbol{\beta}_{{j_1}}^{0}\right)^T(H_j)_{12}\boldsymbol{\beta}_{\cdot j_2}^0-\left( \hat{\boldsymbol{\beta}}_{{j_2}}^{ols}-\boldsymbol{\beta}_{{j_2}}^{0}\right)^T(H_j)_{22}\boldsymbol{\beta}_{\cdot j_2}^0\\
& -\frac{1}{2} \left( \left(\hat{\boldsymbol{\beta}}_{{j_1}}^{ols}-\boldsymbol{\beta}_{{j_1}}^{0}\right)+(H_j)^{-1}_{11}(H_j)_{12}\left( \hat{\boldsymbol{\beta}}_{{j_2}}^{ols}\right)\right)^T(H_{j}^0)_{11} \left( \left(\hat{\boldsymbol{\beta}}_{{j_1}}^{ols}-\boldsymbol{\beta}_{{j_1}}^{0}\right)+(H_j)^{-1}_{11}(H_j)_{12}\left( \hat{\boldsymbol{\beta}}_{{j_2}}^{ols}\right)\right)\\
&-\frac{1}{2} \left( \boldsymbol{\beta}_{\cdot j_2}^0\right)^T(H_{j}^0)_{22} \boldsymbol{\beta}_{\cdot j_2}^0+ \left( \left(\hat{\boldsymbol{\beta}}_{{j_1}}^{ols}-\boldsymbol{\beta}_{{j_1}}^{0}\right)+(H_j)^{-1}_{11}(H_j)_{12}\left( \hat{\boldsymbol{\beta}}_{{j_2}}^{ols}\right)\right)^T(H_{j}^0)_{12} \boldsymbol{\beta}_{\cdot j_2}^0\\
&+o(\|\hat{\boldsymbol{\beta}}^{H_{0}}_{\cdot j}-\hat{\boldsymbol{\beta}}^{0}_{\cdot j}\|^2)\\
=&\left(\hat{\boldsymbol{\beta}}_{{j_1}}^{ols}-\boldsymbol{\beta}_{{j_1}}^{0}\right)^T(H_j)_{11} \left(\hat{\boldsymbol{\beta}}_{{j_1}}^{ols}-\boldsymbol{\beta}_{{j_1}}^{0}\right)+\left(\hat{\boldsymbol{\beta}}_{{j_1}}^{ols}-\boldsymbol{\beta}_{{j_1}}^{0}\right)^T(H_j)_{12}\left( \hat{\boldsymbol{\beta}}_{{j_2}}^{ols}\right)\\
&+\left( \hat{\boldsymbol{\beta}}_{{j_2}}^{ols}-\boldsymbol{\beta}_{{j_2}}^{0}\right)^T(H_j)_{21} \left(\hat{\boldsymbol{\beta}}_{{j_1}}^{ols}-\boldsymbol{\beta}_{{j_1}}^{0}\right)+\left( \hat{\boldsymbol{\beta}}_{{j_2}}^{ols}-\boldsymbol{\beta}_{{j_2}}^{0}\right)^T(H_j)_{21}(H_j)^{-1}_{11}(H_j)_{12}\left( \hat{\boldsymbol{\beta}}_{{j_2}}^{ols}\right)\\
&-\left(\hat{\boldsymbol{\beta}}_{{j_1}}^{ols}-\boldsymbol{\beta}_{{j_1}}^{0}\right)^T(H_j)_{12}\boldsymbol{\beta}_{\cdot j_2}^0-\left( \hat{\boldsymbol{\beta}}_{{j_2}}^{ols}-\boldsymbol{\beta}_{{j_2}}^{0}\right)^T(H_j)_{22}\boldsymbol{\beta}_{\cdot j_2}^0\\
& -\frac{1}{2} \left(\hat{\boldsymbol{\beta}}_{{j_1}}^{ols}-\boldsymbol{\beta}_{{j_1}}^{0}\right)^T(H_{j})_{11}\left(\hat{\boldsymbol{\beta}}_{{j_1}}^{ols}-\boldsymbol{\beta}_{{j_1}}^{0}\right)
-(\hat{\boldsymbol{\beta}}_{{j_2}}^{ols})^T (H_{j})_{12}  \left(\hat{\boldsymbol{\beta}}_{{j_1}}^{ols}-\boldsymbol{\beta}_{{j_1}}^{0}\right)\\
&-\frac{1}{2}\left( \hat{\boldsymbol{\beta}}_{{j_2}}^{ols}\right)^T(H_j)_{21}(H_j)^{-1}_{11}(H_j)_{12}\left( \hat{\boldsymbol{\beta}}_{{j_2}}^{ols}\right)-\frac{1}{2} \left( \boldsymbol{\beta}_{\cdot j_2}^0\right)^T(H_{j}^0)_{22} \boldsymbol{\beta}_{\cdot j_2}^0\\
&-\frac{1}{2} \left( \boldsymbol{\beta}_{\cdot j_2}^0\right)^T(H_{j}^0)_{22} \boldsymbol{\beta}_{\cdot j_2}^0+  \left(\hat{\boldsymbol{\beta}}_{{j_1}}^{ols}-\boldsymbol{\beta}_{{j_1}}^{0}\right)^T(H_{j}^0)_{12} \boldsymbol{\beta}_{\cdot j_2}^0+\left( \hat{\boldsymbol{\beta}}_{{j_2}}^{ols}\right)^T(H_j)_{21}(H_j)^{-1}_{11}(H_{j}^0)_{12} \boldsymbol{\beta}_{\cdot j_2}^0\\
&+o(\|\hat{\boldsymbol{\beta}}^{H_{0}}_{\cdot j}-\hat{\boldsymbol{\beta}}^{0}_{\cdot j}\|^2)\\
=&\frac{1}{2} \left(\hat{\boldsymbol{\beta}}_{{j_1}}^{ols}-\boldsymbol{\beta}_{{j_1}}^{0}\right)^T(H_{j})_{11}\left(\hat{\boldsymbol{\beta}}_{{j_1}}^{ols}-\boldsymbol{\beta}_{{j_1}}^{0}\right)+\frac{1}{2}\left( \hat{\boldsymbol{\beta}}_{{j_2}}^{ols}\right)^T(H_j)_{21}(H_j)^{-1}_{11}(H_j)_{12}\left( \hat{\boldsymbol{\beta}}_{{j_2}}^{ols}\right)\\
&+\left( \hat{\boldsymbol{\beta}}_{{j_2}}^{ols}-\boldsymbol{\beta}_{{j_2}}^{0}\right)^T(H_j)_{21} \left(\hat{\boldsymbol{\beta}}_{{j_1}}^{ols}-\boldsymbol{\beta}_{{j_1}}^{0}\right)-\left( \hat{\boldsymbol{\beta}}_{{j_2}}^{ols}-\boldsymbol{\beta}_{{j_2}}^{0}\right)^T(H_j)_{22}\boldsymbol{\beta}_{\cdot j_2}^0\\
&-\frac{1}{2} \left( \boldsymbol{\beta}_{\cdot j_2}^0\right)^T(H_{j}^0)_{22} \boldsymbol{\beta}_{\cdot j_2}^0+o(\|\hat{\boldsymbol{\beta}}^{H_{0}}_{\cdot j}-\hat{\boldsymbol{\beta}}^{0}_{\cdot j}\|^2).
\end{align*}
Then under $ H_{\alpha} $, the $ j $-th part of log-likelihood ratio $ lr_j $ can be rewrite as
\begin{align*}
lr_j =&\frac{1}{2} \left( \hat{\boldsymbol{\beta}}_{\cdot j_1}^{ols}-\boldsymbol{\beta}_{\cdot j_1}^0\right)^T\left( H_{j}\right)_{11} \left( \hat{\boldsymbol{\beta}}_{\cdot j_1}^{ols}-\boldsymbol{\beta}_{\cdot j_1}^0\right)+\left( \hat{\boldsymbol{\beta}}_{\cdot j_1}^{ols}-\boldsymbol{\beta}_{\cdot j_1}^0\right)^T(H_j)_{12}\left( \hat{\boldsymbol{\beta}}_{{j_2}}^{ols}-\boldsymbol{\beta}_{{j_2}}^{0}\right)\\
&+\frac{1}{2} \left( \hat{\boldsymbol{\beta}}_{\cdot j_2}^{ols}-\boldsymbol{\beta}_{\cdot j_2}^0\right)^T\left( H_{j}\right)_{22} \left( \hat{\boldsymbol{\beta}}_{\cdot j_2}^{ols}-\boldsymbol{\beta}_{\cdot j_2}^0\right)\\
&-\frac{1}{2} \left(\hat{\boldsymbol{\beta}}_{{j_1}}^{ols}-\boldsymbol{\beta}_{{j_1}}^{0}\right)^T(H_{j})_{11}\left(\hat{\boldsymbol{\beta}}_{{j_1}}^{ols}-\boldsymbol{\beta}_{{j_1}}^{0}\right)-\frac{1}{2}\left( \hat{\boldsymbol{\beta}}_{{j_2}}^{ols}\right)^T(H_j)_{21}(H_j)^{-1}_{11}(H_j)_{12}\left( \hat{\boldsymbol{\beta}}_{{j_2}}^{ols}\right)\\
&-\left( \hat{\boldsymbol{\beta}}_{{j_2}}^{ols}-\boldsymbol{\beta}_{{j_2}}^{0}\right)^T(H_j)_{21} \left(\hat{\boldsymbol{\beta}}_{{j_1}}^{ols}-\boldsymbol{\beta}_{{j_1}}^{0}\right)+\left( \hat{\boldsymbol{\beta}}_{{j_2}}^{ols}-\boldsymbol{\beta}_{{j_2}}^{0}\right)^T(H_j)_{22}\boldsymbol{\beta}_{\cdot j_2}^0\\
&-\frac{1}{2} \left( \boldsymbol{\beta}_{\cdot j_2}^0\right)^T(H_{j}^0)_{22} \boldsymbol{\beta}_{\cdot j_2}^0+o(\|\hat{\boldsymbol{\beta}}^{H_{\alpha}}_{\cdot j}-\hat{\boldsymbol{\beta}}^{H_0}_{\cdot j}\|^2)\\
=&\frac{1}{2}\left( \hat{\boldsymbol{\beta}}_{{j_2}}^{ols}\right)^T\left[ (H_j)_{22}-(H_j)^T_{12}(H_j)_{11}^{-1}(H_j)_{12}\right] \left( \hat{\boldsymbol{\beta}}_{{j_2}}^{ols}\right)+o(\|\hat{\boldsymbol{\beta}}^{H_{\alpha}}_{\cdot j}-\hat{\boldsymbol{\beta}}^{H_0}_{\cdot j}\|^2).
\end{align*}
For (i), if $ |D^0| =0$, for any linkage in $ F $ it form a cycle with the linkages in $ E^0 $, while since $ \boldsymbol{\beta}^n $ is acyclic, then when $ \boldsymbol{\beta}^n=\boldsymbol{\beta}^0 $ and $ h=0 , \pi_n(h)=1$.
For (ii), assume $ |D^0|>0 $ is fixed, Let $ V_j=\{i:i\in E^0\cup D^0,\boldsymbol{\beta}_{ji}\neq 0\} $, $ V_{j_1}=\{i:i\in E^0\setminus D^0,\boldsymbol{\beta}_{ji}\neq 0\} $ and $ V_{j_2}=\{i:i\in  D^0,\boldsymbol{\beta}_{ji}\neq 0\} $.
	From the proof of Theorem \ref{thm1}, we know that when $ |D^0|>0 $ is fixed, under $ H_{\alpha} $ is true, 
	\[
	\left( \hat{\boldsymbol{\beta}}_{{j_2}}^{ols}\right)^T\left[ (H_j)_{22}-(H_j)^T_{12}(H_j)_{11}^{-1}(H_j)_{12}\right] \left( \hat{\boldsymbol{\beta}}_{{j_2}}^{ols}\right)\xrightarrow{d}\chi^2_{|D^0|}(u^2),
	\]
	which is $ 2lr$ follows a non-central chi-square distribution with degree of freedom $ |D^0| $ and non-central parameter $ u^2= \sum_{j=1}^{p}{\boldsymbol{\beta}}_{D_j^0}^{0T}\left[(H_j)_{22}-(H_j)^T_{12}(H_j)_{11}^{-1}(H_j)_{12}\right] \boldsymbol{\beta}_{D_j^0}^{0}$.\\
	When analysis the power of the linkage testing, we have $ H_{\alpha}: \boldsymbol{\beta}_{jk}=\boldsymbol{\beta}^0_{jk}+\boldsymbol{\delta}_{jk}^n  $ for $ (j,k)\in F $, where $ \boldsymbol{\delta}^n=(\delta_{jk}^n)_{p\times p} $ satisfies: $ \boldsymbol{\delta}_{F^c}^n=0 $, and $ ||\boldsymbol{\delta}^n||_F=n^{-1/2}h $ if $ |D^0| $ is fixed and $ ||\boldsymbol{\delta}^n||_F=|D^0|^{1/4}n^{-1/2}h $ if $ |D^0|\to \infty $.
	Under Assumption \ref{asm2},we have that 
	\[
	\begin{array}{ll}
	u^2=\sum_{j=1}^{p}(\sqrt{n}{\boldsymbol{\beta}}_{D_j^0}^{0T})\left[(H_j/n)_{22}-(H_j/n)^T_{12}(H_j/n)_{11}^{-1}(H_j/n)_{12}\right] \sqrt{n}\boldsymbol{\beta}_{D_j^0}^{0}\geq c_2h^2 & |D^0|>0 \text{ is fixed,}\\
	u^2=\sum_{j=1}^{p}(\sqrt{n}{\boldsymbol{\beta}}_{D_j^0}^{0T})\left[ (H_j/n)_{22}-(H_j/n)^T_{12}(H_j/n)_{11}^{-1}(H_j/n)_{12}\right] \sqrt{n}\boldsymbol{\beta}_{D_j^0}^{0}\geq |D^0|^{1/2}c_2h^2 & |D^0|\to \infty.\\
	\end{array}
	\]
	Then when $ |D^0|>0$ is fixed and $ h\to \infty $ the local power of the peoposed test is
	\[
	\lim\inf_{n\to \infty}P_{\boldsymbol{\beta}^n}\left( 2lr >\chi^2_{|D^0|,1-\alpha}\right)\geq P\left(\chi^2_{|D^0|}(c_2h^2)\geq \chi^2_{|D^0|,1-\alpha} \right) \to 1.
	\]
	When $ |D^0|\to \infty $, as $ h\to \infty $,
	\begin{align*}
	\lim\inf_{n\to \infty}&P_{\boldsymbol{\beta}^n}\left( (2|D^0|)^{-1/2}\left( 2lr-|D^0|\right) >z_{1-\alpha}\right) \\
	&\geq P\left(\chi^2_{|D^0|}(c_2|D^0|^{1/2}h^2)\geq (2|D^0|)^{1/2}z_{1-\alpha} +|D^0|\right) \\
	&=P\left( N(0,1)> \frac{ (2|D^0|)^{1/2}z_{1-\alpha} +|D^0|-|D^0|-c_2 |D^0|^{1/2}h^2}{\sqrt{2\left( |D^0|+c_2 |D^0|^{1/2}h^2\right) }}\right) \\
	&\geq P\left( N(0,1)> z_{1-\alpha}-c_2h^2\right) \to 1,
	\end{align*}
	under the fact that 
	\[
	\frac{\chi^2_r(u^2)-r-u^2}{\sqrt{2(r+u^2)}}\xrightarrow{d}N(0,1).
	\]
	This completes the proof.
\end{prf}

\begin{prf}[Proof of theorem \ref{thm4}]
	Suppose that $ E\cup F $ is cyclic, then $ \boldsymbol{\beta}^n=\boldsymbol{\beta}^0 $ and $ h=0 $, then proof of (i) is established.\\
	When $ d>0 $ is fixed, we have 
	\[ 
 lr(k)=l\left( \hat{\boldsymbol{\beta}}_{H_{\alpha}}-l\left( \hat{\boldsymbol{\beta}}_{H_0}(k)\right)  \right); \quad (i_k,i_{k+1})\in A.
 \]
	Set $ B=\{k:(i_k,i_{k+1})\in A\} $ and assume $ B=\{1,\cdots,|A|\} $, then for $ k\in B $, we have
	\begin{align*}
	lr(k)=l(\hat{\boldsymbol{\beta}}^{H_{\alpha}})-l(\tilde{\boldsymbol{\beta}}^{H_{0}}_{-i_{k}i_{k+1}},\boldsymbol{\delta}^n_{i_{k}i_{k+1}})+l(\tilde{\boldsymbol{\beta}}^{H_{0}}_{-i_{k}i_{k+1}},\boldsymbol{\delta}^n_{i_{k}i_{k+1}})-l(\hat{\boldsymbol{\beta}}^{H_{0}}_{-i_{k}i_{k+1}},0),
	\end{align*}
	where $ 2l(\hat{\boldsymbol{\beta}}^{H_{\alpha}})-2l(\tilde{\boldsymbol{\beta}}^{H_{0}}_{-i_{k}i_{k+1}},\boldsymbol{\delta}^n_{i_{k}i_{k+1}}) \xrightarrow{d}\chi^2_1$ and $ 2l(\tilde{\boldsymbol{\beta}}^{H_{0}}_{-i_{k}i_{k+1}},\boldsymbol{\delta}^n_{i_{k}i_{k+1}})-2l(\hat{\boldsymbol{\beta}}^{H_{0}}_{-i_{k}i_{k+1}},0)=(\frac{1}{n}H)_{i_ki_k}h^2 $. Then for the local power of pathway test, as $ h \to \infty $ we have
	\begin{align*}
	\lim\inf_{n\to \infty} &P_{\boldsymbol{\beta}^n}\left(2lr>\Gamma_{d,1-\alpha} \right) \\
	&\geq \lim\inf_{n\to \infty} P\left(g(\chi^2_1(h^2(\frac{1}{n}H)_{i_1i_1}),\cdots,\chi^2_1(h^2(\frac{1}{n}H)_{i_di_d}))>\Gamma_{d,1-\alpha} \right)  \to 1,
	\end{align*}
	where $ g({\bf x})=\min_{1\leq k\leq d}x_k; {\bf x}=(x_1,\cdots,x_d)\in \mathbb{R}^d $.
	
	Similarly, when $ d\to \infty $ and $ h\to \infty $,
	\begin{align*}
	\lim\inf_{n\to \infty} &P_{\boldsymbol{\beta}^n}\left(2lr>d^{-2}\Gamma_{d,1-\alpha} \right) \\
	&\geq \lim\inf_{n\to \infty} P\left(d^{2}g(\chi^2_1(h^2(\frac{1}{n}H)_{i_1i_1}),\cdots,\chi^2_1(h^2(\frac{1}{n}H)_{i_di_d}))>\Gamma_{d,1-\alpha} \right)  \to 1.
	\end{align*}
	We complete the proof.
\end{prf}
\begin{prf}[Proof of Theorem \ref{thm5}]
	 In this part we prove the convergence of ADMM and convergence of DC programing.
	Firstly we take a look at Assumption \ref{olsasm1} and Assumption \ref{olsasm2} of \cite{boyd_2011}.
	\begin{assum}
		\label{olsasm1}. The (extended-real-valued) functions $ f:\mathbb{R}^n\to \mathbb{R}\cup {+\infty} $ and $ g:\mathbb{R}^m\to \mathbb{R}\cup {+\infty} $ are closed. proper, and convex.
	\end{assum}
	\begin{assum}
		\label{olsasm2} The unanugmented Lagrangian $ L_0 $ has a saddle point.
	\end{assum}
	Then under Assumption \ref{olsasm1} and Assumption \ref{olsasm2} the ADMM iterates satisfy residual convergence, objective convergence amd dual variable convergence.
	Then for ADMM part, \eqref{iter} is a two-block ADMM. Our convex subproblem at $ m^{th} $ iteration is
	\begin{align*}
	& \min_{\boldsymbol{\beta}}\quad R(\boldsymbol{\beta})+\mu\sum_{(j,k)\in F^c}|A_{jk}|\mathbb{I}\left(|\beta_{jk}^{(m-1)}|\leq \tau\right),\\
	&\text {subj to } \boldsymbol{\beta}_{E_1}=0,\boldsymbol{\beta}-A=\bm{0},\\
	&A_{jk}\mathbb{I}\left(|\beta_{jk}^{(m-1)}|\leq \tau\right)+\xi_{ijk}^1-  \tau\lambda_{ki}-\tau I(i\neq j)+\tau \lambda_{ij}+\tau\mathbb{I}\left(|\beta_{jk}^{(m-1)}|> \tau\right)=0,\\
	&A_{jk}\mathbb{I}\left(|\beta_{jk}^{(m-1)}|\leq \tau\right)-\xi_{ijk}^2+\tau\lambda_{ki}+\tau I(i\neq j)-\tau \lambda_{ij}-\tau\mathbb{I}\left(|\beta_{jk}^{(m-1)}|> \tau\right)=0, \\
	& \xi_{ijk}^1,\xi_{ijk}^2>0, i,j,k=1,\dots,p,k\neq j .
	\end{align*}
	Let $ g(A)=\mu\sum_{(j,k)\in F^c}|A_{jk}|\mathbb{I}\left(|\beta_{jk}^{(m-1)}|\leq \tau\right) $ and let $ h(\boldsymbol{\beta},\Gamma,\xi)=R(\boldsymbol{\beta}) $. Then treating $ (\boldsymbol{\beta},\Gamma,\xi) $ as a variable, \eqref{iter} minimizes the objection function in $ (\boldsymbol{\beta},\Gamma,\xi) $ direction at each ADMM iteration. Note that $ g $ and $ h $ are convex, proper, and closed. The linear contraints implies the existence of Lagrangian multipliers of unaugmented lagrangian $ L_0 $ defined in Section 3.2 of \cite{boyd_2011}, which implies the existence of saddle points. Then \eqref{iter} reduced to a two-block ADMM satisfies Assumption 1 and Assumption \ref{asm2}, Then convergences is established.\\
	For the DC programming part, note that the KKT conditions imply that there exists Lagrangian multipliers $ \mu>0 $ and $ \bm{v}=\{v_{ijk}\leq 0\}_{i,j,k=1\dots,p; k\neq j} $ such that $ (\boldsymbol{\beta}^{(m^{*}-1)},\Gamma^{(m^{*}-1)}) $ minimizes the Lagrangian function at $ m^{*th} $ iteration,
	\begin{align*}
	&f(\boldsymbol{\beta},\Lambda)= R(\boldsymbol{\beta})+\mu\left( \sum_{(j,k)\in F^c}J_{\tau}(\beta_{jk})-\kappa\right) +\sum_{i,j,k=1\dots,p; k\leq j}v_{ijk}\left(J_{\tau}(\beta_{jk})-\lambda_{ki}-\mathbb{I}(i\neq j)+\lambda_{ij} \right) ,
	\end{align*}
	with respect to $ \boldsymbol{\beta} $. For the constrained MLEs of likelihood functions of testing linkages and directed pathways, 
	\[
	f(\boldsymbol{\beta}^{(m)},\Gamma^{(m)})=f^{(m+1)}(\boldsymbol{\beta}^{(m)},\Gamma^{(m)})\leq f^{(m)}(\boldsymbol{\beta}^{(m)},\Gamma^{(m)})\leq f^{(m)}(\boldsymbol{\beta}^{(m-1)},\Gamma^{(m-1)})=f(\boldsymbol{\beta}^{(m-1)},\Gamma^{(m-1)}),
	\]
	where $ f^{(m)} $ is the difference convex objective function at iteration $ m $. By monotonicity, \[
	\lim_{m\to \infty}f(\boldsymbol{\beta}^{(m)},\Gamma^{(m)})=f(\boldsymbol{\beta}^{(m^{*})},\Gamma^{(m^{*})}).
	\]
	Finally, finite step convergence follows from strict decreasingness of $ f(\boldsymbol{\beta}^{(m)},\Gamma^{(m)}) $ in $ m $ and finite possible values of subgradient of the trailing convex function.
	At termination $ f(\boldsymbol{\beta}^{(m^{*})},\Gamma^{(m^{*})})=f(\boldsymbol{\beta}^{(m^{*}-1)},\Gamma^{(m^{*}-1)}) $, otherwise the iteration continues. It can be verified that $ (\boldsymbol{\beta}^{(m)},\Gamma^{(m)}) $ satisfies the local optimality condition, means that the acyclicity requirement for $ \boldsymbol{\beta} $ is met when $ \tau >0$ is sufficiently small, therefore, the estimated graph is a DAG as desired.
\end{prf}
\end{document}